\documentclass[a4paper]{scrartcl}
\usepackage[utf8]{inputenc}
\usepackage{amsmath}
\usepackage{amssymb}
\usepackage{graphicx}
\usepackage{algpseudocode,algorithm,algorithmicx}
\usepackage{float}
\usepackage{url}
\usepackage{booktabs}
\usepackage{array}
\usepackage{paralist}
\usepackage{verbatim}
\usepackage{subfig}
\usepackage{hyperref}
\usepackage[binary-units = true]{siunitx}
\usepackage{csvsimple}
\usepackage{multirow}
\usepackage[capitalise]{cleveref}

\renewcommand\Re{\operatorname{Re}}

\newcommand*{\symDefine}[2]{\newcommand{{#1}}{{#2}}}
\symDefine{\symPrecision}{p}
\symDefine{\symRegRange}{q}
\symDefine{\symRegVal}{k}
\symDefine{\symRegValVariate}{K}
\symDefine{\symNumReg}{m}
\symDefine{\symDataItem}{D}
\symDefine{\symDistanceMeasure}{D}
\symDefine{\symBitRepA}{a}
\symDefine{\symBitRepB}{b}
\symDefine{\symNumBitsMinwiseHashing}{b}
\symDefine{\symImaginary}{\mathrm{i}}
\symDefine{\symIndexI}{i}
\symDefine{\symIndexJ}{j}
\symDefine{\symIndexL}{l}
\symDefine{\symCount}{c}
\symDefine{\symCountVariate}{C}
\symDefine{\symAlpha}{\alpha}
\symDefine{\symBeta}{\beta}
\symDefine{\symA}{a}
\symDefine{\symB}{b}
\symDefine{\symS}{s}
\symDefine{\symX}{x}
\symDefine{\symY}{y}
\symDefine{\symZ}{z}
\symDefine{\symKappa}{\kappa}
\symDefine{\symPhi}{\varphi}
\symDefine{\symXEstimate}{\hat{\symX}}
\symDefine{\symCardinality}{n}
\symDefine{\symCardinalityEstimate}{\hat{\symCardinality}}
\symDefine{\symCardinalityRawEstimate}{\symCardinalityEstimate_\text{raw}}
\symDefine{\symError}{\varepsilon}
\symDefine{\symStopDelta}{\delta}
\symDefine{\symStopEpsilon}{\varepsilon}
\symDefine{\symBigO}{\mathcal{O}}
\symDefine{\symExpectation}{\mathbb{E}}
\symDefine{\symProbability}{P}
\symDefine{\symProbabilityMass}{\rho}
\symDefine{\symRegProbability}{\gamma}
\symDefine{\symLikelihood}{\mathcal{L}}
\symDefine{\symPoissonRate}{\lambda}
\symDefine{\symPoissonRateEstimate}{\hat{\symPoissonRate}}
\symDefine{\symFunc}{f}
\symDefine{\symFuncPrime}{g}
\symDefine{\symHelper}{h}
\symDefine{\symHelperApprox}{\tilde{\symHelper}}
\symDefine{\symSetA}{A}
\symDefine{\symSetB}{B}
\symDefine{\symSetS}{S}
\symDefine{\symSetX}{X}
\symDefine{\symSetASuffix}{a}
\symDefine{\symSetBSuffix}{b}
\symDefine{\symSetXSuffix}{x}
\symDefine{\symPowerSeriesFunc}{\xi}
\symDefine{\symSmallCorrectionFunc}{\sigma}
\symDefine{\symLargeCorrectionFunc}{\tau}
\symDefine{\symEpsPowerSeriesFunc}{\varepsilon_\symPowerSeriesFunc}
\symDefine{\symBiasCorrectionFunc}{w_{\text{corr}}}

\newcommand{\numformat}[1]{{\num[scientific-notation = true,round-mode = places,round-precision = 3, output-exponent-marker = \ensuremath{\mathrm{E}}]{#1}}}

\newcommand{\numformattwo}[1]{{\num[scientific-notation = fixed,round-mode = places,round-precision = 3,round-integer-to-decimal=true]{#1}}}

\newcommand{\comm}[2]{{\Comment{\parbox[t]{#1\linewidth}{{#2}}}}}

\title{New cardinality estimation algorithms for HyperLogLog sketches}
\author{Otmar Ertl \\ otmar.ertl@gmail.com}
\begin{document}
\maketitle
\let\thefootnote\relax\footnotetext{This paper together with source code for all presented algorithms and simulations is available at \\ \url{https://github.com/oertl/hyperloglog-sketch-estimation-paper}. A first version of this paper was published there on April 17, 2016.}
\begin{abstract}
This paper presents new methods to estimate the cardinalities of data sets recorded by HyperLogLog sketches. A theoretically motivated extension to the original estimator is presented that eliminates the bias for small and large cardinalities. Based on the maximum likelihood principle a second unbiased method is derived together with a robust and efficient numerical algorithm to calculate the estimate. The maximum likelihood approach can also be applied to more than a single HyperLogLog sketch. In particular, it is shown that it gives more precise cardinality estimates for union, intersection, or relative complements of two sets that are both represented by HyperLogLog sketches compared to the conventional technique using the inclusion-exclusion principle. All the new methods are demonstrated and verified by extensive simulations.
\end{abstract}

\section{Introduction}
Counting the number of distinct elements in a data stream or large datasets is a common problem in big data processing. In principle, finding the number of distinct elements $\symCardinality$ with a maximum relative error $\symError$  in a data stream requires $\symBigO(\symCardinality)$ space \cite{Alon1999}. However, probabilistic algorithms that achieve the requested accuracy only with high probability are able to drastically reduce space requirements. Many different probabilistic algorithms have been developed over the past two decades \cite{Metwally2008,Ting2014} until a theoretically optimal algorithm was finally found \cite{Kane2010}. Although this algorithm achieves the optimal space complexity of $\symBigO(\symError^{-2}+\log \symCardinality)$ \cite{Alon1999, Indyk2003}, it is not very efficient in practice \cite{Ting2014}.

More practicable and already widely used in many applications is the HyperLogLog algorithm \cite{Flajolet2007} with a near-optimal space complexity $\symBigO(\symError^{-2} \log\log\symCardinality +\log \symCardinality)$. It has the nice property that partial results can be easily merged, which is a necessity for distributed environments. Unfortunately, the originally proposed estimation method has some problems to guarantee the same accuracy over the full cardinality range. Therefore, a couple of variants have been developed to correct the original estimate by empirical means \cite{Heule2013,Rhodes2015,Sanfilippo2014}. 

A more theoretically profound estimator for HyperLogLog sketches that does not depend on empirical data and significantly improves the estimation error is the historic inverse probability estimator \cite{Ting2014, Cohen2014}. It trades memory efficiency for mergeability. The estimator needs to be continuously updated while inserting elements and the estimate depends on the element insertion order. Moreover, the estimator cannot be further used after merging two sketches, which limits its application to single data streams. If this restriction is acceptable, the self-learning bitmap \cite{Chen2011} 
could also be used, which provides a similar trade-off and also needs less space than the original HyperLogLog method to achieve the same precision.

Sometimes not only the number of distinct elements but also a sample of them is needed in order to allow later filtering according to some predicate and estimating the cardinalities of corresponding subsets. In this case the k-minimum values algorithm \cite{Beyer2007} is the method of choice which also allows set manipulations like the construction of intersections, relative complements, or unions \cite{Dasgupta2015}. The latter operation is the only one that is natively supported by HyperLogLog sketches. A sketch that represents the set operation result is not always needed. One approach to estimate the corresponding result cardinality directly is based on the inclusion-exclusion principle, which however can become quite inaccurate, especially for small Jaccard indices of the input sets \cite{Dasgupta2015}. Alternatively, the HyperLogLog sketches can be combined with some additional data structure that allows estimating the Jaccard index. Together with the HyperLogLog cardinality estimates the error of the intersection size estimate can be reduced \cite{Pascoe2013, Cohen2016}, however, at the expense of a significantly larger memory footprint due to the additional data structure.

\subsection{Outline}
This paper presents new algorithms to extract cardinality information from HyperLogLog sketches. In \cref{sec:hyperloglog_data_structure} we start with an introduction to HyperLogLog sketches and the corresponding update algorithm. We also discuss how the resulting state can be statistically described and approximated by a Poisson model.
In \cref{sec:cardinality_estimation} we describe the original cardinality estimation algorithm, its shortcomings at low and large cardinalities, and previous approaches to overcome them. We present a simple derivation of the original raw estimator and analyze the root cause for its limited operating range. We derive an improved version of the raw estimator, which leads to a new fast cardinality estimation algorithm that works over the full cardinality range as demonstrated by extensive simulations.
In \cref{sec:max_likelihood_estimation} we derive a second even more precise cardinality estimation algorithm based on the maximum likelihood principle which is again verified by simulations. As presented in \cref{sec:cardinality_estimation_set_intersections} the same approach can be generalized to two HyperLogLog sketches, which allows result cardinality estimation of set operations like intersections or complements. The simulation results show that the estimates are significantly better than those of the conventional approach using the inclusion-exclusion principle.
Finally, in \cref{sec:future_work} we discuss open problems and ideas for future work including HyperLogLog's potential application as locality-sensitive hashing algorithm before we conclude in \cref{sec:conclusion}.

\section{HyperLogLog data structure}
\label{sec:hyperloglog_data_structure}
The HyperLogLog algorithm collects information of incoming elements into a 
very compact sketching data structure, that finally allows the estimation of the number of distinct elements. The data structure consists of $\symNumReg = 2^\symPrecision$ registers whose number is chosen to be a power of two for performance reasons. $\symPrecision$ is the precision parameter that directly controls the relative estimation error which scales like $1/\sqrt{\symNumReg}$ \cite{Flajolet2007}. All registers start with zero initial value. Each element insertion potentially increases the value of one of these registers. The maximum value a register can reach is a natural bound given either by the output size of the used hash algorithm or the space that is reserved for a single register. Common implementations allocate up to 8 bits per register.

\subsection{Data element insertion}
\label{sec:data_element_insertion}
The insertion of a data element into a HyperLogLog data structure requires the calculation of a $(\symPrecision+\symRegRange)$-bit hash value. The leading $\symPrecision$ bits of the hash value are used to select one of the $2^\symPrecision$ registers. Among the next following $\symRegRange$ bits, the position of the first 1-bit is determined which is a value in the range $[1,\symRegRange+1]$. The value $\symRegRange+1$ is used, if all $\symRegRange$ bits are zeros. If the position of the first 1-bit exceeds the current value of the selected register, the register value is replaced. \cref{alg:insert} shows the update procedure for inserting a data element into the HyperLogLog sketch.

\begin{algorithm}
\caption{Insertion of a data element $\symDataItem$ into a HyperLogLog data structure that consists of $\symNumReg=2^\symPrecision$ registers. All registers $\boldsymbol{\symRegValVariate} = (\symRegValVariate_1,\ldots,\symRegValVariate_\symNumReg)$ start from zero. $\langle\ldots\rangle_2$ denotes the binary representation of an integer.}
\label{alg:insert}
\begin{algorithmic}
\Procedure {InsertElement}{\symDataItem}
\State $\langle \symBitRepA_1, \ldots, \symBitRepA_\symPrecision,\symBitRepB_1,\ldots,\symBitRepB_\symRegRange\rangle_2 \gets$ $(\symPrecision + \symRegRange)$-bit hash value of $\symDataItem$
\comm{0.25}{$\symBitRepA_\symIndexI,\symBitRepB_\symIndexI\in\{0,1\}$}
\State $\symRegVal \gets \min(\{\symS\mid \symBitRepB_\symS = 1\}\cup {\{\symRegRange+1\}} )$
\comm{0.25}{$\symRegVal\in\lbrace 1,2,\ldots,\symRegRange+1\rbrace$}
\State $\symIndexI \gets 1+ \langle \symBitRepA_1, \ldots, \symBitRepA_\symPrecision\rangle_2$
\comm{0.25}{$\symIndexI\in\lbrace 1,2,\ldots,\symNumReg\rbrace$}
\If{$\symRegVal>\symRegValVariate_\symIndexI$}
\State $\symRegValVariate_\symIndexI \gets\symRegVal$
\EndIf
\EndProcedure
\end{algorithmic}
\end{algorithm}

The described element insertion algorithm makes use of what is known as stochastic averaging \cite{Flajolet1985}. Instead of updating each of all $\symNumReg$ registers using $\symNumReg$ independent hash values, which would be an $\symBigO(\symNumReg)$ operation, only one register is selected and updated, which requires only a single hash function and reduces the complexity to $\symBigO(1)$.

A HyperLogLog sketch can be characterized by a parameter pair $(\symPrecision, \symRegRange)$. The precision parameter $\symPrecision$ controls the relative error while the second parameter defines the possible value range of a register. A register can take all values starting from 0 to $\symRegRange+1$, inclusively. The sum $\symPrecision+\symRegRange$ corresponds to the number of consumed hash value bits and defines the maximum cardinality that can be tracked. Obviously, if the cardinality reaches values in the order of $2^{\symPrecision+\symRegRange}$, hash collisions will become more apparent and the estimation error will increase drastically.

\cref{alg:insert} has some properties which are especially useful for distributed data streams. First, the insertion order of elements has no influence on the final HyperLogLog sketch. Furthermore, any two HyperLogLog sketches with same parameters $(\symPrecision, \symRegRange)$ representing two different sets can be easily merged. The HyperLogLog sketch that represents the union of both sets can be easily constructed by taking the register-wise maximum values as demonstrated by \cref{alg:union}.

\begin{algorithm}
\caption{Merge operation for two HyperLogLog sketches with register values $\boldsymbol{\symRegValVariate}_1 = (\symRegValVariate_{11},\ldots,\symRegValVariate_{1\symNumReg})$ and
$\boldsymbol{\symRegValVariate}_2 = (\symRegValVariate_{21},\ldots,\symRegValVariate_{2\symNumReg})$ representing sets $\symSetS_1$ and $\symSetS_2$, respectively, to obtain the register values $\boldsymbol{\symRegValVariate}' = (\symRegValVariate'_{1},\ldots,\symRegValVariate'_{\symNumReg})$ of a HyperLogLog sketch representing $\symSetS_1\cup\symSetS_2$.}
\label{alg:union}
\begin{algorithmic}
\Function {Merge}{$\boldsymbol{\symRegValVariate}_1$,$\boldsymbol{\symRegValVariate}_2$}
\comm{.4}{$\boldsymbol{\symRegValVariate}_1,\boldsymbol{\symRegValVariate}_2\in\lbrace 0,1,\ldots,\symRegRange+1\rbrace^\symNumReg$}
\State allocate $\boldsymbol{\symRegValVariate'} = (\symRegValVariate'_{1},\ldots,\symRegValVariate'_{\symNumReg})$
\comm{.4}{$\boldsymbol{\symRegValVariate}'\in\lbrace 0,1,\ldots,\symRegRange+1\rbrace^\symNumReg$}
\For{$\symIndexI\gets 1, \symNumReg$}
\State $\symRegValVariate'_{\symIndexI}
\gets\max(\symRegValVariate_{1\symIndexI},\symRegValVariate_{2\symIndexI})$
\EndFor
\State \Return $\boldsymbol{\symRegValVariate'}$
\EndFunction
\end{algorithmic}
\end{algorithm}

At any time a $(\symPrecision, \symRegRange)$-HyperLogLog sketch can be reduced to a $(\symPrecision', \symRegRange')$-HyperLogLog data structure, if $\symPrecision'\leq\symPrecision$ and $\symPrecision'+\symRegRange' \leq\symPrecision + \symRegRange$ is satisfied (see \cref{alg:compress}). This transformation is lossless in a sense that the resulting HyperLogLog sketch is the same as if all elements would have been recorded by a $(\symPrecision', \symRegRange')$-HyperLogLog sketch right from the beginning.

\begin{algorithm}
\caption{Compression of a $(\symPrecision, \symRegRange)$-HyperLogLog sketch with register values $\boldsymbol{\symRegValVariate}=(\symRegValVariate_1,\ldots,\symRegValVariate_\symNumReg)$ into a $(\symPrecision', \symRegRange')$-HyperLogLog sketch with $\symPrecision'\leq\symPrecision$ and $\symPrecision'+\symRegRange' \leq\symPrecision + \symRegRange$.}
\label{alg:compress}
\begin{algorithmic}
\Function {Compress}{$\boldsymbol{\symRegValVariate}$}
\comm{.4}{$\boldsymbol{\symRegValVariate}\in\lbrace 0,1,\ldots,\symRegRange+1\rbrace^\symNumReg$, $\symNumReg = 2^\symPrecision$}
\State allocate $\boldsymbol{\symRegValVariate}' = (\symRegValVariate'_1,\ldots,\symRegValVariate'_{2^{\symPrecision'}})$
\comm{.4}{$\boldsymbol{\symRegValVariate}'\in\lbrace 0,1,\ldots,\symRegRange'+1\rbrace^{\symNumReg'}$, $\symNumReg' = 2^{\symPrecision'}$}
\For{$\symIndexI\gets 1, 2^{\symPrecision'}$}
\State $\symB \gets (\symIndexI-1)\cdot 2^{\symPrecision-\symPrecision'}$
\State $\symIndexJ\gets 1$
\While{$\symIndexJ \leq 2^{\symPrecision-\symPrecision'}
\wedge
\symRegValVariate_{\symB+\symIndexJ}=0
$}
\State $\symIndexJ\gets\symIndexJ+1$
\EndWhile
\If{$\symIndexJ = 1$}
\State $\symRegValVariate'_\symIndexI \gets \min(\symRegValVariate_{\symB+\symIndexJ} + \symPrecision-\symPrecision',\symRegRange'+1)$
\ElsIf{$\symIndexJ\leq2^{\symPrecision-\symPrecision'}$}
\State $\langle \symBitRepA_1, \ldots, \symBitRepA_{\symPrecision-\symPrecision'}\rangle_2 \gets \symIndexJ-1$
\State $\symRegValVariate'_\symIndexI \gets 
\min(\{\symS\mid \symBitRepA_\symS = 1\} \cup \{\symRegRange'+1\})$
\Else
\State $\symRegValVariate'_\symIndexI \gets 0$
\EndIf
\EndFor
\State \Return $\boldsymbol{\symRegValVariate}'$
\EndFunction
\end{algorithmic}
\end{algorithm}

A $(\symPrecision, 0)$-HyperLogLog sketch corresponds to a bit array as used by linear counting \cite{Whang1990}. Each register value can be stored by a single bit in this case. Hence, linear counting can be regarded as a special case of the HyperLogLog algorithm for which $\symRegRange = 0$.

\subsection{Joint probability distribution of register values}

Under the assumption of a uniform hash function, the probability that the register values $\boldsymbol{\symRegValVariate} = (\symRegValVariate_1,\ldots,\symRegValVariate_\symNumReg)$ of a HyperLogLog sketch with parameters $\symPrecision$ and $\symRegRange$ are equal to $\boldsymbol{\symRegVal}=(\symRegVal_1,\ldots,\symRegVal_\symNumReg)$ is given by the corresponding probability mass function
\begin{equation}
\label{equ:multinomialProbabilityMass}
\symProbabilityMass(\boldsymbol{\symRegVal}\vert\symCardinality)
=
\sum_{\symCardinality_1+\ldots+\symCardinality_\symNumReg = \symCardinality} \binom{\symCardinality}{\symCardinality_1,\ldots,\symCardinality_\symNumReg}
\frac{1}{\symNumReg^\symCardinality}\prod_{\symIndexI=1}^\symNumReg \symRegProbability(\symRegVal_\symIndexI\vert\symCardinality_\symIndexI)
\end{equation}
where $\symCardinality$ is the cardinality. The $\symCardinality$ distinct elements are distributed over all $\symNumReg$ registers according to a multinomial distribution with equal probabilities. $\symRegProbability(\symRegVal\vert\symCardinality)$ is the probability that the value of a register is equal to $\symRegVal$, after it was selected $\symCardinality$ times by the insertion algorithm
\begin{equation}
\symRegProbability(\symRegVal\vert\symCardinality) 
:=
\begin{cases}
1 & \symCardinality=0 \wedge \symRegVal = 0\\
0 & \symCardinality=0 \wedge 1\leq\symRegVal\leq\symRegRange+1\\
0 & \symCardinality\geq1 \wedge \symRegVal = 0\\
\left(1-\frac{1}{2^\symRegVal}\right)^\symCardinality - \left(1-\frac{1}{2^{\symRegVal-1}}\right)^\symCardinality & \symCardinality \geq 1 \wedge 1\leq\symRegVal\leq\symRegRange\\
1 - \left(1-\frac{1}{2^{\symRegRange}}\right)^\symCardinality & \symCardinality\geq 1 \wedge \symRegVal = \symRegRange +1.
\end{cases}
\end{equation}

The order of register values $\symRegValVariate_1,\ldots,\symRegValVariate_\symNumReg$ is not important for the estimation of the cardinality. More formally, the multiset $\lbrace\symRegValVariate_1,\ldots,\symRegValVariate_\symNumReg\rbrace$ is a sufficient statistic for $\symCardinality$.
Since the values of the multiset are all in the range $[0, \symRegRange+1]$ the multiset can also be represented as $\lbrace\symRegValVariate_1,\ldots,\symRegValVariate_\symNumReg\rbrace = 0^{\symCountVariate_0}1^{\symCountVariate_1}\cdots\symRegRange^{\symCountVariate_{\symRegRange}}(\symRegRange+1)^{\symCountVariate_{\symRegRange+1}}$ where $\symCountVariate_\symRegVal$ is the multiplicity of value $\symRegVal$. As a consequence, the multiplicity vector $\boldsymbol{\symCountVariate} := (\symCountVariate_0,\ldots,\symCountVariate_{\symRegRange+1})$ is also a sufficient statistic for the cardinality. In addition, this vector contains all the information about the HyperLogLog sketch that is required for cardinality estimation. The two HyperLogLog parameters can be obtained by $\symPrecision = \log_2 (\sum_{\symRegVal=0}^{\symRegRange+1}\symCountVariate_\symRegVal
)$ and $\symRegRange=\dim\boldsymbol{\symCountVariate}-2$, respectively. \cref{alg:sufficient_statistic} shows the calculation of the multiplicity vector $\boldsymbol{\symCountVariate}$ for a given HyperLogLog sketch with register values $\boldsymbol{\symRegValVariate}$.

\begin{algorithm}
\caption{Multiplicity vector extraction from a $(\symPrecision, \symRegRange)$-HyperLogLog sketch with $\symNumReg=2^\symPrecision$ registers having values $\boldsymbol{\symRegValVariate} = (\symRegValVariate_1,\ldots,\symRegValVariate_\symNumReg)$.} 
\label{alg:sufficient_statistic}
\begin{algorithmic}
\Function {ExtractCounts}{$\boldsymbol{\symRegValVariate}$}
\comm{.4}{$\boldsymbol{\symRegValVariate}\in\lbrace 0,1,\ldots,\symRegRange+1\rbrace^\symNumReg$}
\State allocate $\boldsymbol{\symCountVariate} = (\symCountVariate_0,\ldots,\symCountVariate_{\symRegRange+1})$
\State $\boldsymbol{\symCountVariate}\gets \left(0,0,\ldots, 0\right)$
\For{$\symIndexI\gets 1, \symNumReg$}
\State $\symCountVariate_{\symRegValVariate_\symIndexI} \gets \symCountVariate_{\symRegValVariate_\symIndexI} + 1$
\EndFor
\State \Return $\boldsymbol{\symCountVariate}$
\EndFunction
\end{algorithmic}
\end{algorithm}

\subsection{Poisson approximation}
\label{sec:poisson_approximation}
Due to the statistical dependence of the register values, the probability mass function \eqref{equ:multinomialProbabilityMass} makes further analysis difficult. Therefore, a Poisson model can be used \cite{Flajolet2007}, which assumes that the cardinality itself is distributed according to a Poisson distribution
\begin{equation}
\symCardinality \sim \text{Poisson}(\symPoissonRate).
\end{equation}
Under the Poisson model the distribution of the register values is
\begin{align}
\symProbabilityMass(\boldsymbol{\symRegVal}\vert\symPoissonRate) 
&= 
\sum_{\symCardinality=0}^\infty \symProbabilityMass(\boldsymbol{\symRegVal}\vert\symCardinality) e^{-\symPoissonRate}\frac{\symPoissonRate^\symCardinality}{\symCardinality!}
\label{equ:poissonization}
\\
&= 
\sum_{\symCardinality_1=0}^\infty
\cdots
\sum_{\symCardinality_\symNumReg=0}^\infty
\prod_{\symIndexI=1}^\symNumReg
\symRegProbability(\symRegVal_\symIndexI\vert\symCardinality_\symIndexI)e^{-\frac{\symPoissonRate}{\symNumReg}}\frac{\symPoissonRate^{\symCardinality_\symIndexI}}{\symCardinality_\symIndexI!\symNumReg^{\symCardinality_\symIndexI}}
\nonumber\\
&= 
\prod_{\symIndexI=1}^\symNumReg \sum_{\symCardinality=0}^\infty\symRegProbability(\symRegVal_\symIndexI\vert\symCardinality)e^{-\frac{\symPoissonRate}{\symNumReg}}\frac{\symPoissonRate^\symCardinality}{\symCardinality!\symNumReg^\symCardinality}
\nonumber\\
&= 
\prod_{\symRegVal=0}^{\symRegRange+1} \left(
\sum_{\symCardinality=0}^\infty
\symRegProbability(\symRegVal\vert\symCardinality)e^{-\frac{\symPoissonRate}{\symNumReg}}\frac{\symPoissonRate^\symCardinality}{\symCardinality!\symNumReg^\symCardinality}
\right)^{\!\!\symCount_\symRegVal}
\nonumber\\
\label{equ:poisson_pmf}
&=
e^{-\symCount_0\frac{\symPoissonRate}{\symNumReg}}
\left(\prod_{\symRegVal=1}^{\symRegRange}\left(e^{-\frac{\symPoissonRate}{\symNumReg 2^\symRegVal}}\left(1-e^{-\frac{\symPoissonRate}{\symNumReg 2^\symRegVal}}\right)\right)^{\!\symCount_\symRegVal}\right)
\left(1-e^{-\frac{\symPoissonRate}{\symNumReg 2^\symRegRange}}\right)^{\!\symCount_{\symRegRange+1}}.
\end{align}
Here $\symCount_\symRegVal$ denotes the multiplicity of value $\symRegVal$  in the multiset $\lbrace\symRegVal_1,\ldots,\symRegVal_\symNumReg\rbrace$.
The final factorization shows that the register values $\symRegValVariate_1,\ldots,\symRegValVariate_\symNumReg$ are independent and identically distributed under the Poisson model. The probability that a register has a value less than or equal to $\symRegVal$ for a given rate $\symPoissonRate$ is given by
\begin{equation}
\label{equ:register_value_distribution}
\symProbability(\symRegValVariate \leq \symRegVal\vert\symPoissonRate)
=
\begin{cases}
0 & \symRegVal < 0 \\
e^{-\frac{\symPoissonRate}{\symNumReg 2^\symRegVal}} & 0\leq \symRegVal \leq \symRegRange \\
1 & \symRegVal > \symRegRange.
\end{cases}
\end{equation}

\subsection{Depoissonization}
\label{sec:depoissonization}
Due to the simpler probability mass function \eqref{equ:poisson_pmf}, it is easier to find an estimator $\symPoissonRateEstimate = \symPoissonRateEstimate(\boldsymbol{\symRegValVariate})$ for the Poisson rate $\symPoissonRate$ than for the cardinality $\symCardinality$ under the fixed-size model \eqref{equ:multinomialProbabilityMass}. Depoissonization \cite{Jacquet1998} finally allows to translate the estimates back to the fixed-size model. Assume we have found an unbiased estimator for the Poisson rate
\begin{equation}
\symExpectation(\symPoissonRateEstimate\vert\symPoissonRate) = \symPoissonRate
\quad
\text{for all $\symPoissonRate\geq 0$}.
\end{equation}
We know from \eqref{equ:poissonization} 
\begin{equation}
\symExpectation(\symPoissonRateEstimate\vert\symPoissonRate) = 
\sum_{\symCardinality=0}^\infty \symExpectation(\symPoissonRateEstimate\vert\symCardinality) e^{-\symPoissonRate}\frac{\symPoissonRate^\symCardinality}{\symCardinality!}
\end{equation}
and therefore
\begin{equation}
\sum_{\symCardinality=0}^\infty \symExpectation(\symPoissonRateEstimate\vert\symCardinality) e^{-\symPoissonRate}\frac{\symPoissonRate^\symCardinality}{\symCardinality!}
=
\symPoissonRate
\end{equation}
holds for all $\symPoissonRate\geq0$. The unique solution of this equation is given by
\begin{equation}
\symExpectation(\symPoissonRateEstimate\vert\symCardinality) = \symCardinality.
\end{equation}
Hence, the unbiased estimator $\symPoissonRateEstimate$ conditioned on $\symCardinality$ is also an unbiased estimator for $\symCardinality$, which motivates us to use $\symPoissonRateEstimate$ directly as estimator for the cardinality $\symCardinalityEstimate := \symPoissonRateEstimate$. As simulation results will show later, the Poisson approximation works well over the entire cardinality range, even for estimators that are not exactly unbiased.

\section{Original cardinality estimation method}
\label{sec:cardinality_estimation}
The original cardinality estimator \cite{Flajolet2007} is based on the idea that the number of distinct element insertions a register needs to reach the value $\symRegVal$ is proportional to $\symNumReg  2^{\symRegVal}$. Given that, a rough cardinality estimate can be obtained by averaging the values $\lbrace \symNumReg 2^{\symRegValVariate_1},\ldots,\symNumReg2^{\symRegValVariate_\symNumReg}\rbrace$. 
In the history of the HyperLogLog algorithm different averaging techniques have been proposed. First, there was the LogLog algorithm using the geometric mean and the SuperLogLog algorithm that enhanced the estimate by truncating the largest register values before applying the geometric mean \cite{Durand2003}. Finally, the harmonic mean was found to give even better estimates as it is inherently less sensitive to outliers. The result is the so-called raw estimator which is given by
\begin{equation}
\label{equ:raw_estimator}
\symCardinalityRawEstimate
=
\alpha_\symNumReg
\frac{\symNumReg}
{\frac{1}{\symNumReg 2^{\symRegValVariate_1}}+\ldots+\frac{1}{\symNumReg2^{\symRegValVariate_\symNumReg}}}
= 
\frac{\symAlpha_\symNumReg \symNumReg^2}{\sum_{\symIndexI=1}^{\symNumReg}2^{-\symRegValVariate_\symIndexI}}
= 
\frac{\symAlpha_\symNumReg \symNumReg^2}{\sum_{\symRegVal=0}^{\symRegRange+1}\symCountVariate_\symRegVal 2^{-\symRegVal}}.
\end{equation}
Here $\alpha_\symNumReg$ is a bias correction factor which was derived for a given number of registers $\symNumReg$ as \cite{Flajolet2007}
\begin{equation}
\symAlpha_\symNumReg := \left(
\symNumReg
\int_0^\infty
\left(
\log_2\!\left(
\frac{2+\symX}{1+\symX}
\right)
\right)^{\!\symNumReg}
d\symX
\right)^{\!\!-1}.
\end{equation}
Numerical approximations of $\symAlpha_\symNumReg$ for various values of $\symNumReg$ have been listed in \cite{Flajolet2007}. These approximations are used in many HyperLogLog implementations. However, since the published constants have been rounded to 4 significant digits, these approximations even introduce some bias for very high precisions $\symPrecision$. For HyperLogLog sketches that are used in practice with 256 or more registers ($\symPrecision\geq 8$), it is completely sufficient to use 
\begin{equation}
\label{equ:alpha_inf}
\symAlpha_\infty := \lim_{\symNumReg\rightarrow\infty} \symAlpha_\symNumReg = \frac{1}{2\log 2} 
= 0.72134752\ldots
\end{equation}
as approximation for $\symAlpha_\symNumReg$ in \eqref{equ:raw_estimator}, because the additional bias is negligible compared to the overall estimation error.

\begin{figure}
\centering
\includegraphics[width=1\textwidth]{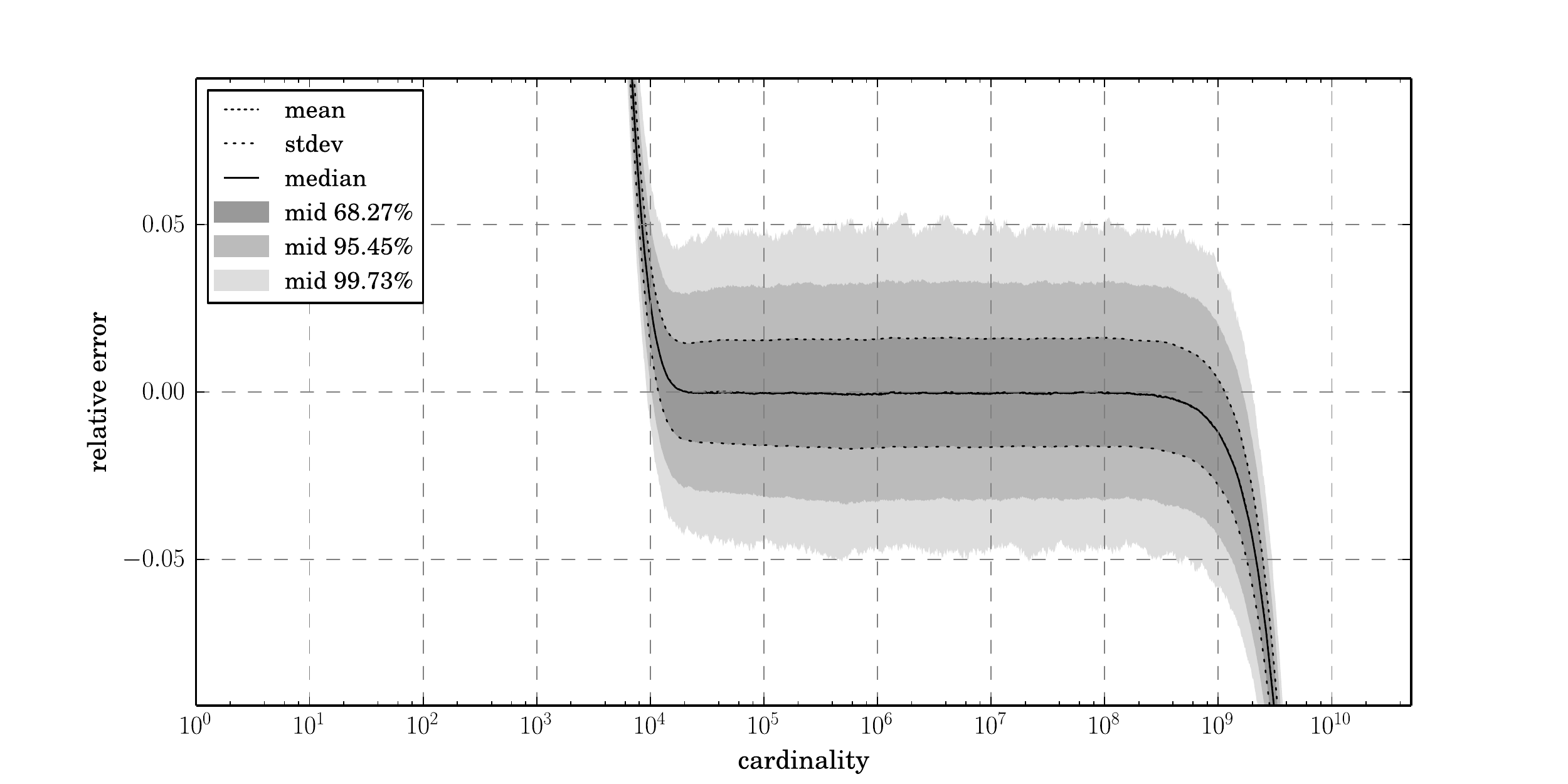}
\caption{The distribution of the relative estimation error over the cardinality for the raw estimator after evaluation of \num{10000} randomly generated HyperLogLog data structures with parameters $\symPrecision = 12$ and $\symRegRange=20$.}
\label{fig:raw_estimate}
\end{figure}

\cref{fig:raw_estimate} shows the distribution of the relative error for the raw estimator as function of the cardinality. The chart is based on \num{10000} randomly generated HyperLogLog sketches. More details of the experimental setup will be explained later in \cref{sec:experimental_setup}. Obviously, the raw estimator is biased for small and large cardinalities where it fails to return accurate estimates. In order to cover the entire cardinality range, corrections for small and large cardinalities have been proposed.

\begin{figure}
\centering
\includegraphics[width=1\textwidth]{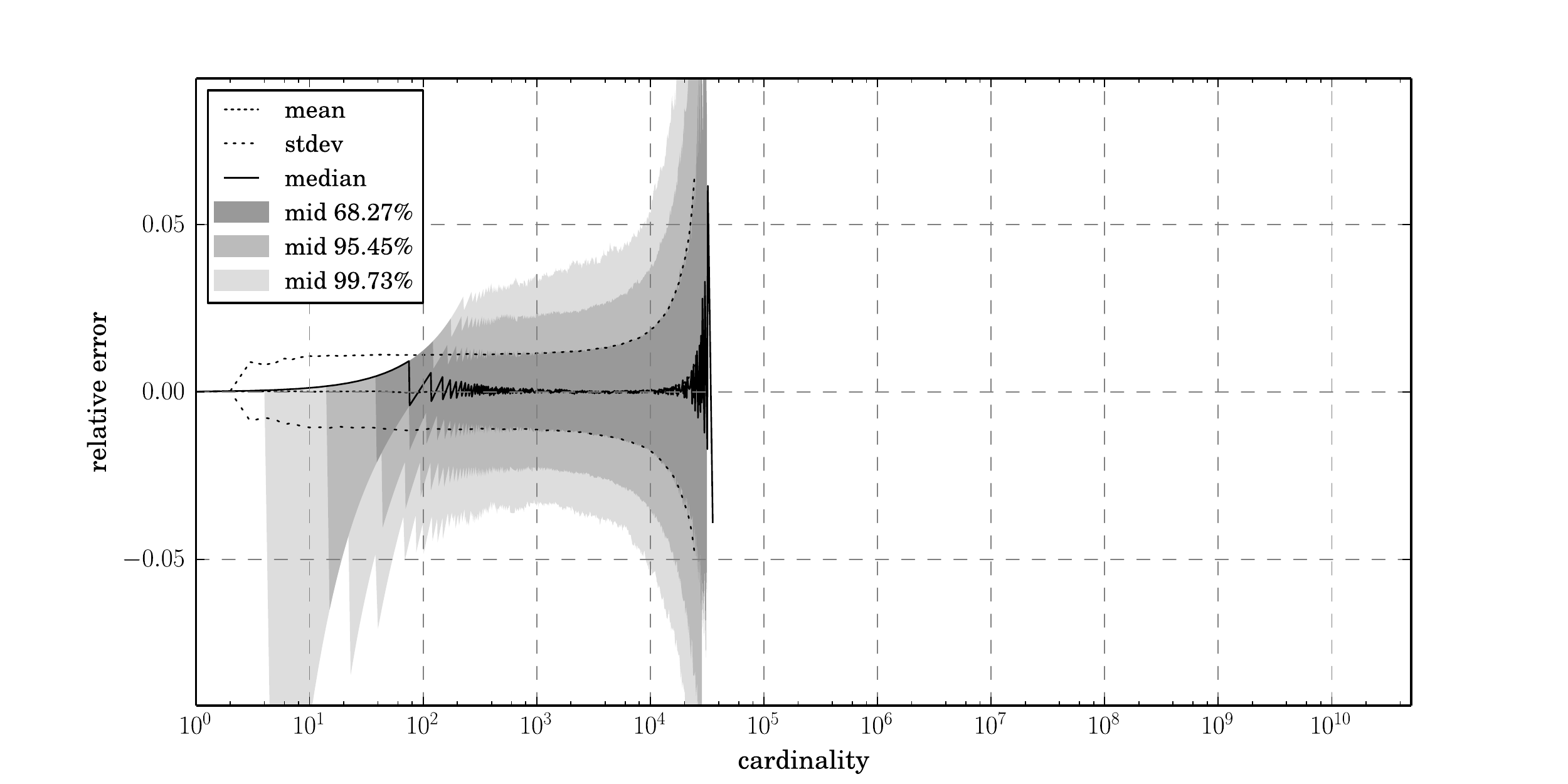}
\caption{The distribution of the relative estimation error for the linear counting estimator after evaluation of \num{10000} randomly generated bitmaps of size $2^{12}$ which correspond to HyperLogLog sketches with parameters $\symPrecision = 12$ and $\symRegRange=0$.}
\label{fig:small_range_estimate}
\end{figure}

As mentioned in \cref{sec:data_element_insertion}, a HyperLogLog sketch with parameters $(\symPrecision, \symRegRange)$ can be mapped to a $(\symPrecision, 0)$-HyperLogLog sketch. Since $\symRegRange=0$ corresponds to linear counting and the reduced HyperLogLog sketch corresponds to a bitset with $\symCountVariate_0$ zeros, the linear counting cardinality estimator \cite{Whang1990} can be used
\begin{equation}
\label{equ:linear_counting_estimator}
\symCardinalityEstimate_\text{small} = \symNumReg \log(\symNumReg/\symCountVariate_0).
\end{equation}
The corresponding relative estimation error as depicted in \cref{fig:small_range_estimate} shows that this estimator is convenient for small cardinalities. It was proposed to use this estimator for small cardinalities as long as $\symCardinalityRawEstimate \leq \frac{5}{2}\symNumReg$ where the factor $\frac{5}{2}$ was empirically determined \cite{Flajolet2007}. 

For large cardinalities in the order of $2^{\symPrecision+\symRegRange}$, for which a lot of registers are already in a saturated state, meaning that they have reached the maximum possible value $\symRegRange + 1$, the raw estimator underestimates the cardinalities. For the 32-bit hash value case $(\symPrecision+\symRegRange=32)$, which was considered in \cite{Flajolet2007}, following correction formula was proposed to take these saturated registers into account
\begin{equation}
\label{equ:large_range_estimate}
\symCardinalityEstimate_\text{large}
=
-2^{32}\log\!\left(1-\symCardinalityRawEstimate/2^{32}\right).
\end{equation}

\begin{figure}
\centering
\includegraphics[width=1\textwidth]{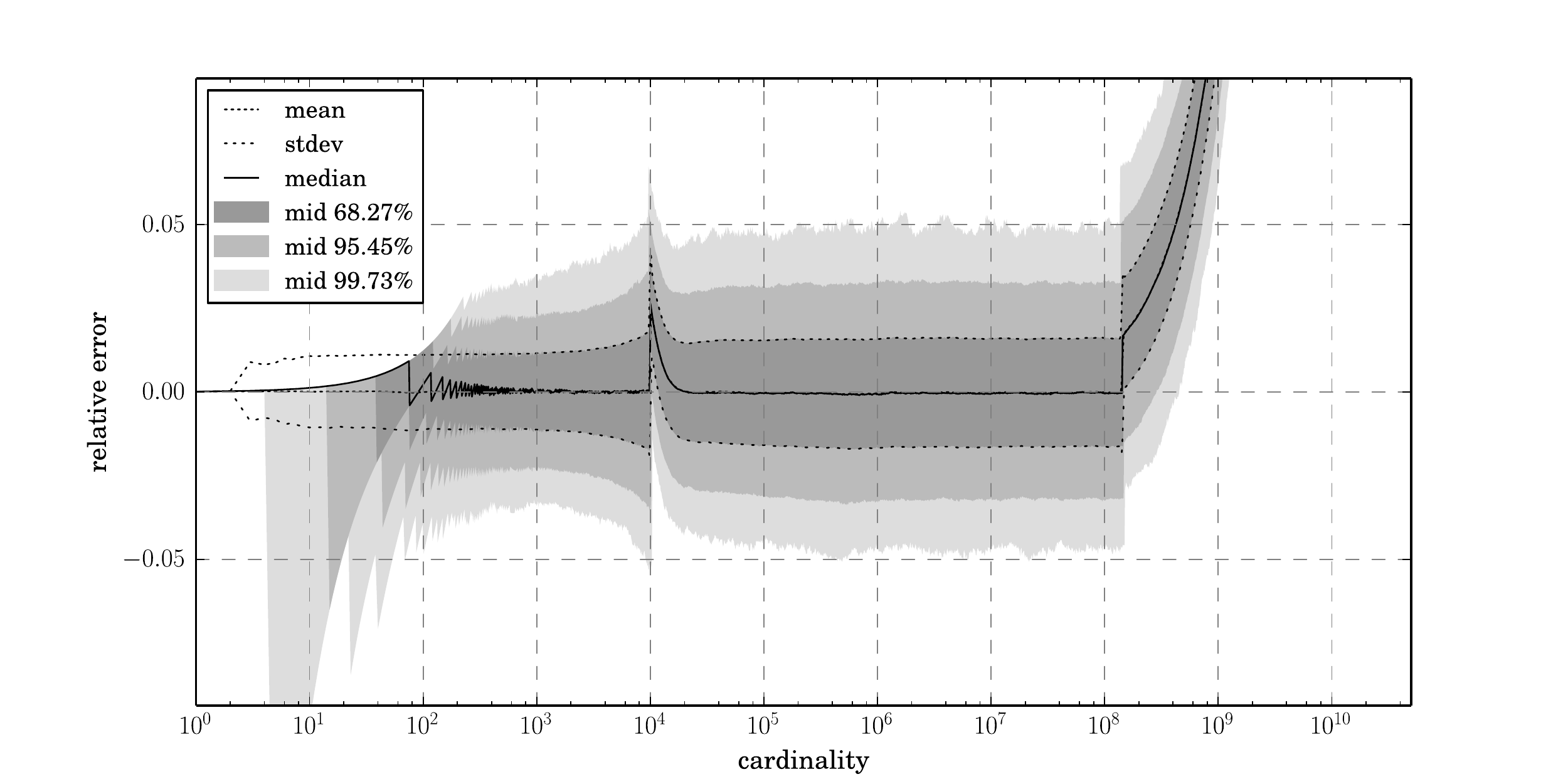}
\caption{The distribution of the relative estimation error over the cardinality for the original estimation algorithm after evaluation of \num{10000} randomly generated HyperLogLog data structures with parameters $\symPrecision = 12$ and $\symRegRange=20$.}
\label{fig:original_estimate}
\end{figure}

The original estimation algorithm as presented in \cite{Flajolet2007} including corrections for small and large cardinalities is summarized by \cref{alg:estimate_original}. 
\begin{algorithm}
\caption{Original cardinality estimation algorithm for HyperLogLog sketches that use 32-bit hash values ($\symPrecision+\symRegRange = 32$) for insertion of data items \cite{Flajolet2007}.}
\label{alg:estimate_original}
\begin{algorithmic}
\Function {EstimateCardinality}{$\boldsymbol{\symRegValVariate}$}
\comm{.4}{$\boldsymbol{\symRegValVariate}\in\lbrace 0,1,\ldots,\symRegRange+1\rbrace^\symNumReg$, $\symNumReg = 2^\symPrecision$}
\State $\symCardinalityRawEstimate
\gets
\symAlpha_\symNumReg \symNumReg^2\left(\sum_{\symIndexI=1}^\symNumReg 2^{-\symRegValVariate_\symIndexI}\right)^{\!-1}$
\comm{.4}{raw estimate \eqref{equ:raw_estimator}}
\If{$\symCardinalityRawEstimate\leq \frac{5}{2}\symNumReg$}
\State $\symCountVariate_0\gets\left|\{\symIndexI \vert\symRegValVariate_\symIndexI=0\}\right|$
\If{$\symCountVariate_0\neq0$}
\State \Return $\symNumReg \log(\symNumReg / \symCountVariate_0)$
\comm{.4}{small range correction \eqref{equ:linear_counting_estimator}}
\Else
\State \Return $\symCardinalityRawEstimate$
\EndIf
\ElsIf{$\symCardinalityRawEstimate\leq \frac{1}{30}2^{32}$}
\State \Return $\symCardinalityRawEstimate$
\Else
\State\Return $-2^{32}\log(1-\symCardinalityRawEstimate/2^{32})$
\comm{.4}{large range correction \eqref{equ:large_range_estimate}}
\EndIf
\EndFunction
\end{algorithmic}
\end{algorithm}
The relative estimation error for the original method is shown in \cref{fig:original_estimate}. Unfortunately, as can be clearly seen, the ranges where the estimation error is small for $\symCardinalityRawEstimate$ and $\symCardinalityEstimate_\text{small}$ do not overlap. Therefore, the estimation error is much larger near the transition region. To reduce the estimation error for cardinalities close to this region, it was proposed to correct $\symCardinalityRawEstimate$ for bias. Empirically collected bias correction data is either stored as set of interpolation points \cite{Heule2013}, as lookup table \cite{Rhodes2015}, or as best-fitting polynomial \cite{Sanfilippo2014}. However, all these empirical approaches treat the symptom and not the cause.

The large range correction formula \eqref{equ:large_range_estimate} is not satisfying either as it does not reduce the estimation error. Quite the contrary, it even makes the bias worse. However, instead of underestimating the cardinalities, they are now overestimated. Another indication for the incorrectness of the proposed large range correction is the fact that it is not even defined for all possible states. For instance, consider a $(\symPrecision,\symRegRange)$-HyperLogLog sketch with $\symPrecision+\symRegRange=32$ for which all registers are equal to the maximum possible value $\symRegRange+1$. The raw estimate would be $\symCardinalityRawEstimate = \symAlpha_\symNumReg 2^{33}$, which is greater than $2^{32}$ and outside of the domain of the large range correction formula.

A simple approach to avoid the need of any large range correction is to extend the operating range of the raw estimator to larger cardinalities. This can be easily accomplished by increasing $\symPrecision+\symRegRange$, which corresponds to using hash values with more bits. Each additional bit doubles the operating range which scales like $2^{\symPrecision+\symRegRange}$. However, in case $\symRegRange \geq 31$ the number of possible register values, which are $\lbrace 0, 1, \ldots, \symRegRange+1\rbrace$, is greater than 32 which is no longer representable by 5-bit registers. Therefore, it was proposed to use 6 bits per register in combination with 64-bit hash values \cite{Heule2013}. Even larger hash values are needless in practice, because it is unrealistic to encounter cardinalities of order $2^{64}$.

\subsection{Derivation of the raw estimator}
\label{sec:derivation_raw_estimator}
In order to better understand why the raw estimator fails for small and large cardinalities, we start with a brief and simple derivation without the restriction to large cardinalities ($\symCardinality\rightarrow\infty$) and without using complex analysis as in \cite{Flajolet2007}.

\begin{figure}
\centering
\includegraphics[width=0.6\textwidth]{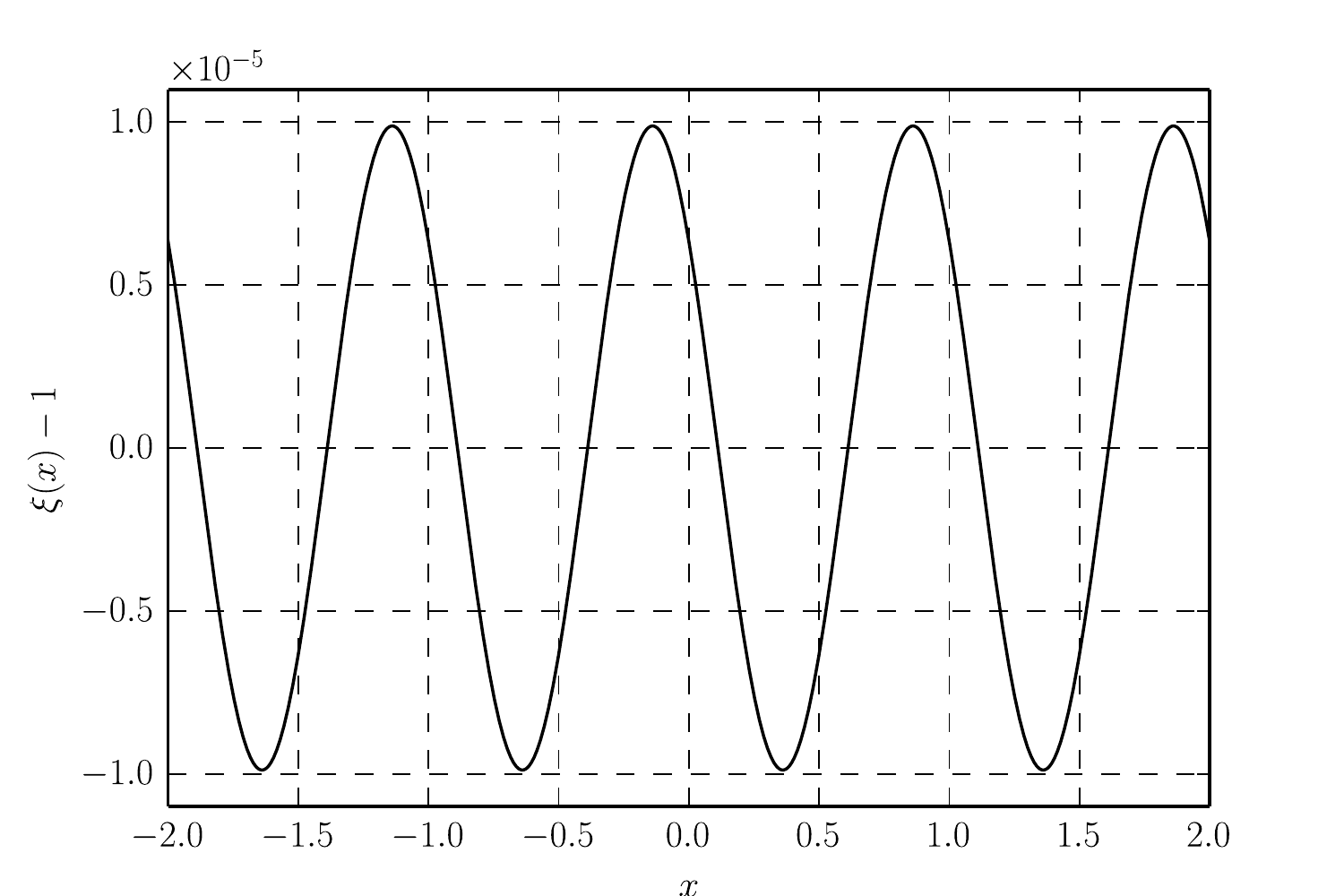}
\caption{The deviation of $\symPowerSeriesFunc(\symX)$ from 1.}
\label{fig:power_series_func}
\end{figure}

Let us assume that the register values have following cumulative distribution function
\begin{equation}
\label{equ:assumed_register_val_distribution}
\symProbability(\symRegValVariate \leq \symRegVal\vert\symPoissonRate) = e^{-\frac{\symPoissonRate}{\symNumReg 2^{ \symRegVal}}}.
\end{equation}
For now we ignore that this distribution has infinite support and differs from the register value distribution under the Poisson model \eqref{equ:register_value_distribution}, whose support is limited to the range $[0, \symRegRange+1]$. For a random variable $\symRegValVariate$ obeying \eqref{equ:assumed_register_val_distribution} the expectation of $2^{-\symRegValVariate}$ is given by
\begin{multline}
\label{equ:expectation_power_two}
\symExpectation(2^{-\symRegValVariate})
=
\\
\sum_{\symRegVal = -\infty}^\infty
2^{-\symRegVal}
\left(
e^{-\frac{\symPoissonRate}{\symNumReg 2^{\symRegVal}}}
-e^{-\frac{\symPoissonRate}{\symNumReg 2^{\symRegVal-1}}}
\right)
=
\frac{1}{2}
\sum_{\symRegVal = -\infty}^\infty
2^{\symRegVal}
e^{-\frac{\symPoissonRate}{\symNumReg} 2^{\symRegVal}}
=
\frac{\symAlpha_\infty\,\symNumReg\,\symPowerSeriesFunc\!\left(\log_2\!\left(\symPoissonRate/\symNumReg\right)\right)}{\symPoissonRate},
\end{multline}
where the function 
\begin{equation}
\label{equ:power_series_function}
\symPowerSeriesFunc(\symX):= \log(2) \sum_{\symRegVal = -\infty}^\infty
2^{\symRegVal+\symX}
e^{-2^{\symRegVal+\symX}}
\end{equation}
is a smooth periodic function with period 1. 
Numerical evaluations indicate that this function can be bounded by $1 - \symEpsPowerSeriesFunc 
\leq \symPowerSeriesFunc(\symX) \leq 1 + \symEpsPowerSeriesFunc$ with $\symEpsPowerSeriesFunc:=\num{9.885e-6}$ (see \cref{fig:power_series_func}). This value can also be found using Fourier analysis as shown in \cref{app:analysis_xi}.

Let $\symRegValVariate_1,\ldots,\symRegValVariate_\symNumReg$ be a sample distributed according to \eqref{equ:assumed_register_val_distribution}. For large sample sizes $\symNumReg\rightarrow \infty$ we have asymptotically
\begin{equation}
\symExpectation\!\left(
\frac{1}{2^{-\symRegValVariate_1}+\ldots+2^{-\symRegValVariate_\symNumReg}}
\right)
\underset{\symNumReg\rightarrow \infty}{=}
\frac{1}{\symExpectation
\!\left(2^{-\symRegValVariate_1}+\ldots+2^{-\symRegValVariate_\symNumReg}\right)}
=
\frac{1}{\symNumReg \symExpectation\!\left(2^{-\symRegValVariate}\right)}.
\end{equation}
Together with \eqref{equ:expectation_power_two} we obtain
\begin{equation}
\symPoissonRate
=
\symExpectation\!\left(
\frac{\symAlpha_\infty\,\symNumReg^2\,\symPowerSeriesFunc\!\left(\log_2\!\left( \symPoissonRate/\symNumReg\right)\right)}{2^{-\symRegValVariate_1}+\ldots+2^{-\symRegValVariate_\symNumReg}}
\right)
\quad
\text{for $\symNumReg\rightarrow \infty$}.
\end{equation}
Therefore, the asymptotic relative bias of 
\begin{equation}
\label{equ:rawestimator_2}
\symPoissonRateEstimate 
= 
\frac{\symAlpha_\infty\,\symNumReg^2}{2^{-\symRegValVariate_1}+\ldots+2^{-\symRegValVariate_\symNumReg}}
\end{equation}
is bounded by $\symEpsPowerSeriesFunc$, which makes this statistic a good estimator for the Poisson parameter. It also corresponds to the raw estimator \eqref{equ:raw_estimator}, if the Poisson parameter estimate is used as cardinality estimate (see \cref{sec:depoissonization}).

\subsection{Limitations of the raw estimator}
The raw estimator is based on two prerequisites. First, the number of registers needs to be sufficiently large ($\symNumReg\rightarrow\infty$). And second, the distribution of register values can be described by \eqref{equ:assumed_register_val_distribution}. However, the latter is not true for small and large cardinalities, which is finally the reason for the bias of the raw estimator.

A random variable $\symRegValVariate'$ with cumulative distribution function \eqref{equ:assumed_register_val_distribution} can be transformed into a random variable $\symRegValVariate$ with cumulative distribution function \eqref{equ:register_value_distribution} using
\begin{equation}
\label{equ:dist_transformation}
\symRegValVariate = \min\!\left(\max\!\left(\symRegValVariate',0\right), \symRegRange+1\right).
\end{equation}
Therefore, register values $\symRegValVariate_1,\ldots,\symRegValVariate_\symNumReg$ can be seen as the result after applying this transformation to a sample $\symRegValVariate'_1,\ldots,\symRegValVariate'_\symNumReg$ of the distribution described by \eqref{equ:assumed_register_val_distribution}. If all registers values fall into the range $[1,\symRegRange]$, they must  be identical to the values $\symRegValVariate'_1,\ldots,\symRegValVariate'_\symNumReg$. In other words, the observed register values are also a plausible sample of the assumed distribution described by \eqref{equ:assumed_register_val_distribution} in this case. Hence, as long as all or at least most register values are in the range $[1,\symRegRange]$, which is the case if $2^\symPrecision \ll \symPoissonRate \ll 2^{\symPrecision
+\symRegRange}$, the approximation of \eqref{equ:register_value_distribution} by \eqref{equ:assumed_register_val_distribution} is valid. This explains why the raw estimator works best for intermediate cardinalities. However, for small and large cardinalities many register values are equal to 0 or $\symRegRange+1$, respectively, which cannot be described by \eqref{equ:assumed_register_val_distribution} and finally leads to the observed bias.

\subsection{Corrections to the raw estimator}
If we knew the values $\symRegValVariate'_1,\ldots,\symRegValVariate'_\symNumReg$ for which transformation \eqref{equ:dist_transformation} led to the observed register values $\symRegValVariate_1,\ldots,\symRegValVariate_\symNumReg$, we would be able to estimate $\symPoissonRate$ using
\begin{equation}
\symPoissonRateEstimate 
= 
\frac{\symAlpha_\infty\,\symNumReg^2}{2^{-\symRegValVariate'_1}+\ldots+2^{-\symRegValVariate'_\symNumReg}}.
\end{equation}
As we have already shown, this estimator is approximately unbiased, because all $\symRegValVariate'_\symIndexI$ follow the assumed distribution. It would be even sufficient, if we knew the multiplicity $\symCountVariate'_\symRegVal$
of each $\symRegVal\in\mathbb{Z}$ in $\lbrace\symRegValVariate'_1,\ldots,\symRegValVariate'_\symNumReg\rbrace$, $\symCountVariate'_\symRegVal := \left|\lbrace \symIndexI\vert \symRegVal = \symRegValVariate'_\symIndexI\rbrace\right|$, because the raw estimator can be also written as
\begin{equation}
\label{equ:raw_estimate_with_multiplicities}
\symPoissonRateEstimate 
= 
\frac{\symAlpha_\infty\,\symNumReg^2}{\sum_{\symRegVal=-\infty}^\infty \symCountVariate'_\symRegVal 2^{-\symRegVal}}.
\end{equation}
Due to \eqref{equ:dist_transformation}, the multiplicities $\symCountVariate'_\symRegVal$ and the multiplicities $\symCountVariate_\symRegVal$ for the observed register values have following relationships
\begin{equation}
\label{equ:multiplicity_transformation}
\begin{aligned}
\symCountVariate_0 &= \textstyle\sum_{\symRegVal = -\infty}^{0} \symCountVariate'_\symRegVal,& \\
\symCountVariate_\symRegVal &=  \symCountVariate'_\symRegVal,& \quad 1\leq\symRegVal\leq\symRegRange,\\
\symCountVariate_{\symRegRange+1} &= \textstyle\sum_{\symRegVal = \symRegRange + 1}^{\infty} \symCountVariate'_\symRegVal.&
\end{aligned}
\end{equation}
The idea is now to find estimates $\hat{\symCount}'_\symRegVal$ for all $\symRegVal\in\mathbb{Z}$ and use them as replacements for $\symCountVariate'_\symRegVal$ in \eqref{equ:raw_estimate_with_multiplicities}. For $\symRegVal\in[1,\symRegRange]$ where $\symCountVariate_\symRegVal = \symCountVariate'_\symRegVal$ we can use the trivial estimators 
\begin{equation}
\hat{\symCount}'_\symRegVal := \symCountVariate_\symRegVal,\quad 1\leq\symRegVal\leq\symRegRange.
\end{equation}
To get estimators for $\symRegVal\leq 0$ and $\symRegVal\geq\symRegRange+1$, we consider the expectation of $\symCountVariate'_\symRegVal$
\begin{equation}
\symExpectation(\symCountVariate'_\symRegVal)
=
\symNumReg
\left(
\symProbability(\symRegValVariate' \leq \symRegVal\vert\symPoissonRate)
-
\symProbability(\symRegValVariate' \leq \symRegVal-1\vert\symPoissonRate)
\right)
=
\symNumReg e^{-\frac{\symPoissonRate}{\symNumReg 2^{\symRegVal}}}\left(1-e^{-\frac{\symPoissonRate}{\symNumReg 2^{\symRegVal}}}\right).
\end{equation}
From \eqref{equ:register_value_distribution} we know that $\symExpectation\!\left(\symCountVariate_0/\symNumReg\right)=
e^{-\frac{\symPoissonRate}{\symNumReg}}$ and $\symExpectation\!\left(1-\symCountVariate_{\symRegRange+1}/\symNumReg\right)=
e^{-\frac{\symPoissonRate}{\symNumReg 2^\symRegRange}}$, and therefore, we can also write
\begin{equation}
\symExpectation(\symCountVariate'_\symRegVal)
=
\symNumReg
\left(\symExpectation\!\left(\symCountVariate_0/\symNumReg\right)\right)^{2^{-\symRegVal}}
\left(1-\left(\symExpectation\!\left(\symCountVariate_0/\symNumReg\right)\right)^{2^{-\symRegVal}}\right)
\end{equation}
and
\begin{equation}
\symExpectation(\symCountVariate'_\symRegVal)
=
\symNumReg
\left(\symExpectation\!\left(1-\symCountVariate_{\symRegRange+1}/\symNumReg\right)\right)^{2^{\symRegRange-\symRegVal}}
\left(1-\left(\symExpectation\!\left(1-\symCountVariate_{\symRegRange+1}/\symNumReg\right)\right)^{2^{\symRegRange-\symRegVal}}\right),
\end{equation}
which motivates us to use
\begin{equation}
\hat{\symCount}'_\symRegVal
=
\symNumReg
\left(\symCountVariate_0/\symNumReg\right)^{2^{-\symRegVal}}
\left(1-\left(\symCountVariate_0/\symNumReg\right)^{2^{-\symRegVal}}\right)
\end{equation}
as estimator for $\symRegVal \leq 0$ and
\begin{equation}
\hat{\symCount}'_\symRegVal
=
\symNumReg
\left(1-\symCountVariate_{\symRegRange+1}/\symNumReg\right)^{2^{\symRegRange-\symRegVal}}
\left(1-\left(1-\symCountVariate_{\symRegRange+1}/\symNumReg\right)^{2^{\symRegRange-\symRegVal}}\right)
\end{equation}
as estimator for $\symRegVal \geq \symRegRange+1$, respectively. This choice of estimators also conserves the mass of zero-valued and saturated registers, because  
\eqref{equ:multiplicity_transformation} is satisfied,
if $\symCountVariate'_\symRegVal$ is replaced by $\hat{\symCount}'_\symRegVal$. 
Plugging all these estimators into \eqref{equ:raw_estimate_with_multiplicities} as replacements for $\symCountVariate'_\symRegVal$ finally gives
\begin{equation}
\label{equ:correctedestimator}
\symPoissonRateEstimate 
= 
\frac{\symAlpha_\infty\symNumReg^2}{\sum_{\symRegVal=-\infty}^\infty \hat{\symCount}'_\symRegVal 2^{-\symRegVal}}
=
\frac{\symAlpha_\infty\symNumReg^2}
{
\symNumReg\,\symSmallCorrectionFunc(\symCountVariate_0/\symNumReg) + \sum_{\symRegVal=1}^\symRegRange \symCountVariate_\symRegVal 2^{-\symRegVal} + \symNumReg\, \symLargeCorrectionFunc(1-\symCountVariate_{\symRegRange+1}/\symNumReg) 2^{-\symRegRange}
}
\end{equation}
which we call the improved raw estimator. Here $\symNumReg\,\symSmallCorrectionFunc(\symCountVariate_0/\symNumReg)$ and $2\symNumReg\,\symLargeCorrectionFunc(1-\symCountVariate_{\symRegRange+1}/\symNumReg)$ are replacements for  $\symCountVariate_0$ and $\symCountVariate_{\symRegRange+1}$ in the raw estimator \eqref{equ:raw_estimator}, respectively. The functions $\symSmallCorrectionFunc$ and $\symLargeCorrectionFunc$ are defined as
\begin{equation}
\label{equ:sigma}
\symSmallCorrectionFunc(\symX) := 
\symX
+
\sum_{\symRegVal=1}^\infty
\symX^{2^\symRegVal} 2^{\symRegVal-1}
\end{equation}
and
\begin{equation}
\label{equ:tau}
\symLargeCorrectionFunc(\symX)
:=
\frac{1}{2}
\left(
-
\symX
+
\sum_{\symRegVal=1}^\infty
\symX^{2^{-\symRegVal}}
2^{-\symRegVal}
\right)
.
\end{equation}

We can cross-check the new estimator for the linear counting case with $\symRegRange=0$. Using the identity 
\begin{equation}
\label{equ:sigma_tau_relationship}
\symSmallCorrectionFunc(\symX) + \symLargeCorrectionFunc(\symX) = \symAlpha_\infty\symPowerSeriesFunc(\log_2(\log(1/\symX)))/\log(1/\symX),
\end{equation}
 we get
\begin{equation}
\symPoissonRateEstimate 
= 
\frac{\symAlpha_\infty\symNumReg}{
\symSmallCorrectionFunc(\symCountVariate_0/\symNumReg)
+
\symLargeCorrectionFunc(\symCountVariate_0/\symNumReg)
}
=
\frac{\symNumReg\log\!\left(\symNumReg/\symCountVariate_0\right)}{\symPowerSeriesFunc\!\left(\log_2\!\left(\log\!\left(\symNumReg/\symCountVariate_0\right)\right)\right)}
\end{equation}
which is as expected almost identical to the linear counting estimator \eqref{equ:linear_counting_estimator}, because $\symPowerSeriesFunc(\symX)\approx 1$ (see \cref{sec:derivation_raw_estimator}).
\subsection{Improved raw estimation algorithm}
The improved raw estimator \eqref{equ:correctedestimator} can be directly translated into a new cardinality estimation algorithm for HyperLogLog sketches as shown in \cref{alg:corrected_raw_estimation}. Since the series \eqref{equ:sigma} converges quadratically for all $\symX\in[0,1)$ and its terms can be recursively calculated using elementary operations, the function $\symSmallCorrectionFunc$ can be quickly calculated to machine precision. The case $\symX=1$ must be handled separately, because the series diverges and causes an infinite denominator in \eqref{equ:correctedestimator} and therefore a vanishing cardinality estimate. As this case only occurs if all register values are zero ($\symCountVariate_0=\symNumReg$), this is exactly what is expected. Among the remaining possible arguments $\lbrace0,\frac{1}{\symNumReg},\frac{2}{\symNumReg},\ldots,\frac{\symNumReg-1}{\symNumReg}\rbrace$ we have slowest convergence for $\symX=\frac{\symNumReg-1}{\symNumReg}$. However, even in this case we have a manageable amount of iteration cycles. For example, if double-precision floating-point arithmetic is used, the routine for calculating  $\symSmallCorrectionFunc$ converges after \num{18} and \num{26} iteration cycles for $\symPrecision=12$ and $\symPrecision=20$, respectively.

The calculation of $\symLargeCorrectionFunc$ is more expensive, because it involves square root evaluations. $\symLargeCorrectionFunc$ can also be written in a numerically more favorable way as 
\begin{equation}
\label{equ:tau2}
\symLargeCorrectionFunc(\symX)
:=
\frac{1}{3}
\left(
1-\symX
-
\sum_{\symRegVal=1}^\infty
\left(
1-
\symX^{2^{-\symRegVal}}
\right)^{\!2}
2^{-\symRegVal}
\right).
\end{equation}
This representation shows faster convergence for $\symX>0$, because $1-\symX^{2^{-\symRegVal}}\leq -\log(\symX) 2^{-\symRegVal}$. Therefore, the convergence speed is comparable to a geometric series with ratio $1/8$. Disregarding the trivial case $\symLargeCorrectionFunc(0)=0$, $\symX=\frac{1}{\symNumReg}$ shows slowest convergence speed among all other possible parameters. For this case and if double-precision floating-point numbers are used, the function $\symLargeCorrectionFunc$ presented in \cref{alg:corrected_raw_estimation} converges after \num{21} and \num{22} iteration cycles for considered precisions $\symPrecision=12$ and $\symPrecision=20$, respectively.

If performance matters $\symSmallCorrectionFunc$ and $\symLargeCorrectionFunc$
can also be precalculated for all possible values of $\symCountVariate_0$ and $\symCountVariate_{\symRegRange+1}$. As their value range is $\{0,1,\ldots, \symNumReg\}$, the function values can be kept in lookup tables of size $\symNumReg+1$. In this way a complete branch-free cardinality estimation can be realized. It is also thinkable to calculate $\symLargeCorrectionFunc$ using the approximation $\symLargeCorrectionFunc(\symX)\approx \symAlpha_\infty/\log(1/\symX)-\symSmallCorrectionFunc(\symX)$ which can be obtained from \eqref{equ:sigma_tau_relationship}. The advantage is that the calculation of $\symSmallCorrectionFunc$ and the additional logarithm is slightly faster than the calculation of $\symLargeCorrectionFunc$. 
However, for arguments close to 1, where $\symAlpha_\infty/\log(1/\symX)$ and $\symSmallCorrectionFunc(\symX)$ are both very large while their difference is very small, special care is needed to avoid numerical cancellation.

The new estimation algorithm is very elegant, because it does neither contain magic numbers nor special cases as the original algorithm. Moreover, the algorithm guarantees monotonicity of the cardinality estimate. The estimate will never become smaller when adding new elements. Although this sounds self-evident, previous approaches that combine estimators for different cardinality ranges or make use of empirical collected data have problems to satisfy this property throughout all cardinalities.

\begin{algorithm}
\caption{Cardinality estimation based on the improved raw estimator. The input is the multiplicity vector $\boldsymbol{\symCountVariate} = (\symCountVariate_0,\ldots,\symCountVariate_{\symRegRange+1})$ as obtained by \cref{alg:sufficient_statistic}.}
\label{alg:corrected_raw_estimation}
\begin{algorithmic}
\Function {EstimateCardinality}{$\boldsymbol{\symCountVariate}$}
\comm{.45}{$\sum_{\symRegVal=0}^{\symRegRange+1}\symCountVariate_\symRegVal = \symNumReg$}
\State $\symZ\gets\symNumReg\cdot\symLargeCorrectionFunc(1 - \symCountVariate_{\symRegRange+1}/\symNumReg)$
\comm{.45}{alternatively, take $\symNumReg\cdot\symLargeCorrectionFunc(1 - \symCountVariate_{\symRegRange+1}/\symNumReg)$ from precalculated lookup table}
\For{$\symRegVal\gets \symRegRange, 1$}
\State $\symZ\gets0.5\cdot\left(\symZ + \symCountVariate_\symRegVal\right)$
\EndFor
\State $\symZ\gets\symZ+\symNumReg\cdot\symSmallCorrectionFunc(\symCountVariate_0/\symNumReg)$
\comm{.45}{alternatively, take $\symNumReg\cdot\symSmallCorrectionFunc(\symCountVariate_0/\symNumReg)$ from precalculated lookup table}
\State \Return$\symAlpha_\infty\symNumReg^2 / \symZ$
\comm{.45}{$\symAlpha_\infty := 1/(2\log(2))$}
\EndFunction
\\
\Function {$\symSmallCorrectionFunc$}{$\symX$}
\comm{.45}{$\symX\in[0,1]$}
\If{$\symX=1$}
\State\Return$\infty$
\EndIf
\State $\symY\gets1$
\State $\symZ\gets\symX$
\Repeat
\State $\symX\gets\symX\cdot\symX$
\State $\symZ'\gets\symZ$
\State $\symZ\gets\symZ + \symX\cdot\symY$
\State $\symY\gets2\cdot\symY$
\Until{$\symZ=\symZ'$}
\State\Return$\symZ$
\EndFunction
\\
\Function {$\symLargeCorrectionFunc$}{$\symX$}
\comm{0.45}{$\symX\in[0,1]$}
\If{$\symX=0\vee \symX=1$}
\State\Return$0$
\EndIf
\State $\symY\gets1$
\State $\symZ\gets 1 - \symX$
\Repeat
\State $\symX\gets\sqrt{\symX}$
\State $\symZ'\gets\symZ$
\State $\symY\gets0.5\cdot\symY$
\State $\symZ\gets\symZ - \left(1-\symX\right)^2\cdot\symY$
\Until{$\symZ=\symZ'$}
\State\Return$\symZ/3$
\EndFunction
\end{algorithmic}
\end{algorithm}

\subsection{Experimental setup}
\label{sec:experimental_setup}
To verify the new estimation algorithm, we generated \num{10000} different HyperLogLog sketches and filled each of them with 50 billion unique elements. During element insertion, we took snapshots of the sketch state by storing the multiplicity vector at predefined cardinality values, which were roughly chosen according to a geometric series with ratio \num{1.01}. In order to achieve that in reasonable time for all the different HyperLogLog parameter combinations  $(\symPrecision,\symRegRange)$ we have been interested in, we applied a couple of optimizations.

First, we assumed a uniform hash function. This allows us to simply generate random numbers instead of creating unique elements and calculating their hash values. We used the Mersenne Twister random number generator with a state size of \num{19937} bits from the C++ standard library. 
Second, we used \cref{alg:insert_fast} for insertions, which does basically the same as \cref{alg:insert}, but additionally keeps track of the multiplicity vector and the minimum register value. In this way the expensive register scan needed to obtain the multiplicity vector as described by \cref{alg:sufficient_statistic} can be avoided. Furthermore, the knowledge of the minimum register value can be used to abort the insertion procedure early, if the hash value is not able to increment any register value. Especially at large cardinalities this saves a great number of register accesses and also reduces the number of cache misses. For example, if $\symRegValVariate_\text{min}=1$ only half of all insertions will access a register. If $\symRegValVariate_\text{min}=2$ it is already just a quarter, and so forth. 
Finally, instead of repeating the simulation for different HyperLogLog parameters, we did that only once using the quite accurate setting $\symPrecision=22$ and $\symRegRange=42$.  Whenever taking a snapshot, we reduced the sketch state exploiting \cref{alg:compress} for desired parameters before calculating the multiplicity vector using \cref{alg:sufficient_statistic}.

After all, we got the multiplicity vectors at predefined cardinalities of \num{10000} different simulation runs for various HyperLogLog parameter settings. All this data constitutes the basis of the presented results that follow. The generated data allows a quick and simple evaluation of cardinality estimation algorithms by just loading the persisted multiplicity vectors from hard disk and passing them as input to the algorithm.

\begin{algorithm}
\caption{Extension of \cref{alg:insert} that keeps track of the multiplicity vector $\boldsymbol{\symCountVariate} = (\symCountVariate_0,\ldots,\symCountVariate_{\symRegRange+1})$ and the minimum register value $\symRegValVariate_\text{min}$ during insertion. Initially, $\symRegValVariate_\text{min}=0$ and $\boldsymbol{\symCountVariate} = (\symNumReg, 0, 0, \dots, 0)$.}
\label{alg:insert_fast}
\begin{algorithmic}
\Procedure {InsertElement}{\symDataItem}
\State $\langle \symBitRepA_1, \ldots, \symBitRepA_\symPrecision,\symBitRepB_1,\ldots,\symBitRepB_\symRegRange\rangle_2 \gets$ $(\symPrecision + \symRegRange)$-bit hash value of $\symDataItem$
\comm{0.25}{$\symBitRepA_\symIndexI,\symBitRepB_\symIndexI\in\{0,1\}$}

\State $\symRegVal \gets \min(\{\symS\mid \symBitRepB_\symS = 1\}\cup {\{\symRegRange+1\}} )$
\comm{0.25}{$\symRegVal\in\lbrace 1,2,\ldots,\symRegRange+1\rbrace$}
\If{$\symRegVal>\symRegValVariate_\text{min}$}
\State $\symIndexI \gets 1+ \langle \symBitRepA_1, \ldots, \symBitRepA_\symPrecision\rangle_2$
\comm{0.25}{$\symIndexI\in\lbrace 1,2,\ldots,\symNumReg\rbrace$}
\If{$\symRegVal>\symRegValVariate_\symIndexI$}
\State $\symCountVariate_{\symRegValVariate_\symIndexI} \gets \symCountVariate_{\symRegValVariate_\symIndexI} - 1$
\State $\symCountVariate_{\symRegVal} \gets \symCountVariate_{\symRegVal} + 1$
\If{$\symRegValVariate_\symIndexI
=
\symRegValVariate_\text{min}$}
\While{$\symCountVariate_{\symRegValVariate_\text{min}}=0$}
\State $\symRegValVariate_\text{min}\gets
\symRegValVariate_\text{min} + 1$
\EndWhile
\EndIf
\State $\symRegValVariate_\symIndexI \gets\symRegVal$

\EndIf
\EndIf
\EndProcedure
\end{algorithmic}
\end{algorithm}

\subsection{Estimation error}
\label{sec:corrected_raw_estimation_error}
\begin{figure}
\centering
\includegraphics[width=1\textwidth]{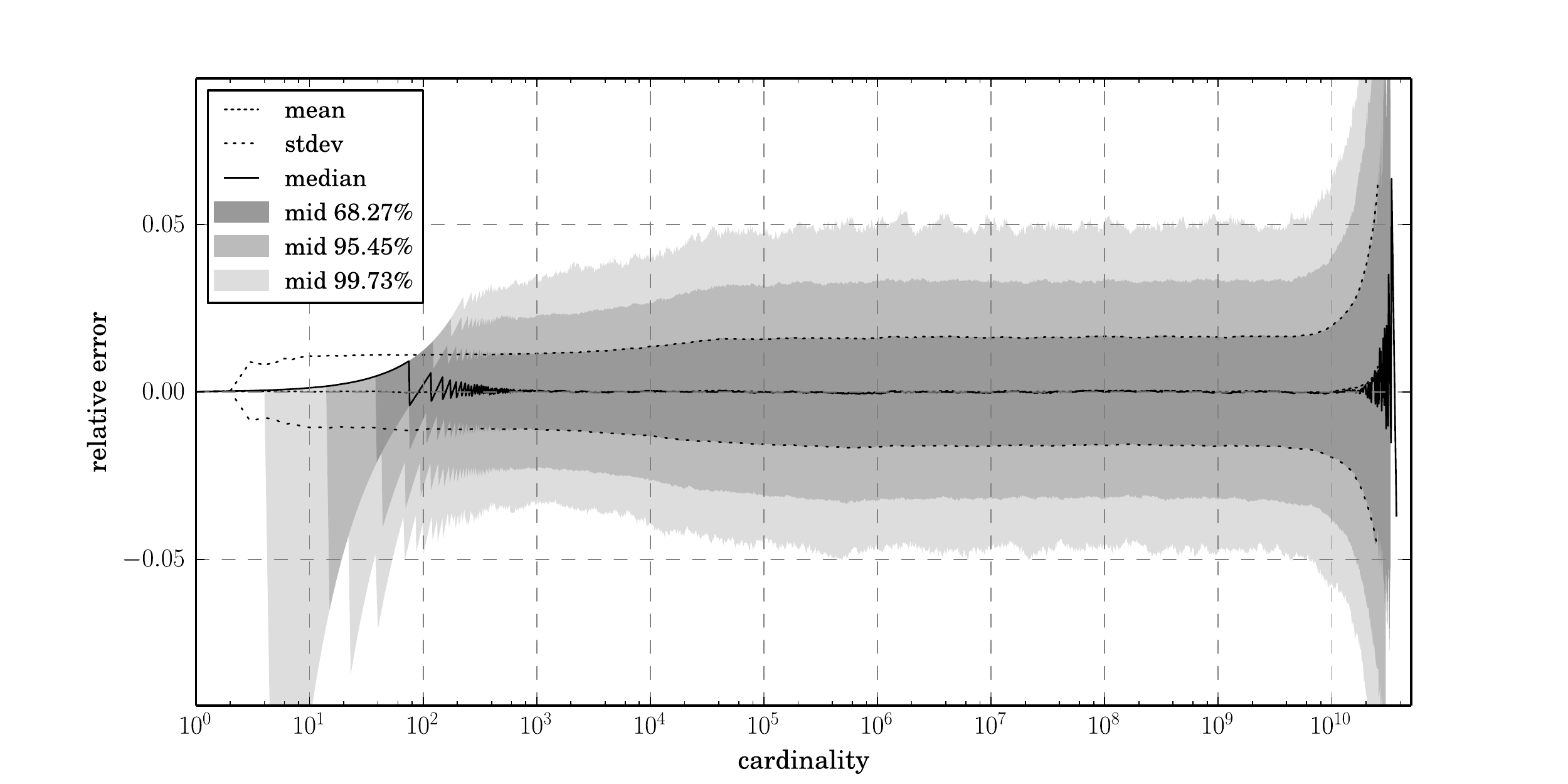}
\caption{Relative error of the improved raw estimator as a function of the true cardinality for a HyperLogLog sketch with parameters $\symPrecision = 12$ and $\symRegRange=20$.}
\label{fig:raw_corrected_estimation_error_12_20}
\end{figure}

\cref{fig:raw_corrected_estimation_error_12_20} shows the distribution of the relative error of the estimated cardinality using \cref{alg:corrected_raw_estimation} compared to the true cardinality for $\symPrecision=12$ and $\symRegRange=20$. As the mean shows, the error is unbiased over the entire cardinality range. The new approach is able to accurately estimate cardinalities up to 4 billions ($\approx 2^{\symPrecision+\symRegRange}$) which is about an order of magnitude larger than the operating range upper bound of the raw estimator (\cref{fig:raw_estimate}).

\begin{figure}
\centering
\includegraphics[width=1\textwidth]{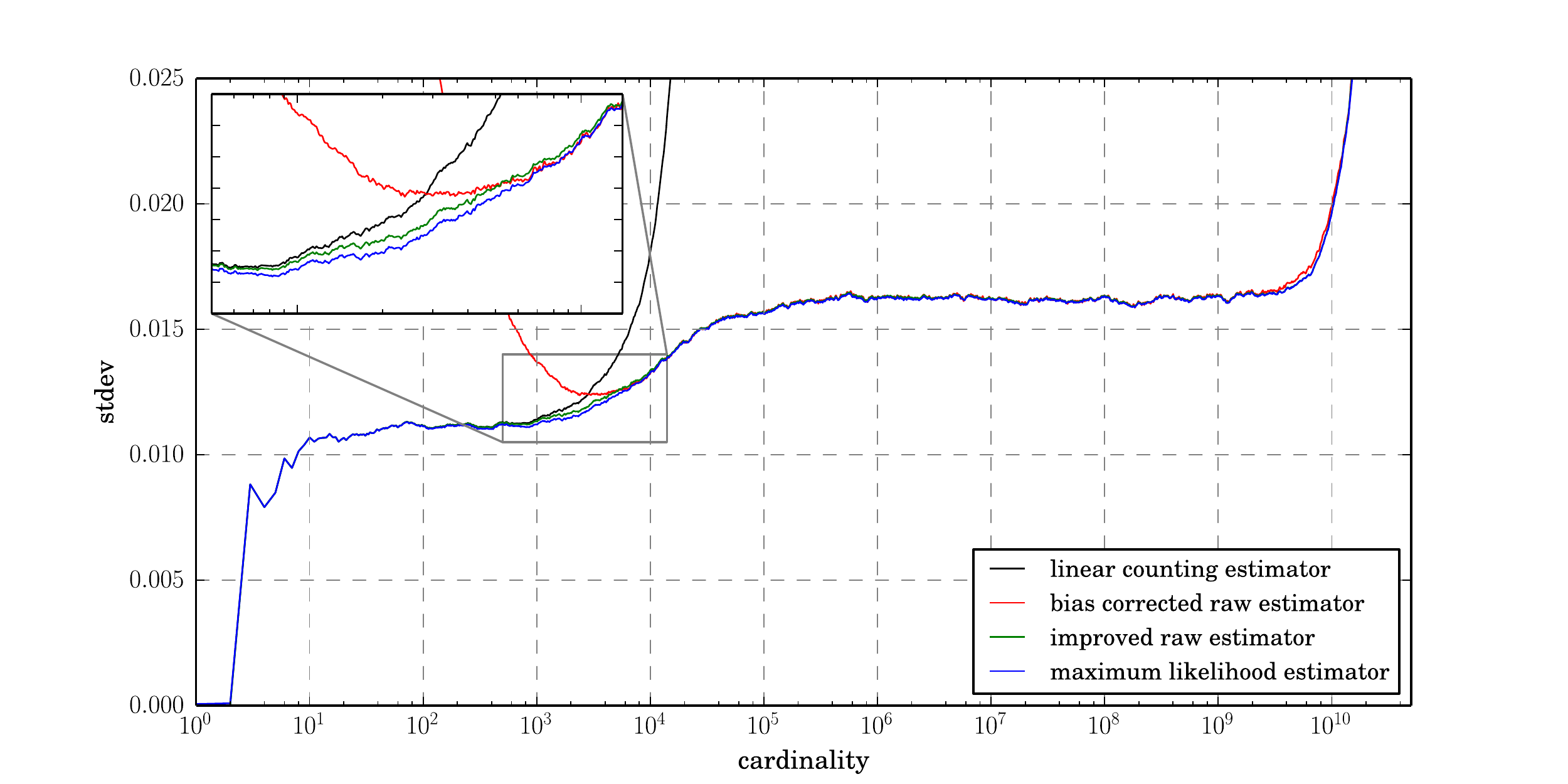}
\caption{Standard deviations of the relative error of different cardinality estimators over the true cardinality for a HyperLogLog sketch with parameters $\symPrecision = 12$ and $\symRegRange=20$ .}
\label{fig:stdev_comparison}
\end{figure}

The improved raw estimator beats the precision of methods that apply bias correction on the raw estimator \cite{Heule2013,Rhodes2015,Sanfilippo2014}.
Based on the simulated data we have empirically determined the bias correction function $\symBiasCorrectionFunc$ for the raw estimator \eqref{equ:raw_estimator} that satisfies $\symCardinality = \symExpectation(\symBiasCorrectionFunc(\symCardinalityRawEstimate)\vert\symCardinality)$ for all cardinalities. By definition, 
the estimator $\symCardinalityRawEstimate':=\symBiasCorrectionFunc(\symCardinalityRawEstimate)$ is unbiased and a function of the raw estimator. Its standard deviation can be compared with that of the improved raw estimator in \cref{fig:stdev_comparison}. For cardinalities smaller than \num{10000} the empirical bias correction approach is not very precise. This is the reason why all previous approaches had to switch over to the linear counting estimator at some point. The standard deviation of the linear counting estimator is also shown in \cref{fig:stdev_comparison}. Obviously, the previous approaches cannot do better than given by the minimum of both curves for linear counting and raw estimator. In practice, the standard deviation is even larger, because the choice between both estimators must be made based on an estimate and not on the true cardinality, for which the intersection point of both curves represents the ideal transition point. In contrast, the improved raw estimator performs well over the entire cardinality range.

\begin{figure}
\centering
\includegraphics[width=1\textwidth]{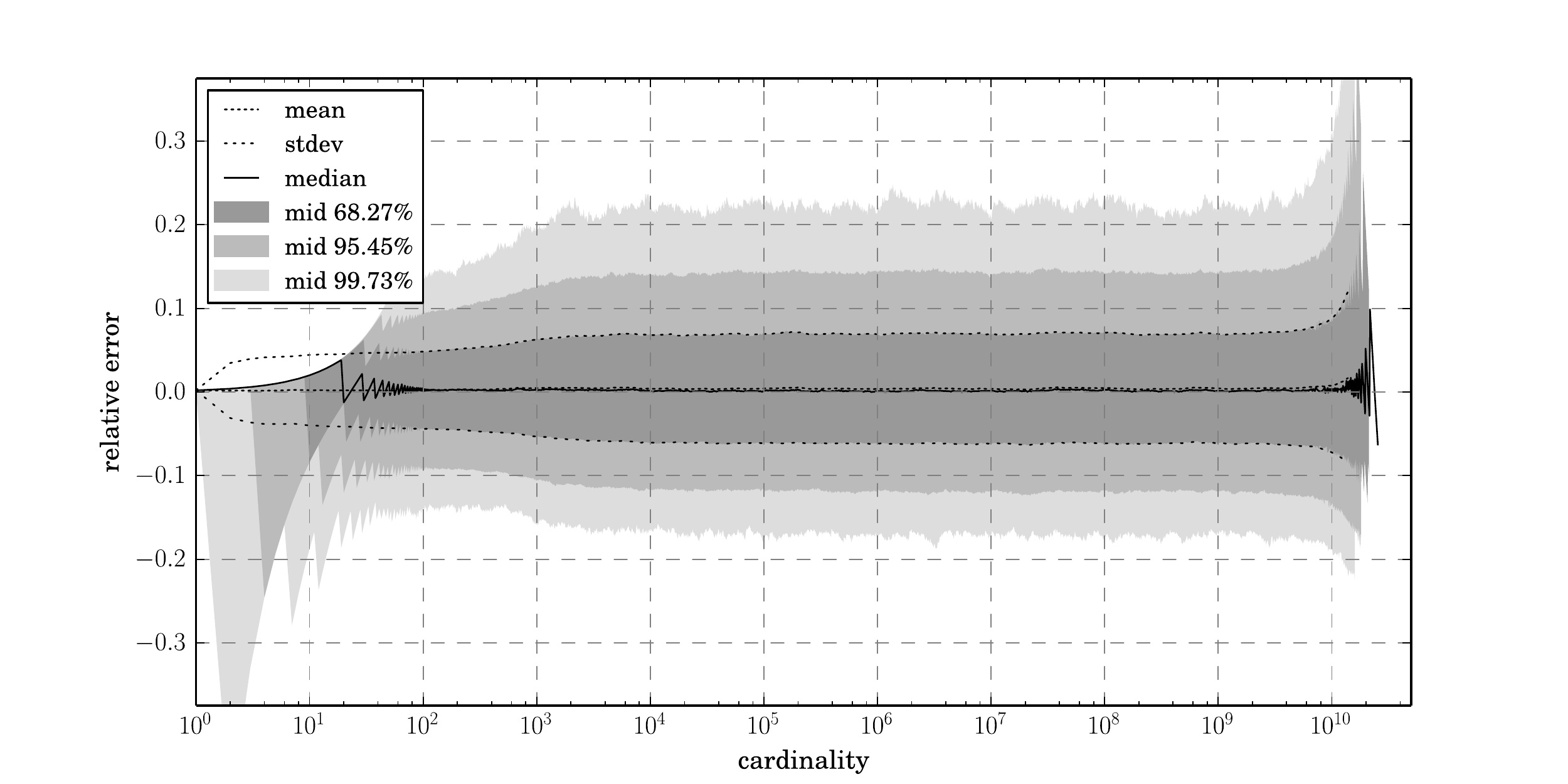}
\caption{Relative error of the improved raw estimator as a function of the true cardinality for a HyperLogLog sketch with parameters $\symPrecision = 8$ and $\symRegRange=24$.}
\label{fig:raw_corrected_estimation_error_8_24}
\end{figure}

\begin{figure}
\centering
\includegraphics[width=1\textwidth]{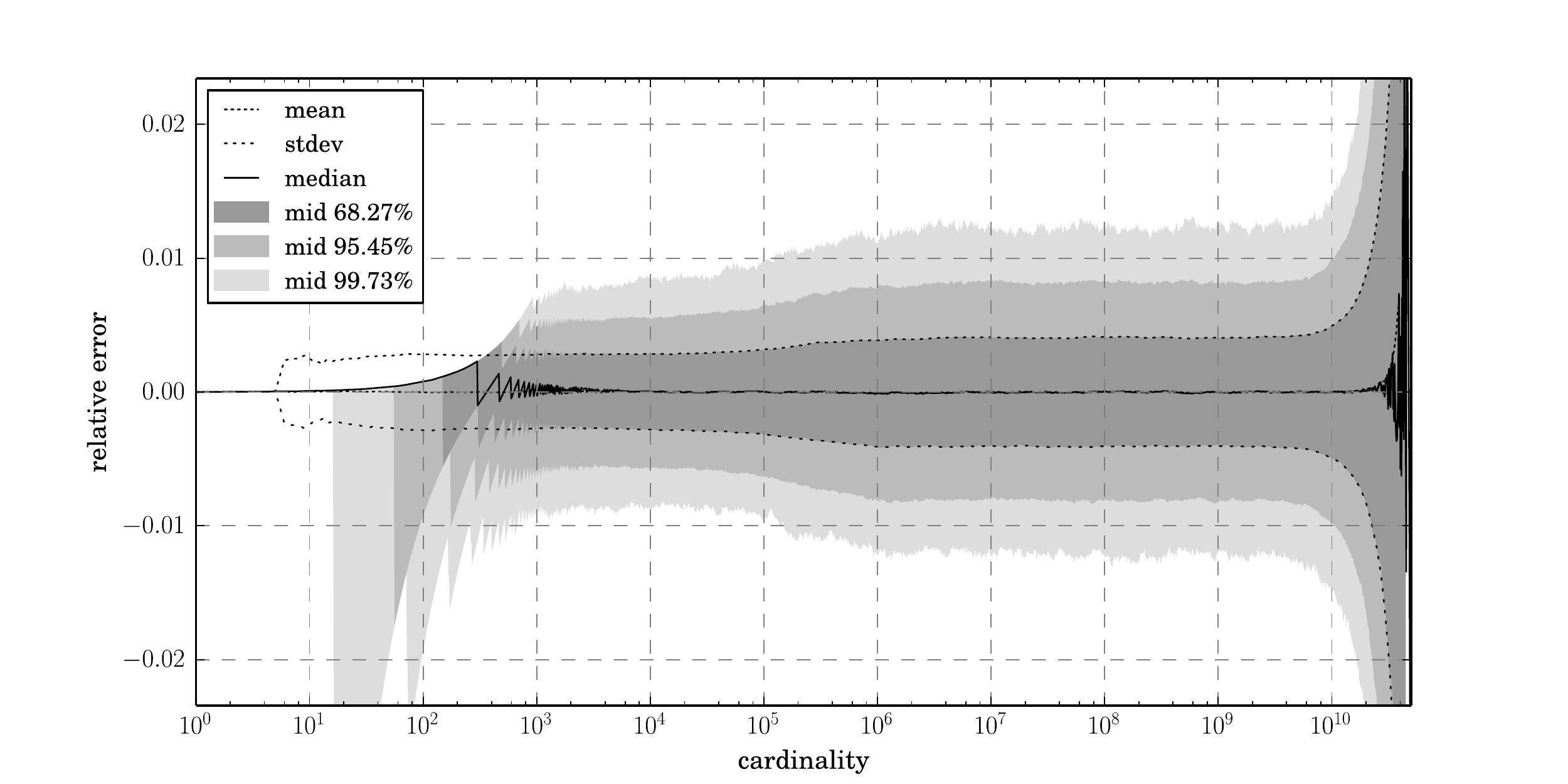}
\caption{Relative error of the improved raw estimator as a function of the true cardinality for a HyperLogLog sketch with parameters $\symPrecision = 16$ and $\symRegRange=16$.}
\label{fig:raw_corrected_estimation_error_16_16}
\end{figure}

\begin{figure}
\centering
\includegraphics[width=1\textwidth]{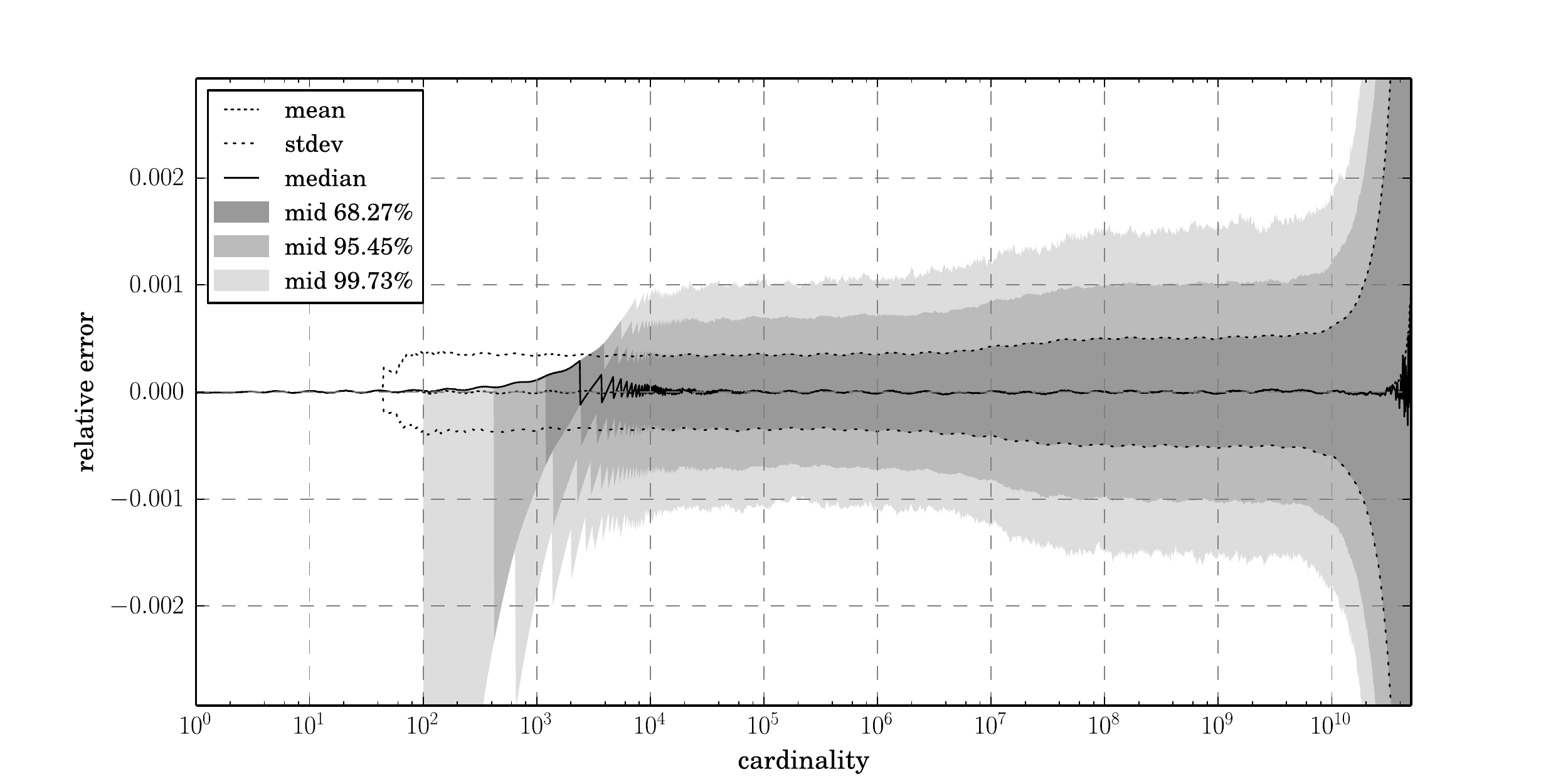}
\caption{Relative error of the improved raw estimator as a function of the true cardinality for a HyperLogLog sketch with parameters $\symPrecision = 22$ and $\symRegRange=10$.}
\label{fig:raw_corrected_estimation_error_22_10}
\end{figure}

The new estimation algorithm also works well for other HyperLogLog configurations. First we considered configurations using a 32-bit hash function ($\symPrecision + \symRegRange = 32$). The relative estimation error for precisions $\symPrecision=8$, $\symPrecision=16$, and $\symPrecision=22$ are shown in \cref{fig:raw_corrected_estimation_error_8_24,fig:raw_corrected_estimation_error_16_16,fig:raw_corrected_estimation_error_22_10}, respectively. As expected, since  $\symPrecision + \symRegRange = 32$ is kept constant, the operating range remains more or less the same, while the relative error decreases with increasing precision parameter $\symPrecision$. Again, the new algorithm gives essentially unbiased estimates. Only for very high precisions, an oscillating bias becomes apparent (compare \cref{fig:raw_corrected_estimation_error_22_10}), that is caused by approximating the periodic function $\symPowerSeriesFunc$ by a constant (see \cref{sec:derivation_raw_estimator}).

\begin{figure}
\centering
\includegraphics[width=1\textwidth]{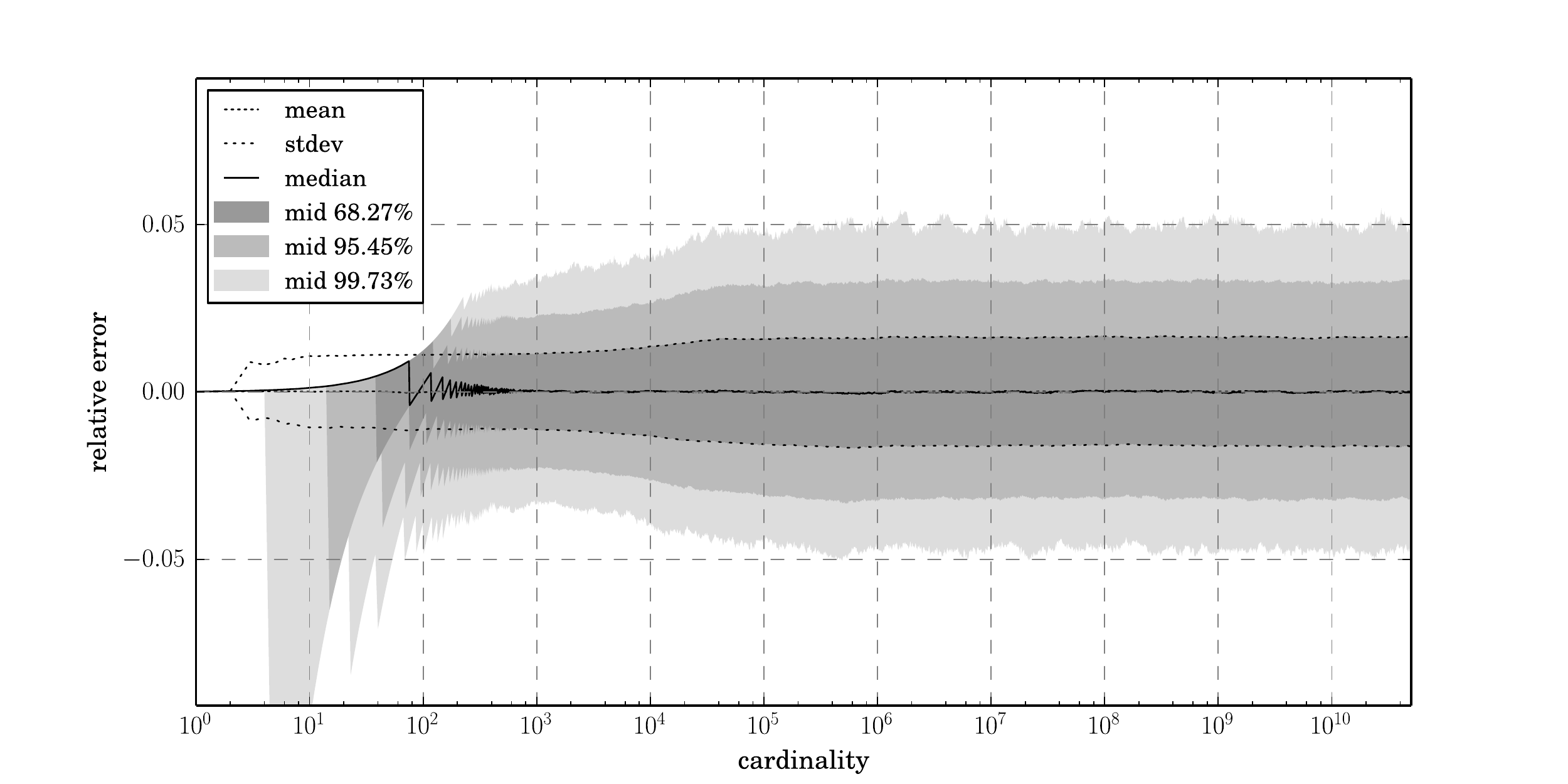}
\caption{Relative error of the improved raw estimator as a function of the true cardinality for a HyperLogLog sketch with parameters $\symPrecision = 12$ and $\symRegRange=52$.}
\label{fig:raw_corrected_estimation_error_12_52}
\end{figure}

As proposed in \cite{Heule2013}, the operating range can be extended by
replacing the 32-bit hash function by a 64-bit hash function. \cref{fig:raw_corrected_estimation_error_12_52} shows the relative error for such a HyperLogLog configuration with parameters $\symPrecision=12$ and $\symRegRange=52$. The doubled hash value size shifts the maximum trackable cardinality value towards $2^{64}$. As \cref{fig:raw_corrected_estimation_error_12_52} shows, when compared to the 32-bit hash value case given in \cref{fig:raw_corrected_estimation_error_12_20}, the estimation error remains constant over the entire simulated cardinality range up to 50 billions.

\begin{figure}
\centering
\includegraphics[width=1\textwidth]{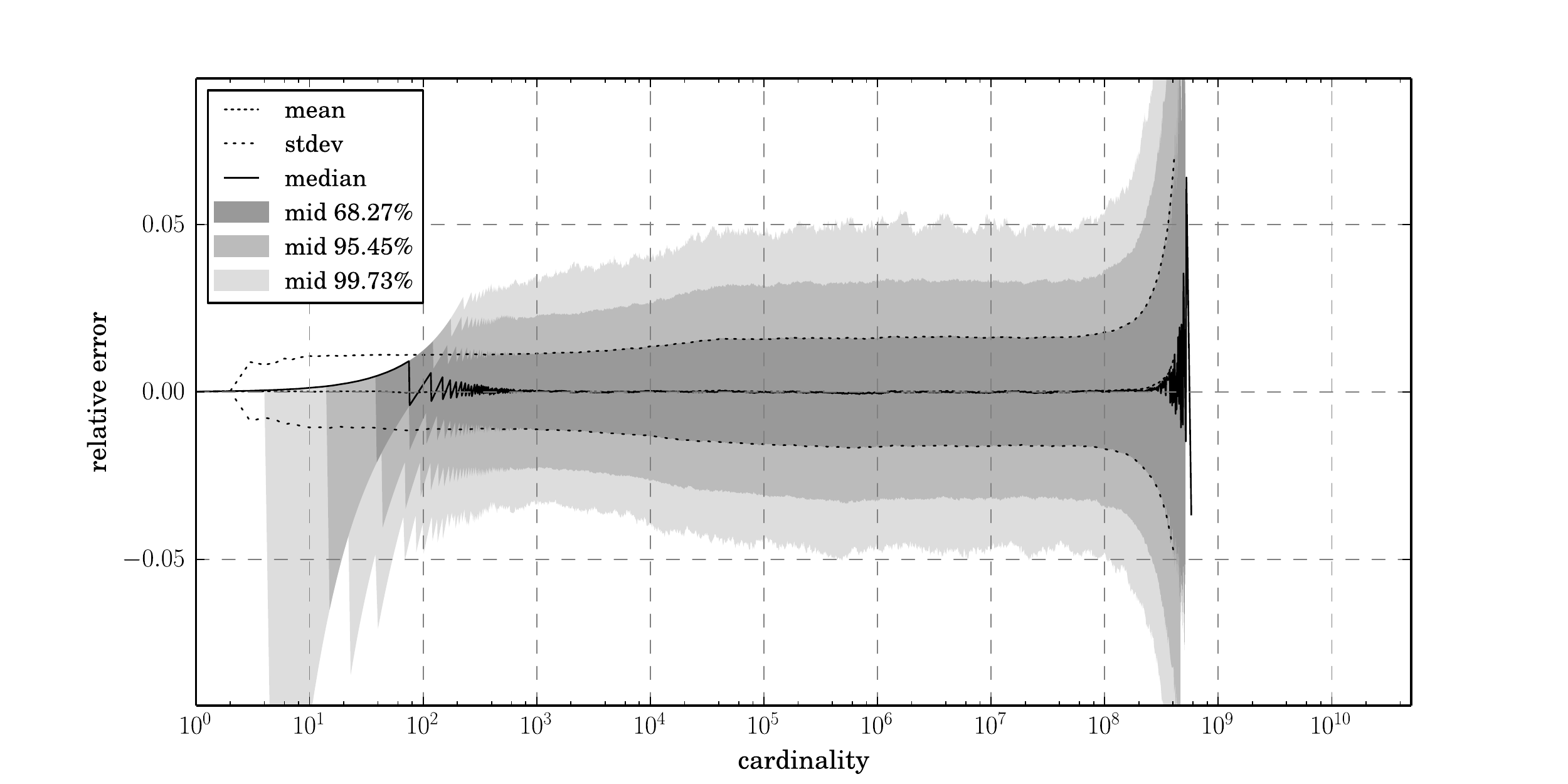}
\caption{Relative error of the improved raw estimator as a function of the true cardinality for a HyperLogLog sketch with parameters $\symPrecision = 12$ and $\symRegRange=14$.}
\label{fig:raw_corrected_estimation_error_12_14}
\end{figure}

We also evaluated the case $\symPrecision = 12$ and $\symRegRange=14$, which is interesting, because the register values are limited to the range $[0, 15]$. As a consequence, 4 bits are sufficient for representing a single register value. This allows two registers to share a single byte, which is beneficial from a performance perspective. Nevertheless, this configuration still allows the estimation of cardinalities up to 100 millions as shown in \cref{fig:raw_corrected_estimation_error_12_14}, which could be enough for many applications.

\subsection{Performance}
\label{sec:corrected_raw_estimation_algorithm}
\begin{figure}
\centering
\includegraphics[width=1\textwidth]{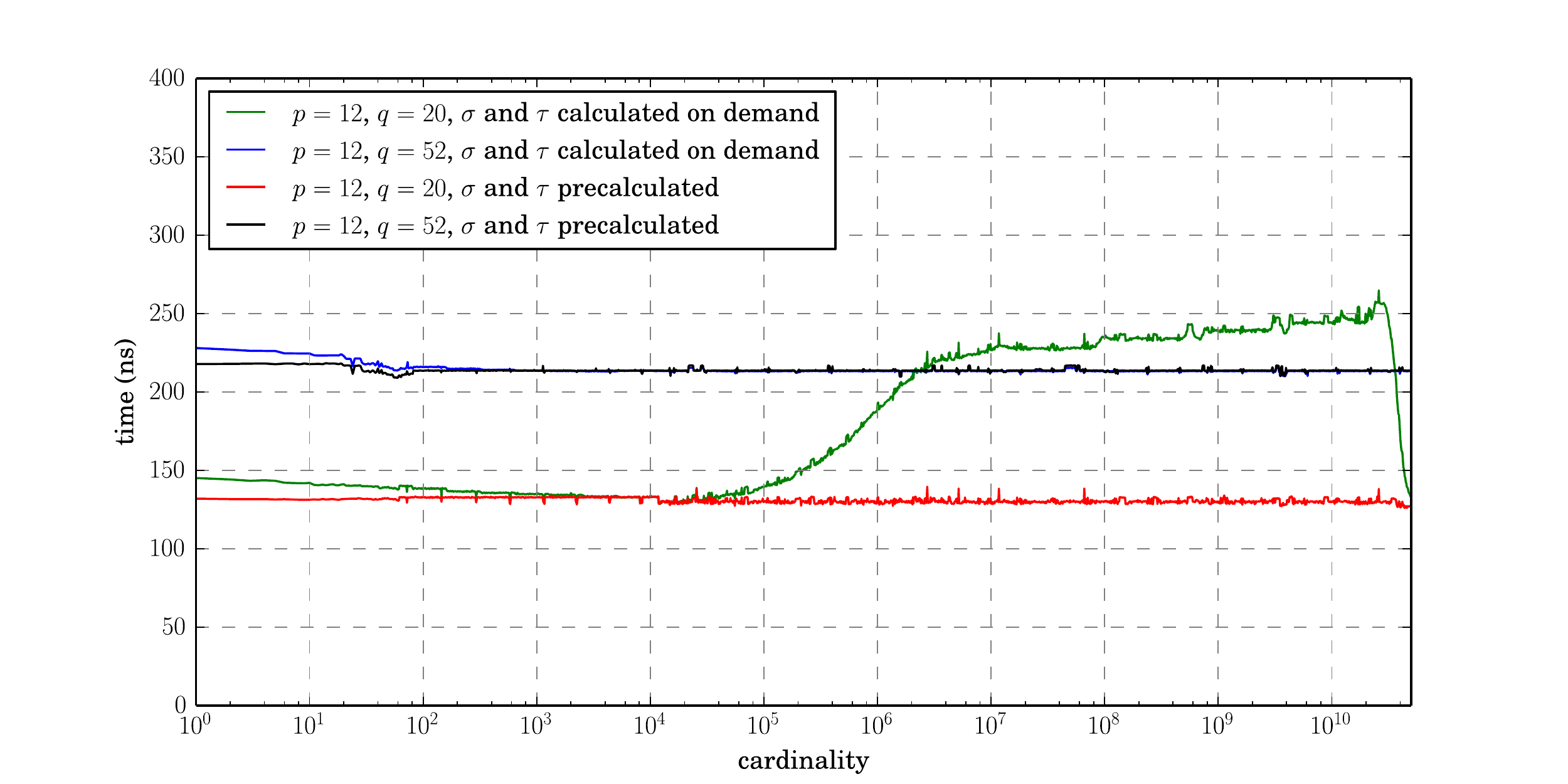}
\caption{Average computation time as a function of the true cardinality with an Intel Core i5-2500K clocking at \SI{3.3}{\giga\hertz} when estimating the cardinality from HyperLogLog sketches with parameters $\symPrecision=12$, $\symRegRange=20$ and $\symPrecision=12$, $\symRegRange=52$, respectively. Both cases, $\symSmallCorrectionFunc$ and $\symLargeCorrectionFunc$ precalculated and calculated on demand have been considered.}
\label{fig:corrected_raw_avg_exec_time}
\end{figure}

To evaluate the performance of the improved raw estimation algorithm, we investigated the average computation time to obtain the cardinality from a given multiplicity vector. For different cardinality values we loaded the precalculated multiplicity vectors (see \cref{sec:experimental_setup}) of \num{1000} randomly generated HyperLogLog sketches into main memory. The average computation time was determined by cycling over these multiplicity vectors and passing them as input to the algorithm. For each evaluated cardinality value the average execution time was calculated after 100 cycles which corresponds to \num{100000} algorithm executions for each cardinality value. The results for HyperLogLog configurations $\symPrecision=12, \symRegRange=20$ and $\symPrecision=12, \symRegRange=52$ are shown in \cref{fig:corrected_raw_avg_exec_time}. Two variants of \cref{alg:corrected_raw_estimation} have been evaluated for which the functions $\symSmallCorrectionFunc$ and $\symLargeCorrectionFunc$ have been either calculated on demand or taken from a lookup table. All these benchmarks where carried out on an Intel Core i5-2500K clocking at \SI{3.3}{\giga\hertz}. 

The results show that the execution times are nearly constant for $\symRegRange=52$. Using a lookup table makes not much difference, because the on-demand calculation for  $\symSmallCorrectionFunc$ is very fast and the calculation of $\symLargeCorrectionFunc$ is rarely needed due to the small probability of saturated registers, $\symCountVariate_{53}=0$ usually holds for realistic cardinalities. If lookup tables are used, the computation time is as expected independent of the cardinality also for the case $\symRegRange=20$. The faster computation times for $\symRegRange=20$ compared to the $\symRegRange=52$ case can be explained by the much smaller dimension of the multiplicity vector which is equal to $\symRegRange+2$. In contrast, if functions $\symSmallCorrectionFunc$ and $\symLargeCorrectionFunc$ are calculated on demand, executions take significantly more time for larger cardinalities. The reason is that beginning at cardinality values in the order of \num{100000} the probability of saturated registers increases. This makes the calculation of $\symLargeCorrectionFunc$ necessary, which is much more expensive than that of $\symSmallCorrectionFunc$, because it requires more iterations and involves square root evaluations. However, the calculation of 
$\symLargeCorrectionFunc$ could be avoided at all, if the HyperLogLog parameters are appropriately chosen. If $2^{\symPrecision+\symRegRange}$ is much larger than the maximum expected cardinality value, the number of saturated registers will be negligible, $\symCountVariate_{\symRegRange+1}\ll \symNumReg$. The calculation of $\symLargeCorrectionFunc$ could be omitted in this case, because $\symLargeCorrectionFunc(1 - \symCountVariate_{\symRegRange+1}/\symNumReg) \approx 0$.

The numbers presented in \cref{fig:corrected_raw_avg_exec_time} do not include the processing time to extract the multiplicity vector out of the HyperLogLog sketch, which requires a complete scan over all registers and counting the different register values into an array as demonstrated by \cref{alg:sufficient_statistic}. A theoretical lower bound for this processing time can be derived using the maximum memory bandwidth of the CPU, which is \SI[per-mode=symbol]{21}{\giga\byte\per\second} for an Intel Core i5-2500K. If we consider a HyperLogLog sketch with precision $\symPrecision=12$ which uses 5 bits per register, the total data size of the HyperLogLog sketch is \SI{2.5}{\kilo\byte} minimum. Consequently, the transfer time from main memory to CPU will be at least \SI{120}{\nano\second}. Having this value in mind, the presented numbers for estimating the cardinality from the multiplicity vector are quite satisfying. 

\section{Maximum likelihood estimation}
\label{sec:max_likelihood_estimation}
We know from \cref{sec:depoissonization} that any unbiased estimator for the Poisson parameter is also an unbiased estimator for the cardinality. Moreover, we know that under suitable regularity conditions of the probability mass function the maximum likelihood estimator is asymptotically efficient \cite{Casella2002}. This means, if the number of registers $\symNumReg$ is large enough, the maximum likelihood method should give us an unbiased estimator for the cardinality.

For HyperLogLog sketches that have been obtained without stochastic averaging (see \cref{sec:data_element_insertion}) the maximum likelihood method was already previously considered \cite{Clifford2012}. The register values are statistically independent by nature in this case, which 
allows factorization of the joint probability mass function and a straightforward calculation of the maximum likelihood estimate. The practically more relevant case which uses stochastic averaging and which is considered in this paper, would lead to a more complicated likelihood function (compare \eqref{equ:multinomialProbabilityMass}). However, the Poisson approximation makes the maximum likelihood method feasible again and we are finally able to derive another new robust and efficient cardinality estimation algorithm. Furthermore, in the course of the derivation we will demonstrate that consequent application of the maximum likelihood method reveals that the cardinality estimate needs to be roughly proportional to the harmonic mean for intermediate cardinality values. The history of the HyperLogLog algorithm shows that the raw estimator \eqref{equ:raw_estimator} was first found after several attempts using the geometric mean \cite{Flajolet2007, Durand2003}.

\subsection{Log-likelihood function}

\begin{figure}
\centering
\includegraphics[width=0.6\textwidth]{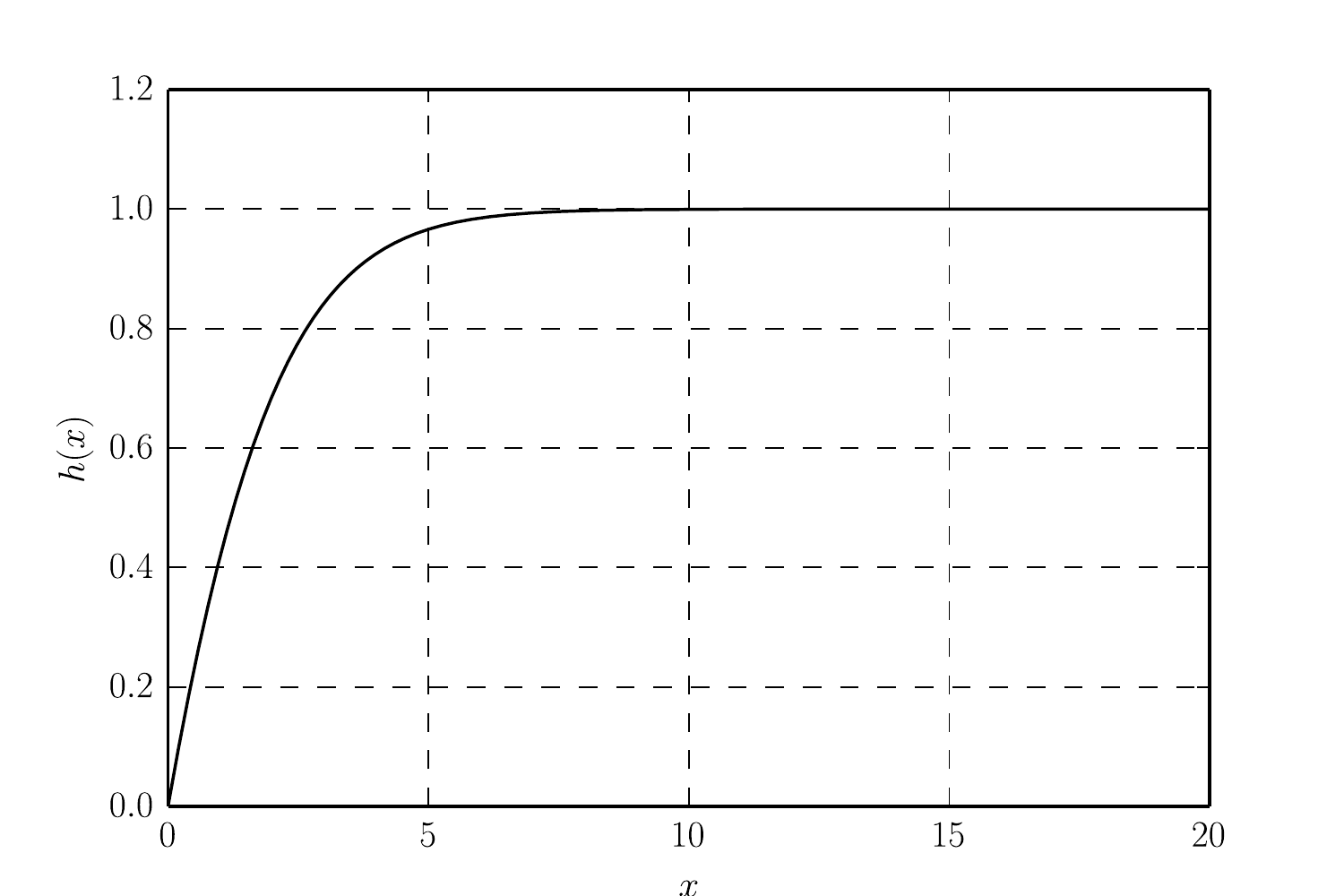}
\caption{The function $\symHelper(\symX)$.}
\label{fig:helper_function}
\end{figure}

Using the probability mass function of the Poisson model \eqref{equ:poisson_pmf} the log-likelihood and its derivative are given by
\begin{equation}
\label{equ:log_likelihood_single}
\log \mathcal{\symLikelihood}(\symPoissonRate\vert\boldsymbol{\symRegValVariate}) = 
-\frac{\symPoissonRate}{\symNumReg}\sum_{\symRegVal=0}^{\symRegRange}\frac{\symCountVariate_\symRegVal}{2^\symRegVal}+ 
\sum_{\symRegVal=1}^{\symRegRange}\symCountVariate_\symRegVal \log\!\left(1-e^{-\frac{\symPoissonRate}{\symNumReg 2^\symRegVal}}\right)
+
\symCountVariate_{\symRegRange+1} \log\!\left(1-e^{-\frac{\symPoissonRate}{\symNumReg 2^{\symRegRange}}}\right)
\end{equation}
and
\begin{equation}
\frac{d}{d\symPoissonRate}\log \mathcal{\symLikelihood}(\symPoissonRate\vert\boldsymbol{\symRegValVariate}) 
=
-\frac{1}{\symPoissonRate}\left(
\frac{\symPoissonRate}{\symNumReg}\sum_{\symRegVal=0}^\symRegRange \frac{\symCountVariate_\symRegVal}{2^\symRegVal}+
\sum_{\symRegVal=1}^\symRegRange \symCountVariate_\symRegVal\frac{\frac{\symPoissonRate}{\symNumReg 2^\symRegVal}}{1-e^{\frac{\symPoissonRate}{\symNumReg 2^\symRegVal}}}
+\symCountVariate_{\symRegRange+1}\frac{\frac{\symPoissonRate}{\symNumReg 2^\symRegRange}}{1-e^{\frac{\symPoissonRate}{\symNumReg 2^\symRegRange}}}
\right).
\end{equation}
As a consequence, the maximum likelihood estimate for the Poisson parameter is given by 
\begin{equation}
\symPoissonRateEstimate = \symNumReg\symXEstimate,
\end{equation}
if $\symXEstimate$ denotes the root of the function
\begin{equation}
\label{equ:funcdef}
\symFunc(\symX)
:=
\symX\sum_{\symRegVal=0}^\symRegRange \frac{\symCountVariate_\symRegVal}{2^\symRegVal}+
\sum_{\symRegVal=1}^\symRegRange \symCountVariate_\symRegVal\frac{\frac{\symX}{2^\symRegVal}}{1-e^{\frac{\symX}{2^\symRegVal}}}
+\symCountVariate_{\symRegRange+1}\frac{\frac{\symX}{2^\symRegRange}}{1-e^{\frac{\symX}{2^\symRegRange}}}.
\end{equation}
This function can also be written as
\begin{equation}
\label{equ:func}
\symFunc(\symX)
:=
\symX\sum_{\symRegVal=0}^\symRegRange \frac{\symCountVariate_\symRegVal}{2^\symRegVal}+
\sum_{\symRegVal=1}^\symRegRange \symCountVariate_\symRegVal\symHelper\!\left(\frac{\symX}{2^\symRegVal}\right)
+
\symCountVariate_{\symRegRange+1}\symHelper\!\left(\frac{\symX}{2^\symRegRange}\right)
-
\left(\symNumReg-\symCountVariate_0\right),
\end{equation}
where the function $\symHelper(\symX)$ is defined as
\begin{equation}
\label{equ:helper}
\symHelper(\symX):=1-\frac{\symX}{e^{\symX}-1}.
\end{equation}
$\symHelper(\symX)$ is strictly increasing and concave as can be seen in \cref{fig:helper_function}. For nonnegative values this function ranges from $\symHelper(0)=0$ to $\symHelper(\symX\rightarrow \infty)=1$.
Since the function $\symFunc(x)$ is also strictly increasing, it is obvious that there exists a unique root $\symXEstimate$ for which $\symFunc(\symXEstimate)=0$. The function is nonpositive at 0 since $\symFunc(0)=\symCountVariate_0-\symNumReg\leq 0$ and, in case $\symCountVariate_{\symRegRange+1}<\symNumReg$ which implies $\sum_{\symRegVal=0}^\symRegRange \frac{\symCountVariate_\symRegVal}{2^\symRegVal}>0$, the function is at least linearly increasing. $\symCountVariate_{\symRegRange+1}=\symNumReg$ corresponds to the case with all registers equal to the maximum value $\symRegRange+1$, for which the maximum likelihood estimate would be positive infinite.

It is easy to see that the estimate $\symPoissonRateEstimate$ remains equal or becomes larger, when inserting an element into the HyperLogLog sketch following \cref{alg:insert}. An update potentially changes the multiplicity vector $(\symCountVariate_0,\ldots,\symCountVariate_{\symRegRange+1})$ to $(\symCountVariate_0,\ldots,\symCountVariate_\symIndexI-1,\ldots,\symCountVariate_\symIndexJ+1,\ldots,\symCountVariate_{\symRegRange+1})$ where $\symIndexI < \symIndexJ$. Writing \eqref{equ:func} as 
\begin{multline}
\symFunc(\symX)
:=
\symCountVariate_0 \symX
+
\symCountVariate_1 \left(
\symHelper\!\left(\frac{x}{2^1}\right)
+\frac{\symX}{2^1}-1
\right)
+
\symCountVariate_2 \left(
\symHelper\!\left(\frac{x}{2^2}\right)
+\frac{\symX}{2^2}-1
\right)
+
\ldots
\\
\dots
+
\symCountVariate_\symRegRange
\left(
\symHelper\!\left(\frac{\symX}{2^\symRegRange}\right)
+\frac{\symX}{2^\symRegRange}-1
\right)
+
\symCountVariate_{\symRegRange+1}
\left(
\symHelper\!\left(\frac{\symX}{2^{\symRegRange}}\right)
-1
\right).
\end{multline}
shows that the coefficient of $\symCountVariate_\symIndexI$ is larger than the coefficient of $\symCountVariate_\symIndexJ$ in case $\symIndexI < \symIndexJ$. Keeping $\symX$ fixed during an update decreases $\symFunc(\symX)$. As a consequence, since $\symFunc(\symX)$ is increasing, the new root and hence the estimate must be larger than prior the update.

For the special case $\symRegRange=0$, which corresponds to the already mentioned linear counting algorithm, \eqref{equ:funcdef} can be solved analytically. In this case, the maximum likelihood method under the Poisson model leads directly to the linear counting estimator \eqref{equ:linear_counting_estimator}. Due to this fact we could expect that maximum likelihood estimation under the Poisson model also works well for the more general HyperLogLog case.

\subsection{Inequalities for the maximum likelihood estimate}
In the following we derive lower and upper bounds for $\symXEstimate$.
Applying Jensen's inequality on $\symHelper$ in \eqref{equ:func} gives an upper bound for $\symFunc(\symX)$:
\begin{equation}
\symFunc(\symX)
\leq
\symX
\sum_{\symRegVal=0}^\symRegRange \frac{\symCountVariate_\symRegVal}{2^\symRegVal}
+
\left(\symNumReg-\symCountVariate_0\right)\cdot
\symHelper\!\left(
\symX\cdot
\frac{
\sum_{\symRegVal=1}^{\symRegRange}
\frac{\symCountVariate_\symRegVal}{2^\symRegVal}
+
\frac{\symCountVariate_{\symRegRange+1}}{2^\symRegRange}
}
{\symNumReg-\symCountVariate_0}\right)
-
\left(\symNumReg-\symCountVariate_0\right).
\end{equation}
The left-hand side is zero, if $\symXEstimate$ is inserted. Resolution for $\symXEstimate$ finally gives the lower bound
\begin{equation}
\label{equ:strong_lower_bound}
\symXEstimate\geq \frac{\symNumReg-\symCountVariate_0}{\sum_{\symRegVal=1}^{\symRegRange}
\frac{\symCountVariate_\symRegVal}{2^\symRegVal}
+
\frac{\symCountVariate_{\symRegRange+1}}{2^\symRegRange}
}
\log\!\left(
1
+
\frac{\sum_{\symRegVal=1}^{\symRegRange}
\frac{\symCountVariate_\symRegVal}{2^\symRegVal}
+
\frac{\symCountVariate_{\symRegRange+1}}{2^\symRegRange}
}
{
\sum_{\symRegVal=0}^{\symRegRange}
\frac{\symCountVariate_\symRegVal}{2^\symRegVal}
}
\right).
\end{equation}
This bound can be weakened using $\log(1+x) \geq \frac{2x}{x+2}$ for $x\geq0$ which results in
\begin{equation}
\label{equ:weak_lower_bound}
\symXEstimate
\geq
\frac{\symNumReg-\symCountVariate_0}
{\symCountVariate_0+\frac{3}{2}\sum_{\symRegVal=1}^{\symRegRange}\frac{\symCountVariate_\symRegVal}{2^\symRegVal} + \frac{\symCountVariate_{\symRegRange+1}}{2^{\symRegRange+1}}}.
\end{equation}
Using the monotonicity of $\symHelper$, the lower bound
\begin{equation}
\symFunc(\symX)
\geq
\symX\sum_{\symRegVal=0}^\symRegRange \frac{\symCountVariate_\symRegVal}{2^\symRegVal}+
\sum_{\symRegVal=1}^\symRegRange \symCountVariate_\symRegVal\symHelper\!\left(\frac{\symX}{2^{\symRegValVariate'_\text{max}}}\right)
+
\symCountVariate_{\symRegRange+1}\symHelper\!\left(\frac{\symX}{2^{\symRegValVariate'_\text{max}}}\right)
-
\left(\symNumReg-\symCountVariate_0\right)
\end{equation}
can be found, where $\symRegValVariate'_\text{max} := \min(\symRegValVariate_\text{max}, \symRegRange)$ and
$\symRegValVariate_\text{max} := \max(\lbrace \symRegVal\vert\symCountVariate_\symRegVal>0\rbrace)$.
Again, inserting $\symXEstimate$ and transformation gives
\begin{equation}
\label{equ:strong_upper_bound}
\symXEstimate
\leq
2^{\symRegValVariate'_\text{max}}
\log\!\left(
1+
\frac{\symNumReg-\symCountVariate_0}
{
2^{\symRegValVariate'_\text{max}}
\sum_{\symRegVal=0}^{\symRegRange}
\frac{\symCountVariate_\symRegVal}{2^\symRegVal}
}
\right)
\end{equation}
as upper bound which can be weakened using $\log(1+\symX)\leq \symX$ for $\symX\geq 0$
\begin{equation}
\label{equ:weak_upper_bound}
\symXEstimate
\leq
\frac{\symNumReg-\symCountVariate_0}
{\sum_{\symRegVal=0}^{\symRegRange}
\frac{\symCountVariate_\symRegVal}{2^\symRegVal}}.
\end{equation}
If the HyperLogLog sketch is in the intermediate range, where $\symCountVariate_0=\symCountVariate_{\symRegRange+1}=0$ the bounds \eqref{equ:weak_lower_bound} and \eqref{equ:weak_upper_bound} differ only by a constant factor and both are proportional to the harmonic mean of $2^{\symRegValVariate_1},\ldots,2^{\symRegValVariate_\symNumReg}$. Hence, consequent application of the maximum likelihood method would have directly suggested to use a cardinality estimator that is proportional to the harmonic mean without knowing the raw estimator \eqref{equ:raw_estimator} in advance.

\subsection{Computation of the maximum likelihood estimate}
\label{sec:comp_ml_estimate}
Since $\symFunc$ is concave and increasing, both, Newton-Raphson iteration and the secant method, will converge to the root, provided that the function is negative for the chosen starting points. We will use the secant method to derive the new cardinality estimation algorithm. Even though the secant method has the disadvantage of slower convergence, a single iteration is simpler to calculate as it does not require the evaluation of the first derivative. An iteration step of the secant method can be written as
\begin{equation}
\symX_{\symIndexI} = 
\symX_{\symIndexI-1} -
\left(\symX_{\symIndexI-1}-\symX_{\symIndexI-2}\right)
\frac{\symFunc(\symX_{\symIndexI-1})}{\symFunc(\symX_{\symIndexI-1}) - \symFunc(\symX_{\symIndexI-2})}.
\end{equation}
If we set $\symX_0$ equal to 0, for which  $\symFunc(\symX_0)=-\left(\symNumReg-\symCountVariate_0\right)$, and $\symX_1$ equal to one of the derived lower bounds \eqref{equ:strong_lower_bound} or \eqref{equ:weak_lower_bound}, the sequence $(\symX_0, \symX_1, \symX_2, \ldots)$ is monotone increasing. Using the definitions
\begin{equation}
\Delta\symX_\symIndexI := \symX_\symIndexI-\symX_{\symIndexI-1}
\end{equation}
and
\begin{equation}
\label{equ:funcprime}
\symFuncPrime(\symX):=\symFunc(\symX) + \left(\symNumReg-\symCountVariate_0\right)
=\symX\sum_{\symRegVal=0}^\symRegRange \frac{\symCountVariate_\symRegVal}{2^\symRegVal}+
\sum_{\symRegVal=1}^\symRegRange \symCountVariate_\symRegVal\symHelper\!\left(\frac{\symX}{2^\symRegVal}\right)
+
\symCountVariate_{\symRegRange+1}\symHelper\!\left(\frac{\symX}{2^\symRegRange}\right)
\end{equation}
the iteration scheme can also be written as
\begin{gather}
\label{equ:secant_delta}
\Delta\symX_{\symIndexI} = \Delta\symX_{\symIndexI-1}
\frac{\left(\symNumReg-\symCountVariate_0\right)-\symFuncPrime(\symX_{\symIndexI-1})}{\symFuncPrime(\symX_{\symIndexI-1}) - \symFuncPrime(\symX_{\symIndexI-2})},
\\
\symX_{\symIndexI} = \symX_{\symIndexI-1} + \Delta\symX_{\symIndexI}.
\end{gather}
The iteration can be stopped, if $\Delta\symX_{\symIndexI} \leq \symStopDelta\cdot \symX_{\symIndexI}$. Since the expected statistical error for the HyperLogLog data structure scales according to $\frac{1}{\sqrt{\symNumReg}}$ \cite{Flajolet2007}, it makes sense to choose $\symStopDelta = \frac{\symStopEpsilon}{\sqrt{\symNumReg}}$ with some constant $\symStopEpsilon$. For all results presented later in \cref{sec:maximum_likelihood_estimation_error} we used $\symStopEpsilon = 10^{-2}$.

\subsection{Maximum likelihood estimation algorithm}
To get a fast cardinality estimation algorithm, it is crucial to minimize evaluation costs for \eqref{equ:funcprime}. A couple of optimizations allow significant reduction of computational effort:
\begin{itemize}
\item Only a fraction of all count values $\symCountVariate_\symRegVal$ is nonzero. If we denote $\symRegValVariate_\text{min}:=\min(\lbrace \symRegVal\vert\symCountVariate_\symRegVal>0\rbrace)$ and $\symRegValVariate_\text{max}:=\max(\lbrace \symRegVal\vert\symCountVariate_\symRegVal>0\rbrace)$,  it is sufficient to loop over all indices in the range $[\symRegValVariate_\text{min}, \symRegValVariate_\text{max}]$.
\item The sum $\sum_{\symRegVal=0}^\symRegRange \frac{\symCountVariate_\symRegVal}{2^\symRegVal}$ in \eqref{equ:funcprime} can be precalculated and reused for all function evaluations.
\item Many programming languages allow the efficient multiplication and division by any integral power of two using special functions, such as $\operatorname{ldexp}$ in C/C++ or $\operatorname{scalb}$ in Java.
\item The function $\symHelper(\symX)$ only needs to be evaluated at points $\left\lbrace\frac{\symX}{2^{\symRegValVariate'_\text{max}}},\frac{\symX}{2^{\symRegValVariate'_\text{max}-1}},\ldots,\frac{\symX}{2^{\symRegValVariate_\text{min}}}\right\rbrace$ where $\symRegValVariate'_\text{max} := \min(\symRegValVariate_\text{max}, \symRegRange)$. This series corresponds to a geometric series with ratio two. A straightforward calculation using \eqref{equ:helper} is very expensive because of the exponential function. However, if we know $\symHelper\!\left(\frac{\symX}{2^{\symRegValVariate'_\text{max}}}\right)$ all other required function values can be easily obtained using the identity
\begin{equation}
\label{equ:helper_recursion1}
\symHelper(4\symX) = \frac{\symX+\symHelper(2\symX)\left(1-\symHelper(2\symX)\right)}{\symX+\left(1-\symHelper(2\symX)\right)}.
\end{equation}
This recursive formula is stable in a sense that the relative error of $\symHelper(4\symX)$ is smaller than that of $\symHelper(2\symX)$ as shown in \cref{app:helper_stable}.

\item If $\symX$ is smaller than 0.5, the function $\symHelper(\symX)$ can be well approximated by a Taylor series around $\symX=0$
\begin{equation}
\symHelper(\symX)
=
\frac{\symX}{2} - \frac{\symX^2}{12} +\frac{\symX^4}{720}-\frac{\symX^6}{30240} + \symBigO(\symX^{8}),
\end{equation}
which can be optimized for numerical evaluation using Estrin's scheme and $\symX' := \frac{\symX}{2}$ and $\symX'' := \symX' \symX'$
\begin{equation}
\label{equ:taylor}
\symHelper(\symX)
=
\symX' - \symX''/3 + \left(\symX'' \symX''\right)\left(1/45-\symX''/472.5\right)
+ \symBigO(\symX^{8}).
\end{equation}
The smallest argument for which $\symHelper$ needs to be evaluated is $\frac{\symX}{2^{\symRegValVariate'_\text{max}}}$. If 
$\symCountVariate_{\symRegRange+1}=0$, we can find an upper bound for the smallest argument using \eqref{equ:strong_upper_bound}
\begin{equation}
\frac{\symX}{2^{\symRegValVariate'_\text{max}}} 
\leq
\frac{\symXEstimate}{2^{\symRegValVariate'_\text{max}}} 
\leq
\log\!\left(
1+
\frac{\sum_{\symRegVal=0}^{\symRegValVariate'_\text{max}}\symCountVariate_\symRegVal}
{
2^{\symRegValVariate'_\text{max}}
\sum_{\symRegVal=0}^{\symRegValVariate'_\text{max}}
\frac{\symCountVariate_\symRegVal}{2^\symRegVal}
}
\right)
\leq \log 2 \approx 0.693.
\end{equation}
In practice, $\frac{\symX}{2^{\symRegValVariate'_\text{max}}}\leq0.5$ is satisfied most of the time as long as only a few registers are saturated, that is $\symCountVariate_{\symRegRange+1}\ll \symNumReg$. In case $\frac{\symX}{2^{\symRegValVariate'_\text{max}}} > 0.5$, $\symHelper\!\left(\frac{\symX}{2^{\symKappa}}\right)$ is calculated instead with $\symKappa = 2+\lfloor\log_2(\symX)\rfloor$. By definition, $\frac{\symX}{2^{\symKappa}}\leq 0.5$ which allows using the Taylor series approximation. $\symHelper\!\left(\frac{\symX}{2^{\symRegValVariate'_\text{max}}}\right)$ is finally obtained after $\symKappa -\symRegValVariate'_\text{max}$ iterations using \eqref{equ:helper_recursion1}. As shown in \cref{app:error_approx}, a small approximation error of $\symHelper$ does not have much impact on the error of the maximum likelihood estimate as long as most registers are not saturated.
\end{itemize}

Putting all these optimizations together finally gives the new cardinality estimation algorithm presented as \cref{alg:estimate_ml}. The algorithm requires mainly only elementary operations. For very large cardinalities it makes sense to use the strong \eqref{equ:strong_lower_bound} instead of the weak lower bound \eqref{equ:weak_lower_bound} as second starting point for the secant method. The stronger bound is a much better approximation especially for large cardinalities, where the extra logarithm evaluation is amortized by savings in the number of iteration cycles. Therefore, the presented algorithm switches over to the stronger bound, if 
\begin{equation}
\frac{
\sum_{\symRegVal=1}^{\symRegRange}\frac{\symCountVariate_\symRegVal}{2^\symRegVal} + \frac{\symCountVariate_{\symRegRange+1}}{2^\symRegRange}
}{
\sum_{\symRegVal=0}^{\symRegRange}\frac{\symCountVariate_\symRegVal}{2^\symRegVal}
}
> 1.5
\end{equation}
is satisfied. The threshold value of $1.5$ was found to be a reasonable choice in order to reduce the computation time for large cardinalities significantly.

\begin{algorithm}
\caption{Cardinality estimation based on the maximum likelihood principle. The input is the multiplicity vector $\boldsymbol{\symCountVariate} = (\symCountVariate_0,\ldots,\symCountVariate_{\symRegRange+1})$ as obtained by \cref{alg:sufficient_statistic}.}
\ContinuedFloat
\label{alg:estimate_ml}
\begin{algorithmic}
\Function {EstimateCardinality}{$\boldsymbol{\symCountVariate}$}
\comm{.31}{$\sum_{\symRegVal=0}^{\symRegRange+1}\symCountVariate_\symRegVal = \symNumReg$}
\If{$\symCountVariate_{\symRegRange+1} = \symNumReg$}
\State\Return $\infty$
\EndIf
\State $\symRegValVariate_\text{min} \gets \min(\lbrace \symRegVal\vert\symCountVariate_\symRegVal>0\rbrace)$
\State $\symRegValVariate'_\text{min} \gets \max(\symRegValVariate_\text{min}, 1)$
\State $\symRegValVariate_\text{max} \gets \max(\lbrace \symRegVal\vert\symCountVariate_\symRegVal>0\rbrace)$
\State $\symRegValVariate'_\text{max} \gets \min(\symRegValVariate_\text{max}, \symRegRange)$
\State $\symZ \gets 0$
\For{$\symRegVal \gets \symRegValVariate'_\text{max},\symRegValVariate'_\text{min}$}
\State $\symZ \gets 0.5\cdot\symZ + \symCountVariate_\symRegVal$
\EndFor
\State $\symZ \gets \symZ\cdot 2^{-\symRegValVariate'_\text{min}}$
\comm{0.31}{here $\symZ=\sum_{\symRegVal=1}^{\symRegRange}\frac{\symCountVariate_\symRegVal}{2^\symRegVal}$}
\State $\symCount \gets \symCountVariate_{\symRegRange+1}$
\If{$\symRegRange\geq 1$}
\State $\symCount \gets \symCount + \symCountVariate_{\symRegValVariate'_\text{max}}$
\EndIf
\State $\symFuncPrime_\text{prev}\gets 0$
\State $\symA\gets \symZ + \symCountVariate_0$
\comm{0.31}{$\symA = \sum_{\symRegVal=0}^{\symRegRange}\frac{\symCountVariate_\symRegVal}{2^\symRegVal}$}
\State $\symB\gets \symZ + 
\symCountVariate_{\symRegRange+1}\cdot 2^{-\symRegRange}$
\comm{0.31}{$\symB = \sum_{\symRegVal=1}^{\symRegRange}\frac{\symCountVariate_\symRegVal}{2^\symRegVal} + \frac{\symCountVariate_{\symRegRange+1}}{2^\symRegRange}$}
\State $\symNumReg' \gets \symNumReg - \symCountVariate_0$
\If{$\symB \leq 1.5\cdot\symA$}
\State $\symX \gets \symNumReg'/(0.5\cdot \symB+\symA)$ \comm{0.31}{weak lower bound \eqref{equ:weak_lower_bound}}
\Else
\State $\symX \gets \symNumReg'/\symB \cdot \log(1+\symB/\symA)$ 
\comm{0.31}{strong lower bound \eqref{equ:strong_lower_bound}}
\EndIf
\algstore{myalg}
\end{algorithmic}
\end{algorithm}

\begin{algorithm}
\caption{(continued)}
\begin{algorithmic}
\algrestore{myalg}
\State $\Delta\symX \gets \symX$
\While{$\Delta\symX > \symX\cdot\symStopDelta$} 
\comm{0.31}{secant method iteration\\
$\symStopDelta=\symStopEpsilon/\sqrt{\symNumReg}$, $\symStopEpsilon = 10^{-2}$}
\State $\symKappa \gets 2+\lfloor\log_2(\symX)\rfloor$
\State $\symX' \gets
\symX \cdot2^{-\max(\symRegValVariate'_\text{max}, \symKappa)-1}$
\comm{0.31}{$\symX' \in [0, 0.25]$}
\State $\symX''\gets \symX'\cdot\symX'$
\State $\symHelper \gets
\symX' - \symX''/3 + \left(\symX''\cdot \symX''\right)\cdot\left(1/45-\symX''/472.5\right)$
\comm{0.31}{Taylor approximation \eqref{equ:taylor}}
\For{$\symRegVal\gets (\symKappa - 1),\symRegValVariate'_\text{max}$}
\State $\symHelper \gets \frac{\symX'+\symHelper\cdot(1-\symHelper)}{\symX'+(1-\symHelper)}$
\comm{0.31}{calculate $\symHelper\!\left(\frac{\symX}{2^{\symRegVal}}\right)$, see \eqref{equ:helper_recursion1}, at this point $\symX' = \frac{\symX}{2^{\symRegVal+2}}$}
\State $\symX' \gets 2\symX'$
\EndFor
\State $\symFuncPrime \gets \symCount\cdot\symHelper$
\comm{0.31}{compare \eqref{equ:funcprime}}
\For{$\symRegVal\gets(\symRegValVariate'_\text{max}-1),\symRegValVariate'_\text{min}$}
\State $\symHelper \gets \frac{\symX'+\symHelper\cdot(1-\symHelper)}{\symX'+(1-\symHelper)}$
\comm{0.31}{calculate $\symHelper\!\left(\frac{\symX}{2^{\symRegVal}}\right)$, see \eqref{equ:helper_recursion1}, at this point $\symX' = \frac{\symX}{2^{\symRegVal+2}}$}
\State $\symFuncPrime\gets \symFuncPrime + \symCountVariate_\symRegVal\cdot\symHelper$
\State $\symX' \gets 2\symX'$
\EndFor
\State $\symFuncPrime\gets \symFuncPrime + \symX\cdot\symA$
\If{$\symFuncPrime > \symFuncPrime_\text{prev} \wedge \symNumReg' \geq\symFuncPrime$}
\State $\Delta\symX \gets \Delta\symX \cdot \frac{\symNumReg' - \symFuncPrime}{\symFuncPrime - \symFuncPrime_\text{prev}}$
\comm{0.31}{see \eqref{equ:secant_delta}}
\Else
\State $\Delta\symX \gets 0$
\EndIf
\State $\symX \gets \symX + \Delta\symX$
\State $\symFuncPrime_\text{prev}\gets \symFuncPrime$
\EndWhile
\State \Return $\symNumReg\cdot\symX$
\EndFunction
\end{algorithmic}
\end{algorithm}

\subsection{Estimation error}
\label{sec:maximum_likelihood_estimation_error}
\begin{figure}
\centering
\includegraphics[width=1\textwidth]{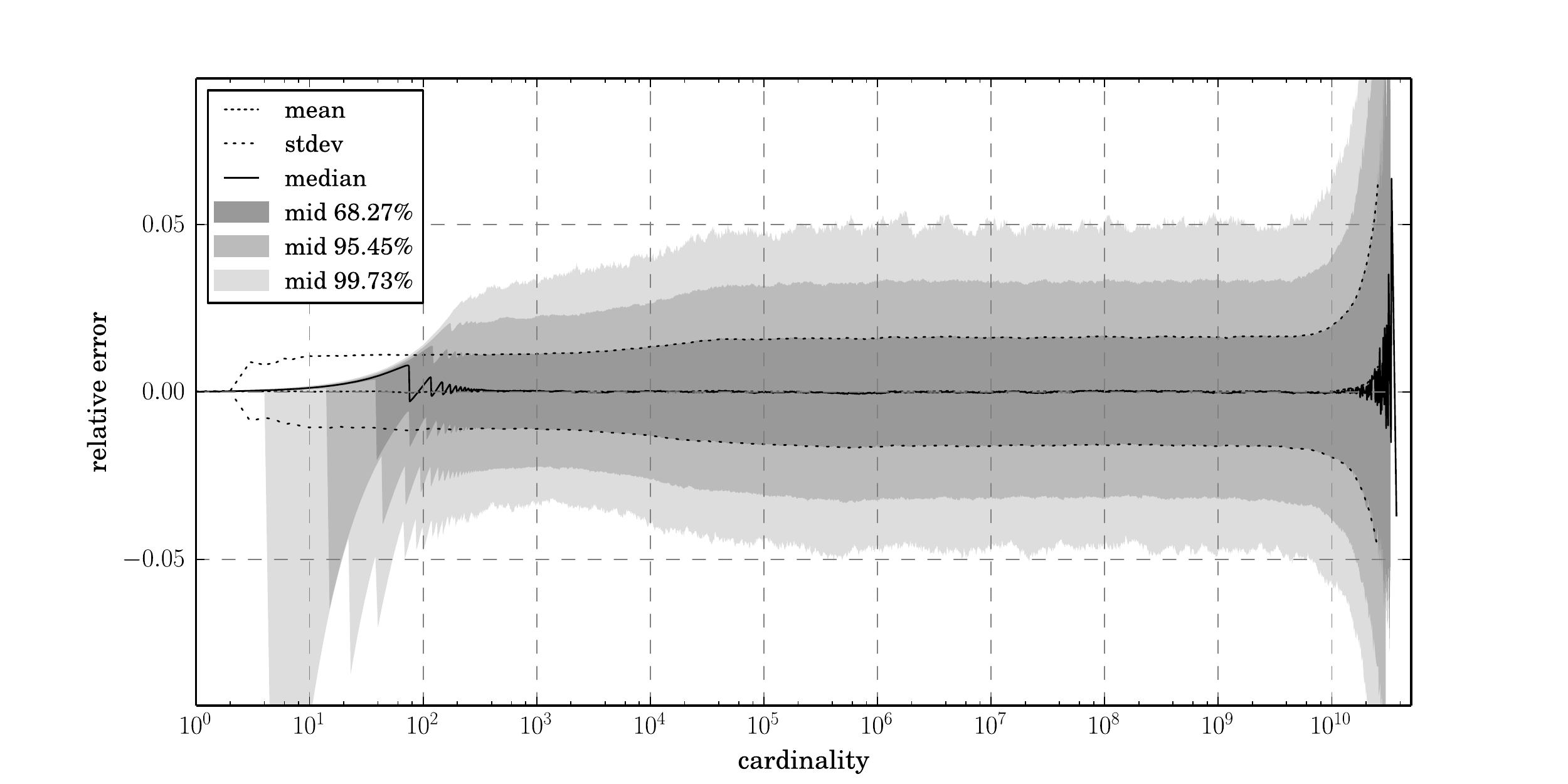}
\caption{Relative error of the maximum likelihood estimates as a function of the true cardinality for a HyperLogLog sketch with parameters $\symPrecision = 12$ and $\symRegRange=20$.}
\label{fig:max_likelihood_estimation_error_12_20}
\end{figure}

\begin{figure}
\centering
\includegraphics[width=1\textwidth]{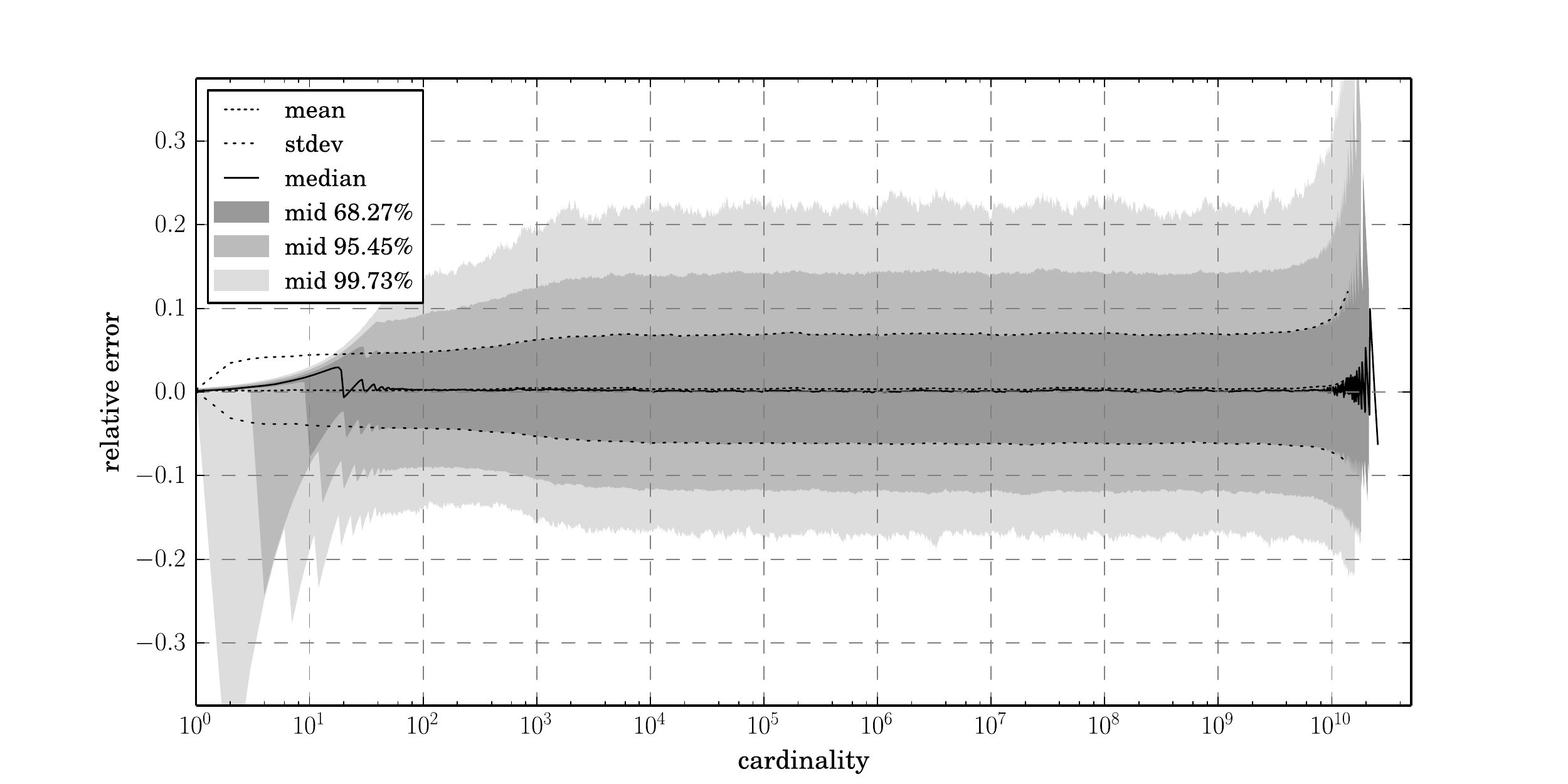}
\caption{Relative error of the maximum likelihood estimates as a function of the true cardinality for a HyperLogLog sketch with parameters $\symPrecision = 8$ and $\symRegRange=24$.}
\label{fig:max_likelihood_estimation_error_8_24}
\end{figure}

\begin{figure}
\centering
\includegraphics[width=1\textwidth]{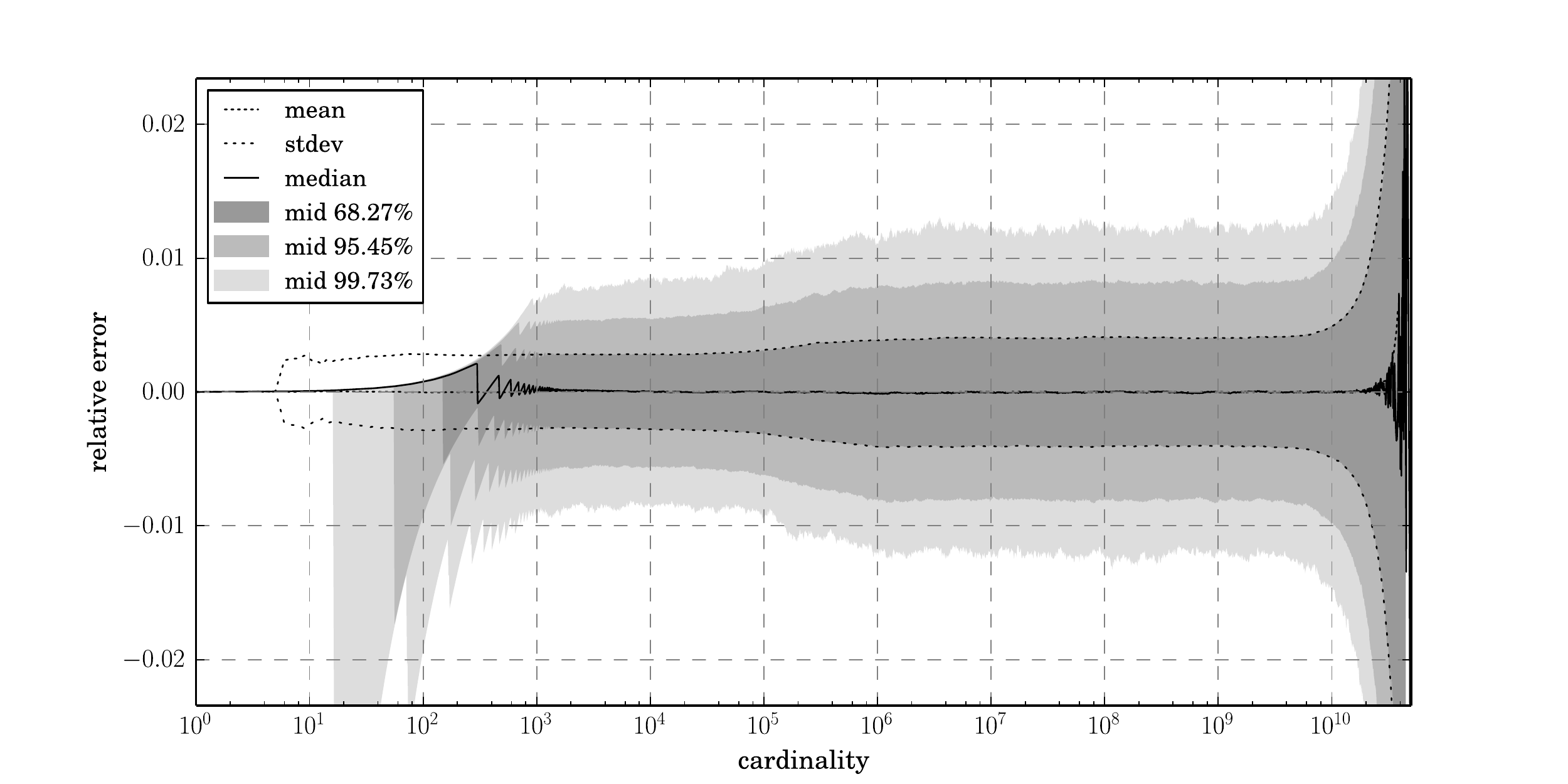}
\caption{Relative error of the maximum likelihood estimates as a function of the true cardinality for a HyperLogLog sketch with parameters $\symPrecision = 16$ and $\symRegRange=16$.}
\label{fig:max_likelihood_estimation_error_16_16}
\end{figure}

\begin{figure}
\centering
\includegraphics[width=1\textwidth]{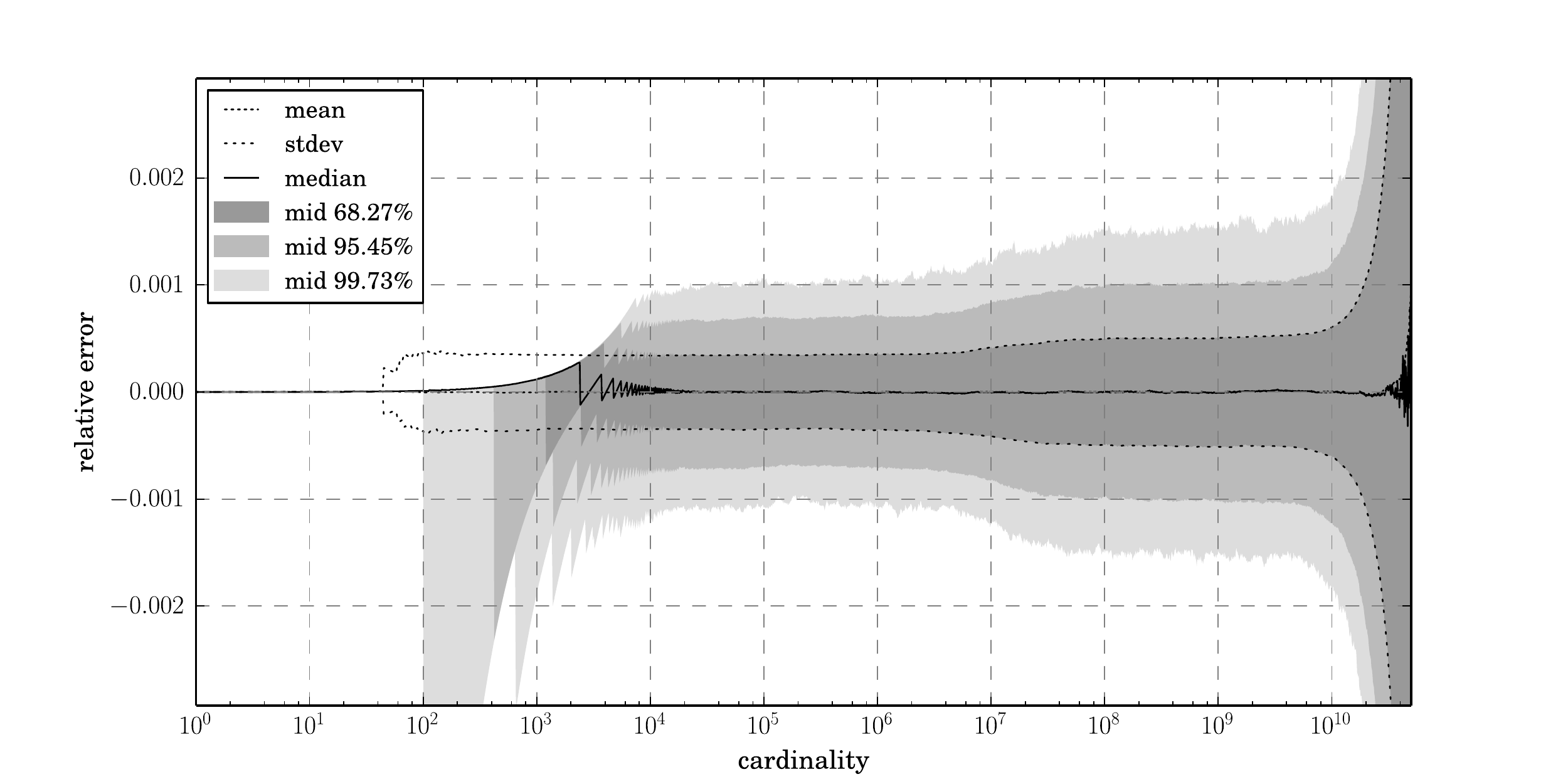}
\caption{Relative error of the maximum likelihood estimates as a function of the true cardinality for a HyperLogLog sketch with parameters $\symPrecision = 22$ and $\symRegRange=10$.}
\label{fig:max_likelihood_estimation_error_22_10}
\end{figure}

\begin{figure}
\centering
\includegraphics[width=1\textwidth]{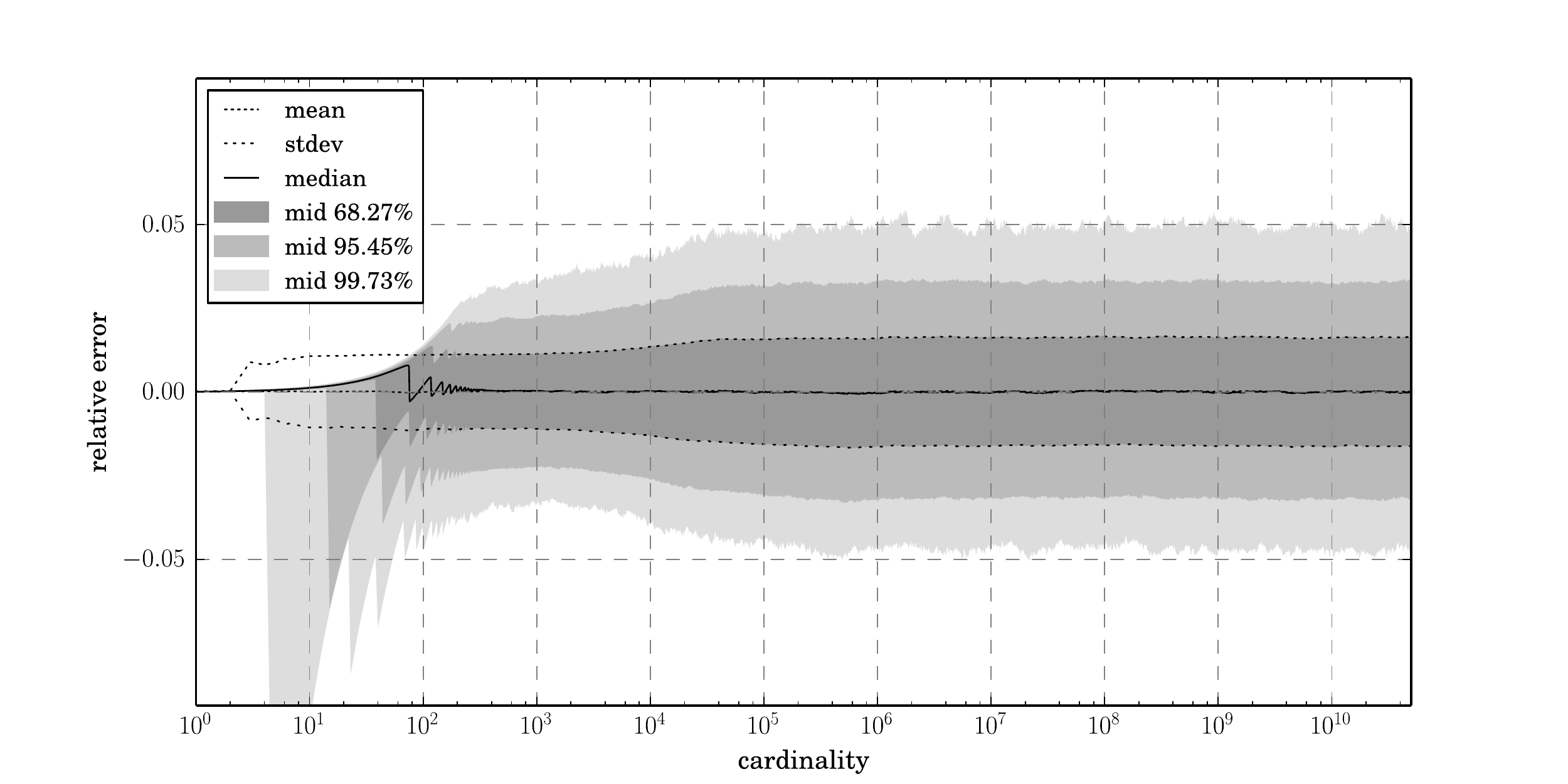}
\caption{Relative error of the maximum likelihood estimates as a function of the true cardinality for a HyperLogLog sketch with parameters $\symPrecision = 12$ and $\symRegRange=52$.}
\label{fig:max_likelihood_estimation_error_12_52}
\end{figure}

\begin{figure}
\centering
\includegraphics[width=1\textwidth]{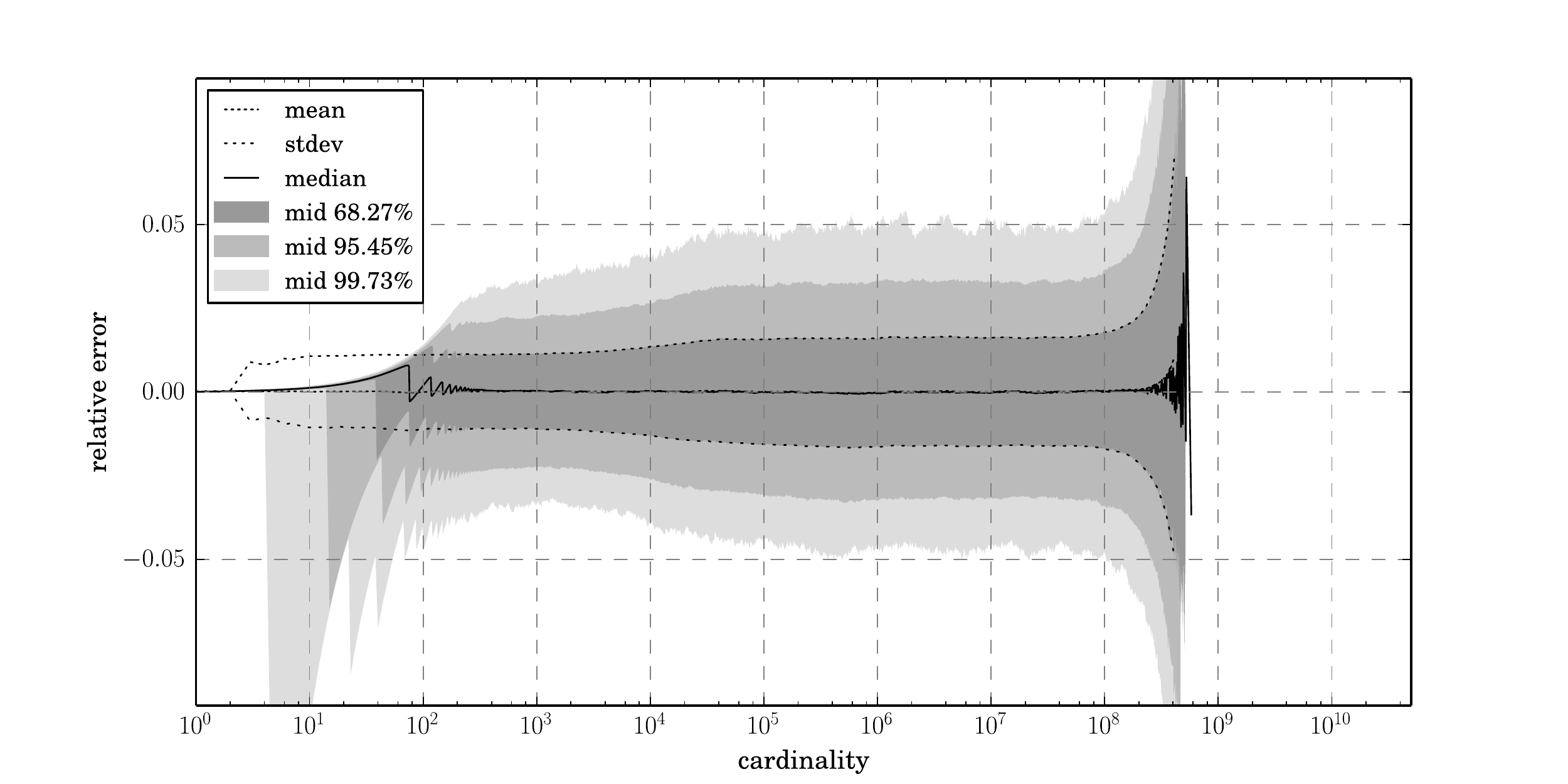}
\caption{Relative error of the maximum likelihood estimates as a function of the true cardinality for a HyperLogLog sketch with parameters $\symPrecision = 12$ and $\symRegRange=14$.}
\label{fig:max_likelihood_estimation_error_12_14}
\end{figure}

We have investigated the estimation error of the maximum likelihood estimation algorithm for the same HyperLogLog configurations as for the improved raw estimation algorithm in \cref{sec:corrected_raw_estimation_error}. 
\cref{fig:max_likelihood_estimation_error_12_20,fig:max_likelihood_estimation_error_8_24,fig:max_likelihood_estimation_error_16_16,fig:max_likelihood_estimation_error_22_10,fig:max_likelihood_estimation_error_12_52,fig:max_likelihood_estimation_error_12_14} show very similar results for same HyperLogLog parameters. 
What is different for the maximum likelihood estimation approach is a somewhat smaller median bias with less oscillations around zero for small cardinalities. The standard deviation of the relative error is also slightly better for the maximum likelihood estimator than for the improved raw estimator as shown in \cref{fig:stdev_comparison}. Furthermore, contrary to the improved raw estimator which reveals a small oscillating bias for the mean (see \cref{fig:raw_corrected_estimation_error_22_10}), the maximum likelihood estimator seems to be completely unbiased (see \cref{fig:max_likelihood_estimation_error_22_10}).

\subsection{Performance}
\begin{figure}
\centering
\includegraphics[width=1\textwidth]{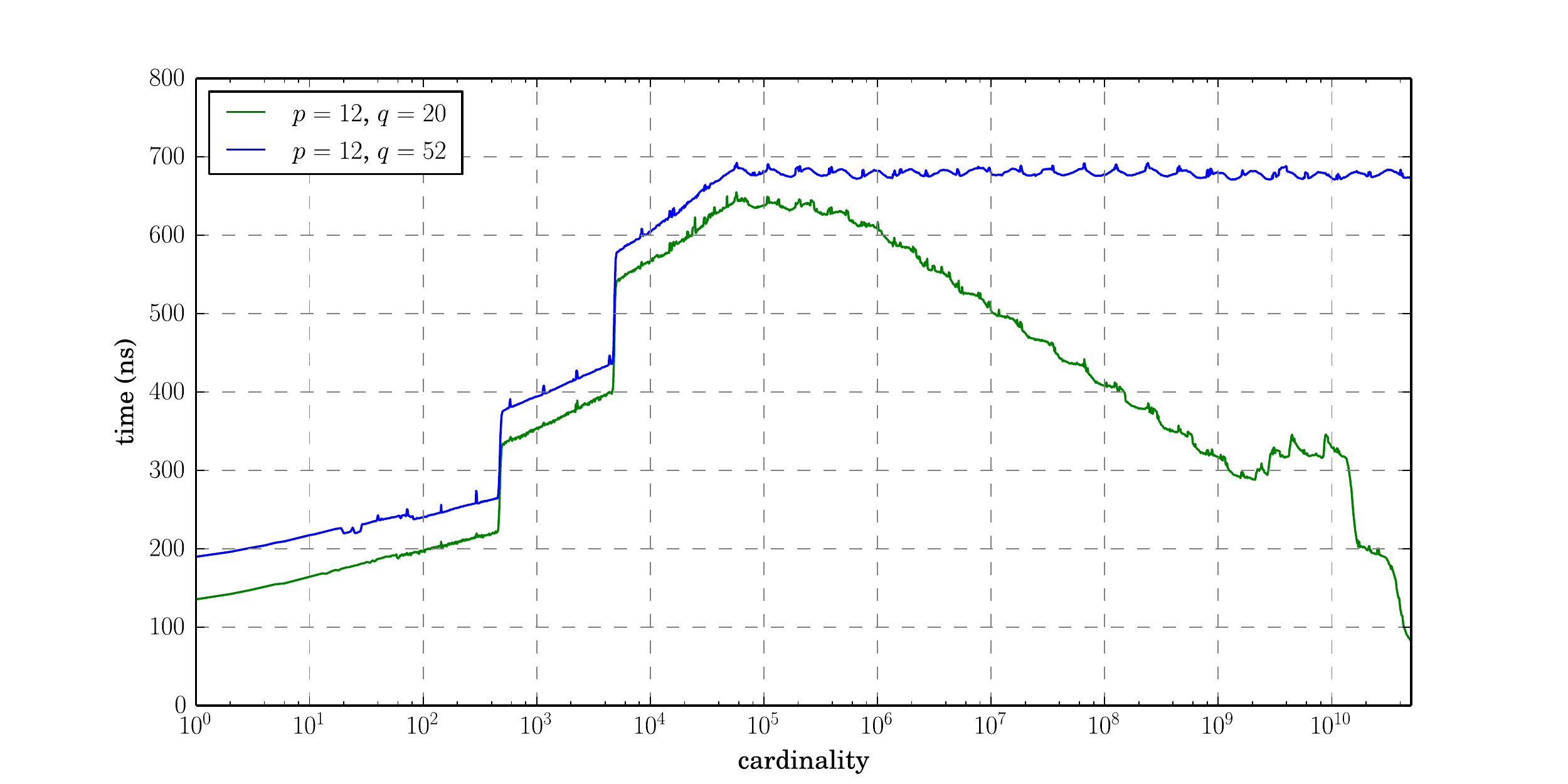}
\caption{Average execution time of the maximum likelihood estimation algorithm as a function of the true cardinality with an Intel Core i5-2500K clocking at \SI{3.3}{\giga\hertz} for HyperLogLog sketches with parameters $\symPrecision=12$, $\symRegRange=20$ and $\symPrecision=12$, $\symRegRange=52$, respectively.}
\label{fig:avg_exec_time}
\end{figure}

We also measured the performance of \cref{alg:estimate_ml} using the same test setup as described in \cref{sec:corrected_raw_estimation_algorithm}. The results for HyperLogLog configurations $\symPrecision=12, \symRegRange=20$ and $\symPrecision=12, \symRegRange=52$ are shown in \cref{fig:avg_exec_time}. The average computation time for the maximum likelihood algorithm shows a different behavior than for the improved raw estimation algorithm (compare \cref{fig:corrected_raw_avg_exec_time}). The average execution time is larger for most cardinalities, but nevertheless, it never exceeds \SI{700}{\nano\second} which is still fast enough for many applications. The steps in the chart can be explained by different numbers of iteration cycles until the secant method is stopped. For example, at a cardinality value around \num{5000} the average number of required cycles until the stop criterion is satisfied increases abruptly from two to three. More than three iteration cycles have never been observed for any cardinality estimate in this performance test.

\section{Cardinality estimation of set intersections and complements}
\label{sec:cardinality_estimation_set_intersections}

While the union of two sets that are represented by HyperLogLog sketches can be straightforwardly computed using \cref{alg:union}, the computation of cardinalities of other set operations like intersections and complements is more challenging. The conventional approach uses the inclusion-exclusion principle 
\begin{equation}
\label{equ:conventional_approach}
\begin{aligned}
\left\vert\symSetS_1 \setminus \symSetS_2\right\vert
&=
\left\vert\symSetS_1 \cup \symSetS_2\right\vert
-
\left\vert\symSetS_2\right\vert,
\\
\left\vert\symSetS_2 \setminus \symSetS_1\right\vert
&=
\left\vert\symSetS_1 \cup \symSetS_2\right\vert
-
\left\vert\symSetS_1\right\vert,
\\
\left\vert\symSetS_1 \cap \symSetS_2\right\vert
&=  
\left\vert\symSetS_1\right\vert  
+
\left\vert\symSetS_2\right\vert
-
\left\vert\symSetS_1 \cup \symSetS_2\right\vert,
\end{aligned}
\end{equation}
and the fact that HyperLogLog sketches can be easily merged using \cref{alg:union}. Unfortunately, the estimation error does not scale well for this approach. Especially for small Jaccard indices, the relative estimation error can become very large \cite{Dasgupta2015}. In the worst case, the estimate could be negative without artificial restriction to nonnegative values. Therefore, it was proposed to combine HyperLogLog sketches with minwise hashing \cite{Pascoe2013, Cohen2016}, which improves the estimation error, even though at the expense of significant more space consumption. 

It was recently pointed out without special focus on HyperLogLog sketches, that the application of the maximum likelihood method to the joint likelihood function of two probabilistic data structures, which represent the intersection operands, gives better intersection size estimates \cite{Ting2016}. For HyperLogLog sketches recorded without stochastic averaging (compare \cref{sec:data_element_insertion}) this was shown in \cite{Cohen2016}, where also an algorithm based on the maximum likelihood principle was outlined. However, in practice, HyperLogLog sketches are recorded using stochastic averaging because of the much cheaper element insertions. Motivated by the good results we have obtained for a single HyperLogLog sketch using the maximum likelihood method in combination with the Poisson approximation, we are tempted to apply this approach also for the estimation of set operation result sizes.

Assume two given  HyperLogLog sketches with register values $\boldsymbol{\symRegValVariate}_1$ and $\boldsymbol{\symRegValVariate}_2$ representing the sets $\symSetS_1$ and $\symSetS_2$, respectively. The goal is to find estimates for the cardinalities of the pairwise disjoint sets $\symSetX = \symSetS_1\cap\symSetS_2$, $\symSetA = \symSetS_1\setminus\symSetS_2$, and $\symSetB = \symSetS_2\setminus\symSetS_1$. The Poisson approximation allows us to assume that pairwise distinct elements are inserted into the HyperLogLog sketches representing $\symSetS_1$ and $\symSetS_2$ at rates $\symPoissonRate_\symSetASuffix$ and $\symPoissonRate_\symSetBSuffix$, respectively. Furthermore, we assume that further unique elements are inserted into both HyperLogLog sketches simultaneously at rate $\symPoissonRate_\symSetXSuffix$. We expect that good estimates $\symPoissonRateEstimate_\symSetASuffix$, $\symPoissonRateEstimate_\symSetBSuffix$, and $\symPoissonRateEstimate_\symSetXSuffix$ for the rates are also good estimates for the cardinalities of $\symSetA$, $\symSetB$, and $\symSetX$. 

\subsection{Joint log-likelihood function}
In order to get maximum likelihood estimators for $\symPoissonRateEstimate_\symSetASuffix$, $\symPoissonRateEstimate_\symSetBSuffix$, and $\symPoissonRateEstimate_\symSetXSuffix$ we need to derive the joint probability distribution of two HyperLogLog sketches. Under the Poisson model the register values are independent and identically distributed. Therefore, we first derive the joint probability distribution for a single register that has value $\symRegValVariate_1$ in the first HyperLogLog sketch representing $\symSetS_1$ and value $\symRegValVariate_2$ in the second HyperLogLog sketch representing $\symSetS_2$. 

The HyperLogLog sketch that represents $\symSetS_1$ can be thought to be constructed from two HyperLogLog sketches representing $\symSetA$ and $\symSetX$ and merging both using \cref{alg:union}. Analogously, the HyperLogLog sketch for $\symSetS_2$ could have been obtained from sketches for $\symSetB$ and $\symSetX$. Let  $\symRegValVariate_\symSetASuffix$, $\symRegValVariate_\symSetBSuffix$, and $\symRegValVariate_\symSetXSuffix$ be the value of the considered register in the HyperLogLog sketch representing $\symSetA$, $\symSetB$, and $\symSetX$, respectively. The corresponding values in sketches for $\symSetS_1$ and $\symSetS_2$ are given by
\begin{equation}
\symRegValVariate_1 = \max\!\left(\symRegValVariate_\symSetASuffix, \symRegValVariate_\symSetXSuffix\right)
,\quad
\symRegValVariate_2 = \max\!\left(\symRegValVariate_\symSetBSuffix, \symRegValVariate_\symSetXSuffix\right).
\end{equation}
Their joint cumulative probability function is given as
\begin{align}
\symProbability\!\left(
\symRegValVariate_1 \leq \symRegVal_1
\wedge
\symRegValVariate_2 \leq \symRegVal_2
\right)
&=
\symProbability\!\left(
\max\!\left(\symRegValVariate_\symSetASuffix, \symRegValVariate_\symSetXSuffix\right) \leq \symRegVal_1
\wedge
\max\!\left(\symRegValVariate_\symSetBSuffix, \symRegValVariate_\symSetXSuffix\right) \leq \symRegVal_2
\right)
\nonumber\\
&=
\symProbability
\!\left(
\symRegValVariate_\symSetASuffix \leq \symRegVal_1
\wedge
\symRegValVariate_\symSetBSuffix \leq \symRegVal_2
\wedge
\symRegValVariate_\symSetXSuffix \leq \min\!\left(\symRegVal_1, \symRegVal_2\right)
\right)
\nonumber\\
&=
\symProbability
\!\left(
\symRegValVariate_\symSetASuffix \leq \symRegVal_1
\right)
\symProbability
\!\left(
\symRegValVariate_\symSetBSuffix \leq \symRegVal_2
\right)
\symProbability
\!\left(
\symRegValVariate_\symSetXSuffix \leq \min\!\left(\symRegVal_1, \symRegVal_2\right)
\right).
\end{align}
Here the last transformation used the independence of $\symRegValVariate_\symSetASuffix$, $\symRegValVariate_\symSetBSuffix$, and $\symRegValVariate_\symSetXSuffix$, because by definition, the sets $\symSetA$, $\symSetB$, and $\symSetX$ are pairwise disjoint. Furthermore, under the Poisson model $\symRegValVariate_\symSetASuffix$, $\symRegValVariate_\symSetBSuffix$, and $\symRegValVariate_\symSetXSuffix$ can be described by \eqref{equ:register_value_distribution}. If we take into account that pairwise distinct elements are added to $\symSetA$, $\symSetB$, and $\symSetX$ at rates $\symPoissonRate_\symSetASuffix$, $\symPoissonRate_\symSetBSuffix$, and $\symPoissonRate_\symSetXSuffix$, respectively, the probability that a certain register has a value less than or equal to $\symRegVal_1$ in the first HyperLogLog sketch and simultaneously a value less than or equal to $\symRegVal_2$ in the second one can be written as
\begin{equation}
\symProbability(
\symRegValVariate_1 \leq \symRegVal_1
\wedge
\symRegValVariate_2 \leq \symRegVal_2
)
=\begin{cases}
0 & \symRegVal_1 < 0 \vee \symRegVal_2 < 0
\\
e^{
-
\frac{\symPoissonRate_{\symSetASuffix}}{\symNumReg 2^{\symRegVal_1}}
-
\frac{\symPoissonRate_{\symSetBSuffix}}{\symNumReg 2^{\symRegVal_2}}
-
\frac{\symPoissonRate_{\symSetXSuffix}}{\symNumReg 2^{\min(\symRegVal_1, \symRegVal_2)}}
}
& 0\leq\symRegVal_1 \leq \symRegRange \wedge 0\leq\symRegVal_2\leq\symRegRange
\\
e^{
-
\frac{\symPoissonRate_{\symSetBSuffix} + \symPoissonRate_{\symSetXSuffix}}{\symNumReg 2^{\symRegVal_2}}
}
& 0\leq\symRegVal_2 \leq \symRegRange < \symRegVal_1
\\
e^{
-
\frac{\symPoissonRate_{\symSetASuffix} + \symPoissonRate_{\symSetXSuffix}}{\symNumReg 2^{\symRegVal_1}}
}
&  0\leq\symRegVal_1 \leq \symRegRange < \symRegVal_2
\\
1
&
\symRegRange < \symRegVal_1 \wedge \symRegRange < \symRegVal_2.
\end{cases}
\end{equation}

The joint probability mass function for both register values can be calculated using
\begin{multline}
\symProbabilityMass(\symRegVal_1,\symRegVal_2)
=
\symProbability(
\symRegValVariate_1 \leq \symRegVal_1
\wedge
\symRegValVariate_2 \leq \symRegVal_2
)
-
\symProbability(
\symRegValVariate_1 \leq \symRegVal_1-1
\wedge
\symRegValVariate_2 \leq \symRegVal_2
)
\\
-\symProbability(
\symRegValVariate_1 \leq \symRegVal_1
\wedge
\symRegValVariate_2 \leq \symRegVal_2-1
)
+\symProbability(
\symRegValVariate_1 \leq \symRegVal_1-1
\wedge
\symRegValVariate_2 \leq \symRegVal_2-1
),
\end{multline}
which finally gives
\begin{multline}
\label{equ:register_value_joint_pmf}
\symProbabilityMass(\symRegVal_1,\symRegVal_2)
=
\\
\begin{cases}
e^{-\frac{\symPoissonRate_\symSetASuffix + \symPoissonRate_\symSetXSuffix}
{\symNumReg }
-\frac{\symPoissonRate_\symSetBSuffix}{\symNumReg 2^{\symRegVal_2}}
}
\left(
1-
e^{
-\frac{\symPoissonRate_\symSetBSuffix}{\symNumReg 2^{\symRegVal_2}}
}
\right)
&
0 = \symRegVal_1 < \symRegVal_2 \leq \symRegRange
\\
e^{-\frac{\symPoissonRate_\symSetASuffix + \symPoissonRate_\symSetXSuffix}
{\symNumReg}
}
\left(
1-
e^{
-\frac{\symPoissonRate_\symSetBSuffix}{\symNumReg 2^{\symRegRange}}
}
\right)
&
0 = \symRegVal_1 < \symRegVal_2 = \symRegRange + 1
\\
e^{-\frac{\symPoissonRate_\symSetASuffix + \symPoissonRate_\symSetXSuffix}
{\symNumReg 2^{\symRegVal_1}}
-\frac{\symPoissonRate_\symSetBSuffix}{\symNumReg 2^{\symRegVal_2}}
}
\left(
1-
e^{-\frac{\symPoissonRate_\symSetASuffix + \symPoissonRate_\symSetXSuffix}{\symNumReg 2^{\symRegVal_1}}
}
\right)
\left(
1-
e^{
-\frac{\symPoissonRate_\symSetBSuffix}{\symNumReg 2^{\symRegVal_2}}
}
\right)
&
1 \leq \symRegVal_1 < \symRegVal_2 \leq \symRegRange
\\
e^{-\frac{\symPoissonRate_\symSetASuffix + \symPoissonRate_\symSetXSuffix}
{\symNumReg 2^{\symRegVal_1}}
}
\left(
1-
e^{-\frac{\symPoissonRate_\symSetASuffix + \symPoissonRate_\symSetXSuffix}{\symNumReg 2^{\symRegVal_1}}
}
\right)
\left(
1-
e^{
-\frac{\symPoissonRate_\symSetBSuffix}{\symNumReg 2^{\symRegRange}}
}
\right)
&
1 \leq \symRegVal_1 < \symRegVal_2 = \symRegRange + 1
\\
e^{-\frac{\symPoissonRate_\symSetBSuffix + \symPoissonRate_\symSetXSuffix}
{\symNumReg }
-\frac{\symPoissonRate_\symSetASuffix}{\symNumReg 2^{\symRegVal_1}}
}
\left(
1-
e^{
-\frac{\symPoissonRate_\symSetASuffix}{\symNumReg 2^{\symRegVal_1}}
}
\right)
&
0 = \symRegVal_2 < \symRegVal_1 \leq \symRegRange
\\
e^{-\frac{\symPoissonRate_\symSetBSuffix + \symPoissonRate_\symSetXSuffix}
{\symNumReg}
}
\left(
1-
e^{
-\frac{\symPoissonRate_\symSetASuffix}{\symNumReg 2^{\symRegRange}}
}
\right)
&
0 = \symRegVal_2 < \symRegVal_1 = \symRegRange + 1
\\
e^{-\frac{\symPoissonRate_\symSetBSuffix + \symPoissonRate_\symSetXSuffix}
{\symNumReg 2^{\symRegVal_2}}
-\frac{\symPoissonRate_\symSetASuffix}{\symNumReg 2^{\symRegVal_1}}
}
\left(
1-
e^{-\frac{\symPoissonRate_\symSetBSuffix + \symPoissonRate_\symSetXSuffix}{\symNumReg 2^{\symRegVal_2}}
}
\right)
\left(
1-
e^{
-\frac{\symPoissonRate_\symSetASuffix}{\symNumReg 2^{\symRegVal_1}}
}
\right)
&
1 \leq \symRegVal_2 < \symRegVal_1 \leq \symRegRange
\\
e^{-\frac{\symPoissonRate_\symSetBSuffix + \symPoissonRate_\symSetXSuffix}
{\symNumReg 2^{\symRegVal_2}}
}
\left(
1-
e^{-\frac{\symPoissonRate_\symSetBSuffix + \symPoissonRate_\symSetXSuffix}{\symNumReg 2^{\symRegVal_2}}
}
\right)
\left(
1-
e^{
-\frac{\symPoissonRate_\symSetASuffix}{\symNumReg 2^{\symRegRange}}
}
\right)
&
1 \leq \symRegVal_2 < \symRegVal_1 = \symRegRange + 1
\\
e^{-\frac{\symPoissonRate_\symSetASuffix + \symPoissonRate_\symSetBSuffix + \symPoissonRate_\symSetXSuffix}
{\symNumReg}
}
&
0 = \symRegVal_1 = \symRegVal_2
\\
e^{-\frac{\symPoissonRate_\symSetASuffix + \symPoissonRate_\symSetBSuffix + \symPoissonRate_\symSetXSuffix}
{\symNumReg 2^\symRegVal}
}
\left(
1
-
e^{-\frac{\symPoissonRate_\symSetASuffix +  \symPoissonRate_\symSetXSuffix}
{\symNumReg 2^\symRegVal}
}
-
e^{-\frac{\symPoissonRate_\symSetBSuffix + \symPoissonRate_\symSetXSuffix}
{\symNumReg 2^\symRegVal}
}
+
e^{-\frac{\symPoissonRate_\symSetASuffix + \symPoissonRate_\symSetBSuffix + \symPoissonRate_\symSetXSuffix}
{\symNumReg 2^\symRegVal}
}
\right)
&
1 \leq \symRegVal_1 = \symRegVal_2 = \symRegVal\leq \symRegRange
\\
1
-
e^{-\frac{\symPoissonRate_\symSetASuffix +  \symPoissonRate_\symSetXSuffix}
{\symNumReg 2^\symRegRange}
}
-
e^{-\frac{\symPoissonRate_\symSetBSuffix + \symPoissonRate_\symSetXSuffix}
{\symNumReg 2^\symRegRange}
}
+
e^{-\frac{\symPoissonRate_\symSetASuffix + \symPoissonRate_\symSetBSuffix + \symPoissonRate_\symSetXSuffix}
{\symNumReg 2^\symRegRange}
}
&
\symRegVal_1 = \symRegVal_2 = \symRegRange + 1.
\end{cases}
\end{multline}

The logarithm of the joint probability mass function can be written using Iverson bracket notation ($\left[\text{true}\right]:=1$, $\left[\text{false}\right]:=0$) as
\begin{multline}
\label{equ:joint_log_pmf_single_register}
\log(\symProbabilityMass(\symRegVal_1,\symRegVal_2))
=\\
\begin{aligned}
&
\log\!\left(1-e^{-\frac{\symPoissonRate_\symSetASuffix+\symPoissonRate_\symSetXSuffix}{\symNumReg 2^{\symRegVal_1}}}\right)
\left[1\leq\symRegVal_1<\symRegVal_2\right]
+
\log\!\left(1-e^{-\frac{\symPoissonRate_\symSetASuffix}{\symNumReg 2^{\min\left(\symRegVal_1, \symRegRange\right)}}}\right)
\left[\symRegVal_2<\symRegVal_1\right]
\\
+
&
\log\!\left(1-e^{-\frac{\symPoissonRate_\symSetBSuffix+\symPoissonRate_\symSetXSuffix}{\symNumReg 2^{\symRegVal_2}}}\right)
\left[1\leq\symRegVal_2<\symRegVal_1\right]
+
\log\!\left(1-e^{-\frac{\symPoissonRate_\symSetBSuffix}{\symNumReg 2^{\min\left(\symRegVal_2, \symRegRange\right)}}}\right)
\left[\symRegVal_1<\symRegVal_2\right]
\\
+
&
\log\!\left(
1
-e^{-\frac{\symPoissonRate_\symSetASuffix+\symPoissonRate_\symSetXSuffix}{\symNumReg 2^{\min\left(\symRegVal_1, \symRegRange\right)}}}
-
e^{-\frac{\symPoissonRate_\symSetBSuffix+\symPoissonRate_\symSetXSuffix}{\symNumReg 2^{\min\left(\symRegVal_1, \symRegRange\right)}}}
+
e^{-\frac{\symPoissonRate_\symSetASuffix+\symPoissonRate_\symSetBSuffix+\symPoissonRate_\symSetXSuffix}{\symNumReg 2^{\min\left(\symRegVal_1, \symRegRange\right)}}}
\right)
\left[1\leq\symRegVal_1=\symRegVal_2\right]
\\
-
&
\frac{\symPoissonRate_\symSetASuffix}{\symNumReg 2^{\symRegVal_1}}
\left[\symRegVal_1\leq\symRegRange\right]
-
\frac{\symPoissonRate_\symSetBSuffix}{\symNumReg 2^{\symRegVal_2}}
\left[\symRegVal_2\leq\symRegRange\right]
-
\frac{\symPoissonRate_\symSetXSuffix}{\symNumReg 2^{\min\left(\symRegVal_1,\symRegVal_2\right)}}
\left[\symRegVal_1\leq\symRegRange\vee\symRegVal_2\leq\symRegRange\right].
\end{aligned}
\end{multline}

Since the values for different registers are independent under the Poisson model, we are now able to write the joint probability mass function for all registers in both HyperLogLog sketches
\begin{equation}
\symProbabilityMass(\boldsymbol{\symRegVal}_1,\boldsymbol{\symRegVal}_2)
=
\prod_{\symIndexI = 1}^{\symNumReg}
\symProbabilityMass(\symRegVal_{1\symIndexI},\symRegVal_{2\symIndexI}).
\end{equation}

The maximum likelihood estimates $\symPoissonRateEstimate_\symSetASuffix$,
 $\symPoissonRateEstimate_\symSetBSuffix$, and  $\symPoissonRateEstimate_\symSetXSuffix$ are obtained by maximization of the log-likelihood function given by
\begin{equation}
\log \symLikelihood(
\symPoissonRate_\symSetASuffix,
\symPoissonRate_\symSetBSuffix,
\symPoissonRate_\symSetXSuffix
\vert
\boldsymbol{\symRegValVariate}_1,
\boldsymbol{\symRegValVariate}_2
)
=
\sum_{\symIndexI = 1}^{\symNumReg}
\log(\symProbabilityMass(\symRegValVariate_{1\symIndexI},\symRegValVariate_{2\symIndexI})).
\end{equation}
Insertion of \eqref{equ:joint_log_pmf_single_register} results in
\begin{multline}
\label{equ:log_likelihood_pair}
\log \symLikelihood(
\symPoissonRate_\symSetASuffix,
\symPoissonRate_\symSetBSuffix,
\symPoissonRate_\symSetXSuffix
\vert
\boldsymbol{\symRegValVariate}_1,
\boldsymbol{\symRegValVariate}_2
)
=
\\
\begin{aligned}
&
\sum_{\symRegVal=1}^{\symRegRange}
\log\!\left(1-e^{-\frac{\symPoissonRate_\symSetASuffix+\symPoissonRate_\symSetXSuffix}{\symNumReg 2^{\symRegVal}}}\right)
\symCountVariate^{<}_{1\symRegVal}
+
\log\!\left(1-e^{-\frac{\symPoissonRate_\symSetBSuffix+\symPoissonRate_\symSetXSuffix}{\symNumReg 2^{\symRegVal}}}\right)
\symCountVariate^{<}_{2\symRegVal}
\\
+
&
\sum_{\symRegVal=1}^{\symRegRange+1}
\log\!\left(1-e^{-\frac{\symPoissonRate_\symSetASuffix}{\symNumReg 2^{\min\left(\symRegVal,\symRegRange\right)}}}\right)
\symCountVariate^{>}_{1\symRegVal}
+
\log\!\left(1-e^{-\frac{\symPoissonRate_\symSetBSuffix}{\symNumReg 2^{\min\left(\symRegVal,\symRegRange\right)}}}\right)
\symCountVariate^{>}_{2\symRegVal}
\\
+
&
\sum_{\symRegVal=1}^{\symRegRange+1}
\log\!\left(
1
-e^{-\frac{\symPoissonRate_\symSetASuffix+\symPoissonRate_\symSetXSuffix}{\symNumReg 2^{\min\left(\symRegVal,\symRegRange\right)}}}
-
e^{-\frac{\symPoissonRate_\symSetBSuffix+\symPoissonRate_\symSetXSuffix}{\symNumReg 2^{\min\left(\symRegVal,\symRegRange\right)}}}
+
e^{-\frac{\symPoissonRate_\symSetASuffix+\symPoissonRate_\symSetBSuffix+\symPoissonRate_\symSetXSuffix}{\symNumReg 2^{\min\left(\symRegVal,\symRegRange\right)}}}
\right)
\symCountVariate^{=}_{\symRegVal}
\\
-
&
\frac{\symPoissonRate_\symSetASuffix}{\symNumReg}
\sum_{\symRegVal=0}^{\symRegRange}
\frac{
  \symCountVariate^{<}_{1\symRegVal}+
  \symCountVariate^{=}_{\symRegVal}+
  \symCountVariate^{>}_{1\symRegVal}
}{2^{\symRegVal}}
-
\frac{\symPoissonRate_\symSetBSuffix}{\symNumReg}
\sum_{\symRegVal=0}^{\symRegRange}
\frac{
  \symCountVariate^{<}_{2\symRegVal}+
  \symCountVariate^{=}_{\symRegVal}+
  \symCountVariate^{>}_{2\symRegVal}
}{2^{\symRegVal}}
-
\frac{\symPoissonRate_\symSetXSuffix}{\symNumReg}
\sum_{\symRegVal=0}^{\symRegRange}
\frac{
  \symCountVariate^{<}_{1\symRegVal}+
  \symCountVariate^{=}_{\symRegVal}+
  \symCountVariate^{<}_{2\symRegVal}
}{2^{\symRegVal}},
\end{aligned}
\end{multline}
where
$\symCountVariate^{<}_{1\symRegVal}$,
$\symCountVariate^{>}_{1\symRegVal}$,
$\symCountVariate^{<}_{2\symRegVal}$,
$\symCountVariate^{>}_{2\symRegVal}$,
and $\symCountVariate^{=}_{\symRegVal}$ are defined as
\begin{equation}
\label{equ:sufficient_joint_statistic}
\begin{aligned}
\symCountVariate^{<}_{1\symRegVal}&:=\left|\lbrace\symIndexI\vert\symRegVal=\symRegValVariate_{1\symIndexI}<
\symRegValVariate_{2\symIndexI}\rbrace\right|,
\\
\symCountVariate^{>}_{1\symRegVal}&:=\left|\lbrace\symIndexI\vert\symRegVal=\symRegValVariate_{1\symIndexI}>
\symRegValVariate_{2\symIndexI}\rbrace\right|,
\\
\symCountVariate^{<}_{2\symRegVal}&:=\left|\lbrace\symIndexI\vert\symRegVal=\symRegValVariate_{2\symIndexI}<
\symRegValVariate_{1\symIndexI}\rbrace\right|,
\\
\symCountVariate^{>}_{2\symRegVal}&:=\left|\lbrace\symIndexI\vert\symRegVal=\symRegValVariate_{2\symIndexI}>
\symRegValVariate_{1\symIndexI}\rbrace\right|,
\\
\symCountVariate^{=}_{\symRegVal}&:=\left|\lbrace\symIndexI\vert\symRegVal=\symRegValVariate_{1\symIndexI}=
\symRegValVariate_{2\symIndexI}\rbrace\right|.
\end{aligned}
\end{equation}
These $5(\symRegRange+2)$ values represent a sufficient statistic for estimating $\symPoissonRate_\symSetASuffix$, $\symPoissonRate_\symSetBSuffix$, and $\symPoissonRate_\symSetXSuffix$. Actually, the number of values could be further reduced, because of the invariants $\symCountVariate^{>}_{10}=\symCountVariate^{>}_{20}=
\symCountVariate^{<}_{1,\symRegRange+1}=\symCountVariate^{<}_{2,\symRegRange+1}=0$, $\sum_{\symRegVal=0}^{\symRegRange+1}{\symCountVariate^{<}_{1\symRegVal}} = \sum_{\symRegVal=0}^{\symRegRange+1}{\symCountVariate^{>}_{2\symRegVal}}$,
$\sum_{\symRegVal=0}^{\symRegRange+1}{\symCountVariate^{<}_{2\symRegVal}} = \sum_{\symRegVal=0}^{\symRegRange+1}{\symCountVariate^{>}_{1\symRegVal}}$,
and 
$\sum_{\symRegVal=0}^{\symRegRange+1}{\symCountVariate^{<}_{1\symRegVal}}
+
{\symCountVariate^{=}_{\symRegVal}}
+
{\symCountVariate^{<}_{2\symRegVal}}=\symNumReg$.

Since \eqref{equ:log_likelihood_pair} is a generalization of \eqref{equ:log_likelihood_single} for two HyperLogLog sketches, the log-likelihood function for a single HyperLogLog sketch can be obtained by considering special cases. For example, assume $\symPoissonRate_\symSetXSuffix=0$, which means that both sketches represent disjoint sets, \eqref{equ:log_likelihood_pair} can be split into the sum of two unary functions with parameters $\symPoissonRate_\symSetASuffix$ and $\symPoissonRate_\symSetBSuffix$ each of which correspond to \eqref{equ:log_likelihood_single} as expected. Or, consider two sketches representing identical sets. In this case all registers are equal, which means  that $\symCountVariate^{>}_{1\symRegVal}=\symCountVariate^{>}_{2\symRegVal}=
\symCountVariate^{<}_{1\symRegVal}=\symCountVariate^{<}_{2\symRegVal}=0$ for all $\symRegVal$, and therefore the maximum likelihood method yields $\symPoissonRateEstimate_\symSetASuffix = \symPoissonRateEstimate_\symSetBSuffix = 0$ and the value for $\symPoissonRateEstimate_\symSetXSuffix$ will be equal to the single HyperLogLog maximum likelihood estimate \eqref{equ:log_likelihood_single}.

The log-likelihood function \eqref{equ:log_likelihood_pair} does not always have a strict global maximum point. For example, if all register values of the first HyperLogLog sketch are larger than the corresponding values in the second HyperLogLog sketch, that is $\symCountVariate_{1\symRegVal}^{<}
=
\symCountVariate_{\symRegVal}^{=}
=
\symCountVariate_{2\symRegVal}^{>}
=
0$ for all $\symRegVal$, the function can be rewritten as sum of two functions, one dependent on $\symPoissonRate_\symSetASuffix$ and the other dependent on $\symPoissonRate_\symSetBSuffix+\symPoissonRate_\symSetXSuffix$. 
The maximum is obtained, if $\symPoissonRate_\symSetASuffix = \symPoissonRateEstimate_1$ and $\symPoissonRate_\symSetBSuffix+\symPoissonRate_\symSetXSuffix = \symPoissonRateEstimate_2$. Here $\symPoissonRateEstimate_1$ and $\symPoissonRateEstimate_2$ are the maximum likelihood cardinality estimates for the first and second HyperLogLog sketch, respectively. This means that the maximum likelihood method makes no clear statement about the intersection size in this case. The estimate for $\symPoissonRate_\symX$ could be anything between 0 and $\symPoissonRateEstimate_2$. For comparison, the inclusion-exclusion approach would give $\symPoissonRateEstimate_2$ as estimate for $\symPoissonRate_\symX$. This result is questionable, because there is no evidence that sets $\symSetS_1$ and $\symSetS_2$ really have common elements in this case.

The inclusion-exclusion method does not use all the available information given by the sufficient statistic \eqref{equ:sufficient_joint_statistic}, because the estimator is a function of the three vectors $(\boldsymbol\symCountVariate^{<}_{1} + \boldsymbol\symCountVariate^{=} + \boldsymbol\symCountVariate^{>}_{1})$, 
$(\boldsymbol\symCountVariate^{<}_{2} + \boldsymbol\symCountVariate^{=} + \boldsymbol\symCountVariate^{>}_{2})$, and 
$(\boldsymbol\symCountVariate^{>}_{1} + \boldsymbol\symCountVariate^{=} + \boldsymbol\symCountVariate^{>}_{2})$. In contrast, the maximum likelihood method uses more information as it incorporates all the individual values of the sufficient statistic.

\subsection{Computation of the maximum likelihood estimates}
The maximum likelihood estimates can be obtained by maximizing \eqref{equ:log_likelihood_pair}. Since the three parameters are all nonnegative, this is a constrained optimization problem. In order to get rid of these constraints, we use the transformation $\symPoissonRate=e^{\symPhi}$. This mapping has also the nice property that relative accuracy limits are translated into absolute ones, because $\Delta\symPhi = \Delta\symPoissonRate/\symPoissonRate$. Many optimization algorithm implementations allow the definition of absolute limits rather than relative ones.

The transformed log-likelihood function can be written as
\begin{multline}
\label{equ:max_likelihood_phi_func}
\symFunc(\symPhi_\symSetASuffix,\symPhi_\symSetBSuffix,\symPhi_\symSetXSuffix):=
\log \symLikelihood(
e^{\symPhi_\symSetASuffix},
e^{\symPhi_\symSetBSuffix},
e^{\symPhi_\symSetXSuffix}
\vert
\boldsymbol{\symRegValVariate}_1,
\boldsymbol{\symRegValVariate}_2
)=
\\
\begin{aligned}
&+
\sum_{\symRegVal=1}^{\symRegRange}
\symCountVariate^{<}_{1\symRegVal}
\log\!\left(
\symZ_{\symSetXSuffix\symRegVal}
+
\symY_{\symSetXSuffix\symRegVal}
\symZ_{\symSetASuffix\symRegVal}
\right)
+
\symCountVariate^{<}_{2\symRegVal}
\log\!\left(
\symZ_{\symSetXSuffix\symRegVal}
+
\symY_{\symSetXSuffix\symRegVal}
\symZ_{\symSetBSuffix\symRegVal}
\right)
\\
&+
\sum_{\symRegVal=1}^{\symRegRange}
\symCountVariate^{>}_{1\symRegVal}
\log\!\left(
\symZ_{\symSetASuffix\symRegVal}
\right)
+
\symCountVariate^{>}_{2\symRegVal}
\log\!\left(
\symZ_{\symSetBSuffix\symRegVal}
\right)
+
\symCountVariate^{=}_{\symRegVal}
\log\!\left(
\symZ_{\symSetXSuffix\symRegVal}
+
\symY_{\symSetXSuffix\symRegVal}
\symZ_{\symSetASuffix\symRegVal}
\symZ_{\symSetBSuffix\symRegVal}
\right)
\\
&+
\symCountVariate^{>}_{1,\symRegRange+1}
\log\!\left(
\symZ_{\symSetASuffix\symRegRange}
\right)
+
\symCountVariate^{>}_{2,\symRegRange+1}
\log\!\left(
\symZ_{\symSetBSuffix\symRegRange}
\right)
+
\symCountVariate^{=}_{\symRegRange+1}
\log\!\left(
\symZ_{\symSetXSuffix\symRegRange}
+
\symY_{\symSetXSuffix\symRegRange}
\symZ_{\symSetASuffix\symRegRange}
\symZ_{\symSetBSuffix\symRegRange}
\right)
\\
&-
\sum_{\symRegVal=0}^{\symRegRange}
\left(
  \symCountVariate^{<}_{1\symRegVal}+
  \symCountVariate^{=}_{\symRegVal}+
  \symCountVariate^{>}_{1\symRegVal}
\right)
\symX_{\symSetASuffix\symRegVal}
+
\left(
  \symCountVariate^{<}_{2\symRegVal}+
  \symCountVariate^{=}_{\symRegVal}+
  \symCountVariate^{>}_{2\symRegVal}
\right)
\symX_{\symSetBSuffix\symRegVal}
+
\left(
  \symCountVariate^{<}_{1\symRegVal}+
  \symCountVariate^{=}_{\symRegVal}+
  \symCountVariate^{<}_{2\symRegVal}
\right)
\symX_{\symSetXSuffix\symRegVal}
.
\end{aligned}
\end{multline}
Here we introduced the following expressions for simplification:
\begin{equation}
\label{equ:definition_xyz}
\symX_{\ast\symRegVal} :=\frac{e^{\symPhi_\ast}}{\symNumReg 2^{\symRegVal}},\quad
\symY_{\ast\symRegVal} := e^{-\symX_{\ast\symRegVal}},\quad
\symZ_{\ast\symRegVal} := 1 - \symY_{\ast\symRegVal}.
\end{equation}

Quasi-Newton methods are commonly used to find the maximum of multi-dimensional functions. They require the calculation of the gradient $\nabla\symFunc = \left(\frac{\partial\symFunc}{\partial\symPhi_\symSetASuffix}, \frac{\partial\symFunc}{\partial\symPhi_\symSetBSuffix},
\frac{\partial\symFunc}{\partial\symPhi_\symSetXSuffix}\right)$ which is, using 
\begin{equation}
\frac{\partial\symX_{\ast\symRegVal}}{\partial\symPhi_\ast} =
\symX_{\ast\symRegVal},
\quad
\frac{\partial\symY_{\ast\symRegVal}}{\partial\symPhi_\ast}
=
-\symX_{\ast\symRegVal}\symY_{\ast\symRegVal},
\quad
\frac{\partial\symZ_{\ast\symRegVal}}{\partial\symPhi_\ast}
=
\symX_{\ast\symRegVal}\symY_{\ast\symRegVal},
\end{equation}
given by
\begin{equation}
\label{equ:df_dphi}
\begin{aligned}
\frac{\partial\symFunc}{\partial\symPhi_\symSetASuffix}
=
&
\sum_{\symRegVal=1}^{\symRegRange}
\symCountVariate^{<}_{1\symRegVal}
\frac{
\symY_{\symSetXSuffix\symRegVal}
\symX_{\symSetASuffix\symRegVal}
\symY_{\symSetASuffix\symRegVal}
}{
\symZ_{\symSetXSuffix\symRegVal}
+
\symY_{\symSetXSuffix\symRegVal}
\symZ_{\symSetASuffix\symRegVal}
}
+
\symCountVariate^{=}_{\symRegVal}
\frac{
\symY_{\symSetXSuffix\symRegVal}
\symX_{\symSetASuffix\symRegVal}
\symY_{\symSetASuffix\symRegVal}
\symZ_{\symSetBSuffix\symRegVal}
}
{
\symZ_{\symSetXSuffix\symRegVal}
+
\symY_{\symSetXSuffix\symRegVal}
\symZ_{\symSetASuffix\symRegVal}
\symZ_{\symSetBSuffix\symRegVal}
}
+
\symCountVariate^{>}_{1\symRegVal}
\frac{
\symX_{\symSetASuffix\symRegVal}
\symY_{\symSetASuffix\symRegVal}
}{
\symZ_{\symSetASuffix\symRegVal}
}
\\
&
+
\symCountVariate^{=}_{\symRegRange+1}
\frac{
\symY_{\symSetXSuffix\symRegRange}
\symX_{\symSetASuffix\symRegRange}
\symY_{\symSetASuffix\symRegRange}
\symZ_{\symSetBSuffix\symRegRange}
}
{
\symZ_{\symSetXSuffix\symRegRange}
+
\symY_{\symSetXSuffix\symRegRange}
\symZ_{\symSetASuffix\symRegRange}
\symZ_{\symSetBSuffix\symRegRange}
}
+
\symCountVariate^{>}_{1,\symRegRange+1}
\frac{
\symX_{\symSetASuffix\symRegRange}
\symY_{\symSetASuffix\symRegRange}
}{
\symZ_{\symSetASuffix\symRegRange}
}
-
\sum_{\symRegVal=0}^{\symRegRange}
\left(
  \symCountVariate^{<}_{1\symRegVal}+
  \symCountVariate^{=}_{\symRegVal}+
  \symCountVariate^{>}_{1\symRegVal}
\right)
\symX_{\symSetASuffix\symRegVal},
\\
\frac{\partial\symFunc}{\partial\symPhi_\symSetBSuffix}
=&
\sum_{\symRegVal=1}^{\symRegRange}
\symCountVariate^{<}_{2\symRegVal}
\frac{
\symY_{\symSetXSuffix\symRegVal}
\symX_{\symSetBSuffix\symRegVal}
\symY_{\symSetBSuffix\symRegVal}
}{
\symZ_{\symSetXSuffix\symRegVal}
+
\symY_{\symSetXSuffix\symRegVal}
\symZ_{\symSetBSuffix\symRegVal}
}
+
\symCountVariate^{=}_{\symRegVal}
\frac{
\symY_{\symSetXSuffix\symRegVal}
\symZ_{\symSetASuffix\symRegVal}
\symX_{\symSetBSuffix\symRegVal}
\symY_{\symSetBSuffix\symRegVal}
}
{
\symZ_{\symSetXSuffix\symRegVal}
+
\symY_{\symSetXSuffix\symRegVal}
\symZ_{\symSetASuffix\symRegVal}
\symZ_{\symSetBSuffix\symRegVal}
}
+
\symCountVariate^{>}_{2\symRegVal}
\frac{
\symX_{\symSetBSuffix\symRegVal}
\symY_{\symSetBSuffix\symRegVal}
}{
\symZ_{\symSetBSuffix\symRegVal}
}
\\
&+
\symCountVariate^{=}_{\symRegRange+1}
\frac{
\symY_{\symSetXSuffix\symRegRange}
\symZ_{\symSetASuffix\symRegRange}
\symX_{\symSetBSuffix\symRegRange}
\symY_{\symSetBSuffix\symRegRange}
}
{
\symZ_{\symSetXSuffix\symRegRange}
+
\symY_{\symSetXSuffix\symRegRange}
\symZ_{\symSetASuffix\symRegRange}
\symZ_{\symSetBSuffix\symRegRange}
}
+
\symCountVariate^{>}_{2,\symRegRange+1}
\frac{
\symX_{\symSetBSuffix\symRegRange}
\symY_{\symSetBSuffix\symRegRange}
}{
\symZ_{\symSetBSuffix\symRegRange}
}
-
\sum_{\symRegVal=0}^{\symRegRange}
\left(
  \symCountVariate^{<}_{2\symRegVal}+
  \symCountVariate^{=}_{\symRegVal}+
  \symCountVariate^{>}_{2\symRegVal}
\right)
\symX_{\symSetBSuffix\symRegVal},
\\
\frac{\partial\symFunc}{\partial\symPhi_\symSetXSuffix}
=&
\sum_{\symRegVal=1}^{\symRegRange}
\symCountVariate^{<}_{1\symRegVal}
\frac{
\symX_{\symSetXSuffix\symRegVal}
\symY_{\symSetXSuffix\symRegVal}
\symY_{\symSetASuffix\symRegVal}
}{
\symZ_{\symSetXSuffix\symRegVal}
+
\symY_{\symSetXSuffix\symRegVal}
\symZ_{\symSetASuffix\symRegVal}
}
+
\symCountVariate^{=}_{\symRegVal}
\frac{
\symX_{\symSetXSuffix\symRegVal}
\symY_{\symSetXSuffix\symRegVal}
\left(
\symY_{\symSetASuffix\symRegVal}
+
\symZ_{\symSetASuffix\symRegVal}
\symY_{\symSetBSuffix\symRegVal}
\right)
}
{
\symZ_{\symSetXSuffix\symRegVal}
+
\symY_{\symSetXSuffix\symRegVal}
\symZ_{\symSetBSuffix\symRegVal}
\symZ_{\symSetASuffix\symRegVal}
}
+
\symCountVariate^{<}_{2\symRegVal}
\frac{
\symX_{\symSetXSuffix\symRegVal}
\symY_{\symSetXSuffix\symRegVal}
\symY_{\symSetBSuffix\symRegVal}
}{
\symZ_{\symSetXSuffix\symRegVal}
+
\symY_{\symSetXSuffix\symRegVal}
\symZ_{\symSetBSuffix\symRegVal}
}
\\
&+
\symCountVariate^{=}_{\symRegRange+1}
\frac{
\symX_{\symSetXSuffix\symRegRange}
\symY_{\symSetXSuffix\symRegRange}
\left(
\symY_{\symSetASuffix\symRegRange}
+
\symZ_{\symSetASuffix\symRegRange}
\symY_{\symSetBSuffix\symRegRange}
\right)
}
{
\symZ_{\symSetXSuffix\symRegRange}
+
\symY_{\symSetXSuffix\symRegRange}
\symZ_{\symSetASuffix\symRegRange}
\symZ_{\symSetBSuffix\symRegRange}
}
-
\sum_{\symRegVal=0}^{\symRegRange}
\left(
  \symCountVariate^{<}_{1\symRegVal}+
  \symCountVariate^{=}_{\symRegVal}+
  \symCountVariate^{<}_{2\symRegVal}
\right)
\symX_{\symSetXSuffix\symRegVal}.
\end{aligned}
\end{equation}

The calculation of \eqref{equ:max_likelihood_phi_func} and its derivatives requires some care when calculating $\symY_{\ast\symRegVal}$ and $\symZ_{\ast\symRegVal}$. Since $\symX_{\ast\symRegVal}$ is nonnegative, we have $\symY_{\ast\symRegVal},\symZ_{\ast\symRegVal}\in[0,1]$. In order to reduce the numerical error of $\symZ_{\ast\symRegVal}$ for small $\symX_{\ast\symRegVal}$, it is essential to use the function $\operatorname{expm1}(\symX):=e^\symX-1$ that is available in most programming languages. If $\symX_{\ast\symRegVal}$ is smaller than $\log(2)$, we calculate $\symY_{\ast\symRegVal}$ and $\symZ_{\ast\symRegVal}$ using $\symZ_{\ast\symRegVal} = 
-\operatorname{expm1}(-\symX_{\ast\symRegVal})$ and $\symY_{\ast\symRegVal} = 1 - \symZ_{\ast\symRegVal}$, respectively, and not as defined in \eqref{equ:definition_xyz}. In this way the numerical error of both, $\symY_{\ast\symRegVal}$ and $\symZ_{\ast\symRegVal}$, is minimized and still only a single exponential function needs to be evaluated.
Apart from that, the numerical evaluation of \eqref{equ:max_likelihood_phi_func} is straightforward. Apparently, the arguments of all logarithms are in range $[0,1]$. The case that some argument vanishes, which would cause the logarithm to be negative infinite, does not occur in practice. Consider for example the logarithm evaluation associated with $\symCountVariate^{<}_{1\symRegVal}$ which is only relevant if $\symCountVariate^{<}_{1\symRegVal} > 0$. In this case however, we can be certain that the cardinality of $\symSetA\cup\symSetX$ is at least 1. Therefore, it is expected that at least one of the two maximum likelihood estimates $\symPoissonRateEstimate_\symSetASuffix$ and $\symPoissonRateEstimate_\symSetXSuffix$ is at least in the order of 1. If the optimization algorithm starts with appropriate initial values, function evaluations for which $\max(\symPoissonRate_\symSetASuffix,\symPoissonRate_\symSetXSuffix)\ll 1$ are not expected.
Furthermore, the argument of the logarithm can be bounded by $\symZ_{\symSetXSuffix\symRegVal}
+
\symY_{\symSetXSuffix\symRegVal}
\symZ_{\symSetASuffix\symRegVal}
\geq
\max(\symZ_{\symSetXSuffix\symRegVal}, \symZ_{\symSetASuffix\symRegVal})
=
1-\exp\!\left(-\frac{\max(\symPoissonRate_\symSetASuffix,\symPoissonRate_\symSetXSuffix)}{\symNumReg 2^{\symRegVal}}\right)
\geq
\min\!\left(\frac{1}{2},\frac{\max(\symPoissonRate_\symSetASuffix,\symPoissonRate_\symSetXSuffix)}{\symNumReg 2^{\symRegVal+1}}
\right)
$.
For the last inequality we used $1-e^{-\symX}\geq\frac{1}{2}\min(1,\symX)$. The derived lower bound shows that the argument of the logarithm is large enough to be accurately represented by double-precision floating-point numbers, provided that $\symPoissonRate_\symSetASuffix$ and $\symPoissonRate_\symSetXSuffix$ are not both substantially smaller than 1. Similar argumentation holds for all other logarithmic terms and also all divisions that appear in \eqref{equ:df_dphi}.

\cref{alg:joint_cardinality_estimation} demonstrates the calculation of the estimates given the register values $\boldsymbol\symRegValVariate_{1}$ and 
$\boldsymbol\symRegValVariate_{2}$ of both HyperLogLog sketches. First, the sufficient statistic consisting of $\boldsymbol\symCountVariate^{>}_{1}$,
$\boldsymbol\symCountVariate^{<}_{1}$,
$\boldsymbol\symCountVariate^{>}_{2}$,
$\boldsymbol\symCountVariate^{<}_{2}$, and
$\boldsymbol\symCountVariate^{=}$ is extracted. Next, a case is distinguished, where
all registers have a value equal to zero in at least one of both HyperLogLog sketches, that is $\symCountVariate^{<}_{10}+\symCountVariate^{=}_0+\symCountVariate^{<}_{20} = \symNumReg$. In this case it is certain that the HyperLogLog sketches represent disjoint sets and their corresponding  cardinality estimates can be  used for $\symPoissonRateEstimate_{\symSetASuffix}$ and $\symPoissonRateEstimate_{\symSetBSuffix}$, respectively.

For the general case, the three-dimensional function $\symFunc(\symPhi_\symSetASuffix,\symPhi_\symSetBSuffix,\symPhi_\symSetXSuffix)$ needs to be optimized numerically. A very popular algorithm for such nonlinear optimization problems is the Broyden-Fletcher-Goldfarb-Shanno (BFGS) algorithm \cite{Press2007}. In particular, we used the implementation provided by the Dlib C++ library \cite{King2009}. Good initial guess values are important to ensure fast convergence for the optimization algorithm. An obvious choice are the cardinality estimates obtained by application of the inclusion-exclusion principle \eqref{equ:conventional_approach}. However, in order to ensure that their logarithms are all defined, we require that the initial values are not smaller than 1. Within the optimization loop the function $\symFunc$ and its gradient $\nabla\symFunc$ need to evaluated using \eqref{equ:max_likelihood_phi_func} and \eqref{equ:df_dphi} as needed by the optimization algorithm. The optimization loop is continued as long as the absolute change of at least one parameter is larger than a predefined threshold $\symStopDelta$. The threshold is again set proportional to $1/\sqrt{\symNumReg}$ (compare \cref{sec:comp_ml_estimate}). For the results presented below we have used $\symStopDelta=\symStopEpsilon/\sqrt{\symNumReg}$ with $\symStopEpsilon = 10^{-2}$.

\begin{algorithm}
\caption{Joint cardinality estimation.}
\label{alg:joint_cardinality_estimation}
\begin{algorithmic}
\Function {EstimateCardinalities}
{
$\boldsymbol\symRegValVariate_{1}$,
$\boldsymbol\symRegValVariate_{2}$
}
\comm{.32}{$\boldsymbol{\symRegValVariate}_1,\boldsymbol{\symRegValVariate}_2\in\lbrace 0,1,\ldots,\symRegRange+1\rbrace^\symNumReg$}

\State $\boldsymbol\symCountVariate^{>}_{1}\gets (0,\ldots,0)$
\comm{.32}{$\boldsymbol{\symCountVariate}^{>}_{1} = (\symCountVariate^{>}_{10},\ldots,\symCountVariate^{>}_{1,\symRegRange+1})$}
\State $\boldsymbol\symCountVariate^{<}_{1}\gets (0,\ldots,0)$
\comm{.32}{$\boldsymbol{\symCountVariate}^{<}_{1} = (\symCountVariate^{<}_{10},\ldots,\symCountVariate^{<}_{1,\symRegRange+1})$}
\State $\boldsymbol\symCountVariate^{>}_{2}\gets (0,\ldots,0)$
\comm{.32}{$\boldsymbol{\symCountVariate}^{>}_{2} = (\symCountVariate^{>}_{20},\ldots,\symCountVariate^{>}_{2,\symRegRange+1})$}
\State $\boldsymbol\symCountVariate^{<}_{2}\gets (0,\ldots,0)$
\comm{.32}{$\boldsymbol{\symCountVariate}^{<}_{2} = (\symCountVariate^{<}_{20},\ldots,\symCountVariate^{<}_{2,\symRegRange+1})$}
\State $\boldsymbol\symCountVariate^{=}\gets (0,\ldots,0)$
\comm{.32}{$\boldsymbol{\symCountVariate}^{=} = (\symCountVariate^{=}_{0},\ldots,\symCountVariate^{=}_{\symRegRange+1})$}

\For{$\symIndexI\gets 1, \symNumReg$}
\If{$\symRegValVariate_{1\symIndexI}
<
\symRegValVariate_{2\symIndexI}$}
\State $\symCountVariate^{<}_{1\symRegValVariate_{1\symIndexI}}
\gets
\symCountVariate^{<}_{1\symRegValVariate_{1\symIndexI}}+1$
\State
$\symCountVariate^{>}_{2\symRegValVariate_{2\symIndexI}}
\gets
\symCountVariate^{>}_{2\symRegValVariate_{2\symIndexI}}+1$
\ElsIf{$\symRegValVariate_{1\symIndexI}
>
\symRegValVariate_{2\symIndexI}$}
\State $\symCountVariate^{>}_{1\symRegValVariate_{1\symIndexI}}
\gets
\symCountVariate^{>}_{1\symRegValVariate_{1\symIndexI}}+1$
\State
$\symCountVariate^{<}_{2\symRegValVariate_{2\symIndexI}}
\gets
\symCountVariate^{<}_{2\symRegValVariate_{2\symIndexI}}+1$
\Else
\State $\symCountVariate^{=}_{\symRegValVariate_{1\symIndexI}}
\gets
\symCountVariate^{=}_{\symRegValVariate_{1\symIndexI}}+1$
\EndIf
\EndFor
\State $\symPoissonRateEstimate_{\symSetASuffix\symSetXSuffix} \gets$ \Call{EstimateCardinality}{$\boldsymbol\symCountVariate^{<}_{1} + \boldsymbol\symCountVariate^{=} + \boldsymbol\symCountVariate^{>}_{1}$}
\comm{0.32}{use \cref{alg:estimate_ml}}
\State $\symPoissonRateEstimate_{\symSetBSuffix\symSetXSuffix} \gets$ \Call{EstimateCardinality}{$\boldsymbol\symCountVariate^{<}_{2} + \boldsymbol\symCountVariate^{=} + \boldsymbol\symCountVariate^{>}_{2}$}
\If{$\symCountVariate^{<}_{10}+\symCountVariate^{=}_0+\symCountVariate^{<}_{20} = \symNumReg$}
\comm{0.32}{$\Leftrightarrow\min(\symRegValVariate_{1\symIndexI},\symRegValVariate_{2\symIndexI})=0\ \forall\symIndexI$}
\State $\symPoissonRateEstimate_{\symSetASuffix}\gets
\symPoissonRateEstimate_{\symSetASuffix\symSetXSuffix},\ 
\symPoissonRateEstimate_{\symSetBSuffix}\gets
\symPoissonRateEstimate_{\symSetBSuffix\symSetXSuffix},\
\symPoissonRateEstimate_{\symSetXSuffix}\gets0$
\State \textbf{return} $(
\symPoissonRateEstimate_{\symSetASuffix},
\symPoissonRateEstimate_{\symSetBSuffix},
\symPoissonRateEstimate_{\symSetXSuffix}
)$
\EndIf
\State $\symPoissonRateEstimate_{\symSetASuffix\symSetBSuffix\symSetXSuffix} \gets$ \Call{EstimateCardinality}{$\boldsymbol\symCountVariate^{>}_{1} + \boldsymbol\symCountVariate^{=} + \boldsymbol\symCountVariate^{>}_{2}$}
\State $\symPhi_\symSetASuffix \gets \log(\max(1, \symPoissonRateEstimate_{\symSetASuffix\symSetBSuffix\symSetXSuffix} - \symPoissonRateEstimate_{\symSetBSuffix\symSetXSuffix}))$
\State $\symPhi_\symSetBSuffix \gets \log(\max(1, \symPoissonRateEstimate_{\symSetASuffix\symSetBSuffix\symSetXSuffix} - \symPoissonRateEstimate_{\symSetASuffix\symSetXSuffix}))$
\State $\symPhi_\symSetXSuffix \gets \log(\max(1,
\symPoissonRateEstimate_{\symSetASuffix\symSetXSuffix}+
\symPoissonRateEstimate_{\symSetBSuffix\symSetXSuffix}
-
\symPoissonRateEstimate_{\symSetASuffix\symSetBSuffix\symSetXSuffix}))$
\Repeat
\comm{0.32}{start optimization}
\State 
$
\symPhi'_\symSetASuffix\gets\symPhi_\symSetASuffix,\ \symPhi'_\symSetBSuffix\gets\symPhi_\symSetBSuffix,\  \symPhi'_\symSetXSuffix\gets\symPhi_\symSetXSuffix$
\State
$
\symPhi_\symSetASuffix,
\symPhi_\symSetBSuffix,
\symPhi_\symSetXSuffix
\gets$
\Call{OptimizationStep}{$\symPhi_\symSetASuffix,
\symPhi_\symSetBSuffix,
\symPhi_\symSetXSuffix
$}
\comm{0.32}{calculate $\symFunc$ and
$\nabla\symFunc$ using \eqref{equ:max_likelihood_phi_func} and \eqref{equ:df_dphi} as needed by optimizer
}
\Until{
$\max(
\left|\symPhi_\symSetASuffix - \symPhi'_\symSetASuffix\right|,
\left|\symPhi_\symSetBSuffix - \symPhi'_\symSetBSuffix\right|,
\left|\symPhi_\symSetXSuffix - \symPhi'_\symSetXSuffix\right|) \leq \symStopDelta
$}
\comm{0.32}{stop criterion\\
$\symStopDelta=\symStopEpsilon/\sqrt{\symNumReg}$, $\symStopEpsilon = 10^{-2}$}
\State 
$
\symPoissonRateEstimate_{\symSetASuffix}
\gets e^{\symPhi_\symSetASuffix},\ \symPoissonRateEstimate_{\symSetBSuffix}
\gets e^{\symPhi_\symSetBSuffix},\ 
\symPoissonRateEstimate_{\symSetXSuffix}
\gets e^{\symPhi_\symSetXSuffix}
$
\State \textbf{return} $(
\symPoissonRateEstimate_{\symSetASuffix},
\symPoissonRateEstimate_{\symSetBSuffix},
\symPoissonRateEstimate_{\symSetXSuffix}
)$
\EndFunction
\end{algorithmic}
\end{algorithm}

\subsection{Results}
To evaluate the estimation error of the new joint cardinality estimation approach we applied \cref{alg:joint_cardinality_estimation} to HyperLogLog sketch pairs with parameters $\symPrecision=20$ and $\symRegRange=44$ and with known cardinalities $|\symSetA|$, $|\symSetB|$, and $|\symSetX|$. HyperLogLog sketches with predefined cardinalities can be quickly generated by taking a precalculated multiplicity vector with known cardinality, setting the register values accordingly, and shuffling them. Using this approach we generated three different HyperLogLog sketches with known cardinalities $|\symSetA|$, $|\symSetB|$, and $|\symSetX|$. Then we merged the HyperLogLog sketches representing the disjoint sets $\symSetA$ and $\symSetX$ using \cref{alg:union} in order to get a sketch for $\symSetS_1 = \symSetA\cup\symSetX$. A second sketch for $\symSetS_2=\symSetA\cup\symSetX$ was built analogously. Since three independent multiplicity vectors are needed to construct a single example and we had simulated \num{10000} different HyperLogLog sketches previously as described in \cref{sec:experimental_setup}, we were able to construct \num{3333} different independent HyperLogLog pairs for a given set of known true cardinalities $|\symSetA|$, $|\symSetB|$, and $|\symSetX|$.

\cref{tbl:joint_estimation_cases} lists all the different cardinality configurations for which we have evaluated the joint cardinality estimation  algorithm. Among the considered cases there are also cardinalities that are small compared to the number of registers in order to prove that the new approach also covers the small cardinality range where many register values are equal to zero. The last column of \cref{tbl:joint_estimation_cases} shows the average number of iterations of the BFGS algorithm until the stop criterion was satisfied. Each iteration step involved a function \eqref{equ:max_likelihood_phi_func} and a gradient \eqref{equ:df_dphi} evaluation.

\cref{tbl:joint_estimation_a,tbl:joint_estimation_b,tbl:joint_estimation_x} compare the relative estimation error for cardinalities $|\symSetA|$, $|\symSetB|$, and $|\symSetX|$ to that of the conventional approach using single sketch cardinality estimation together with \eqref{equ:conventional_approach}. The mean, the standard deviation, and the root-mean-square error have been calculated from the cardinality estimates of \num{3333} examples. Furthermore, we calculated an improvement factor which represents the root-mean-square error ratio between both approaches. Since we only observed values greater than one, the new maximum likelihood estimation approach gives better estimates for all investigated cases. For some cases the improvement factor is even clearly greater than two. Due to the  square root scaling of the error, this means that we would need four times more registers to get the same error when using the conventional approach. As the results suggest the joint estimation algorithm works well over the entire cardinality range without the need of special handling of small cardinalities.

We also investigated, if the new approach could give better estimates for the union operation. The conventional approach merges both HyperLogLog sketches using \cref{alg:union} and estimates the union size using single sketch cardinality estimation. The joint cardinality estimation algorithm provides another opportunity. The union can also be estimated by simply summing up the three estimates for sets $|\symSetA|$, $|\symSetB|$, and $|\symSetX|$. The corresponding results are shown in \cref{tbl:joint_estimation_abx}. As can be clearly seen, the joint estimation algorithm is also able to improve the cardinality estimation of unions by a significant amount.

\begin{table}
\centering
\captionabove{List of true cardinalities of the pairwise disjoint sets $\symSetA$, $\symSetB$, and $\symSetX$, for which maximum likelihood estimation from two HyperLogLog sketches with parameters $\symPrecision=20$ and $\symRegRange=44$ representing sets $\symSetS_1 = \symSetA\cup\symSetX$ and  $\symSetS_2 = \symSetB\cup\symSetX$ was investigated. The last column shows the average number of iteration cycles until the stop criterion of the BFGS algorithm was satisfied.
}
\label{tbl:joint_estimation_cases}

\csvreader[before reading=\scriptsize, tabular=r|rrr|rr|rr, table head= & 
\multicolumn{3}{c|}{true cardinalities} &
\multicolumn{1}{c}{Jaccard index} &
\multicolumn{1}{c|}{cardinality ratio} &
\multicolumn{1}{c}{avg number}
\\
\# &  
\multicolumn{1}{c}{$|\symSetA|$} & 
\multicolumn{1}{c}{$|\symSetB|$} & 
\multicolumn{1}{c|}{$|\symSetX|$} & 
\multicolumn{1}{c}{$|\symSetS_1\cap\symSetS_2|/|\symSetS_1\cup\symSetS_2|$} & 
\multicolumn{1}{c|}{$|\symSetS_1|/|\symSetS_2|$} &
\multicolumn{1}{c}{iterations}
\\\hline]
{joint_cardinality_calculation.csv}{
trueCardA=\trueCardA,
trueCardB=\trueCardB,
trueCardX=\trueCardX,
trueJaccardIdx=\trueJaccardIdx,
trueRatio=\trueRatio,
avgNumFunctionEvaluations=\avgNumFunctionEvaluations
}{
\thecsvrow & 
\num{\trueCardA} & 
\num{\trueCardB} & 
\num{\trueCardX} & 
\numformat{\trueJaccardIdx} & 
\numformattwo{\trueRatio} &
\numformattwo{\avgNumFunctionEvaluations}
}
\end{table}

\begin{table}
\centering
\captionabove{The mean, the standard deviation, and the root-mean-square error of the relative estimation error for all cases given in \cref{tbl:joint_estimation_cases} when estimating $|\symSetA|$ using the conventional approach and the maximum likelihood method, respectively.}
\label{tbl:joint_estimation_a}
\csvreader[before reading=\scriptsize, tabular=r|rrr|rrr|r, table head=
 & 
\multicolumn{3}{c|}{conventional approach}
&
\multicolumn{3}{c|}{maximum likelihood method}
&
\multicolumn{1}{c}{improvement}
\\
\# & 
\multicolumn{1}{c}{mean} & 
\multicolumn{1}{c}{stdev} & 
\multicolumn{1}{c|}{rmse} & 
\multicolumn{1}{c}{mean} & 
\multicolumn{1}{c}{stdev} & 
\multicolumn{1}{c|}{rmse} &
\multicolumn{1}{c}{rmse ratio}
\\\hline]
{joint_cardinality_calculation.csv}{
inclExclMeanA=\inclExclMeanA,
inclExclStdDevA=\inclExclStdDevA,
inclExclRMSEA=\inclExclRMSEA,
maxLikeMeanA=\maxLikeMeanA,
maxLikeStdDevA=\maxLikeStdDevA,
maxLikeRMSEA=\maxLikeRMSEA,
improvementRmseA=\improvementRmseA
}{
\thecsvrow & 
\numformat{\inclExclMeanA} & 
\numformat{\inclExclStdDevA} & 
\numformat{\inclExclRMSEA} & 
\numformat{\maxLikeMeanA} & 
\numformat{\maxLikeStdDevA} & 
\numformat{\maxLikeRMSEA} &
\numformattwo{\improvementRmseA}
}
\end{table}

\begin{table}
\centering
\captionabove{The mean, the standard deviation, and the root-mean-square error of the relative estimation error for all cases given in \cref{tbl:joint_estimation_cases} when estimating $|\symSetB|$ using the conventional approach and the maximum likelihood method, respectively.}
\label{tbl:joint_estimation_b}
\csvreader[before reading=\scriptsize, tabular=r|rrr|rrr|r, table head=
 & 
\multicolumn{3}{c|}{conventional approach}
&
\multicolumn{3}{c|}{maximum likelihood method}
&
\multicolumn{1}{c}{improvement}
\\
\# & 
\multicolumn{1}{c}{mean} & 
\multicolumn{1}{c}{stdev} & 
\multicolumn{1}{c|}{rmse} & 
\multicolumn{1}{c}{mean} & 
\multicolumn{1}{c}{stdev} & 
\multicolumn{1}{c|}{rmse} &
\multicolumn{1}{c}{rmse ratio}
\\\hline]
{joint_cardinality_calculation.csv}{
inclExclMeanB=\inclExclMeanB,
inclExclStdDevB=\inclExclStdDevB,
inclExclRMSEB=\inclExclRMSEB,
maxLikeMeanB=\maxLikeMeanB,
maxLikeStdDevB=\maxLikeStdDevB,
maxLikeRMSEB=\maxLikeRMSEB,
improvementRmseB=\improvementRmseB
}{
\thecsvrow & 
\numformat{\inclExclMeanB} & 
\numformat{\inclExclStdDevB} & 
\numformat{\inclExclRMSEB} & 
\numformat{\maxLikeMeanB} & 
\numformat{\maxLikeStdDevB} & 
\numformat{\maxLikeRMSEB} &
\numformattwo{\improvementRmseB}
}
\end{table}

\begin{table}
\centering
\captionabove{The mean, the standard deviation, and the root-mean-square error of the relative estimation error for all cases given in \cref{tbl:joint_estimation_cases} when estimating $|\symSetX|$ using the conventional approach and the maximum likelihood method, respectively.}
\label{tbl:joint_estimation_x}
\csvreader[before reading=\scriptsize, tabular=r|rrr|rrr|r, table head=
 & 
\multicolumn{3}{c|}{conventional approach}
&
\multicolumn{3}{c|}{maximum likelihood method}
&
\multicolumn{1}{c}{improvement}
\\
\# & 
\multicolumn{1}{c}{mean} & 
\multicolumn{1}{c}{stdev} & 
\multicolumn{1}{c|}{rmse} & 
\multicolumn{1}{c}{mean} & 
\multicolumn{1}{c}{stdev} & 
\multicolumn{1}{c|}{rmse} &
\multicolumn{1}{c}{rmse ratio}
\\\hline]
{joint_cardinality_calculation.csv}{
inclExclMeanX=\inclExclMeanX,
inclExclStdDevX=\inclExclStdDevX,
inclExclRMSEX=\inclExclRMSEX,
maxLikeMeanX=\maxLikeMeanX,
maxLikeStdDevX=\maxLikeStdDevX,
maxLikeRMSEX=\maxLikeRMSEX,
improvementRmseX=\improvementRmseX
}{
\thecsvrow & 
\numformat{\inclExclMeanX} & 
\numformat{\inclExclStdDevX} & 
\numformat{\inclExclRMSEX} & 
\numformat{\maxLikeMeanX} & 
\numformat{\maxLikeStdDevX} & 
\numformat{\maxLikeRMSEX} &
\numformattwo{\improvementRmseX}
}
\end{table}

\begin{table}
\centering
\captionabove{The mean, the standard deviation, and the root-mean-square error of the relative estimation error for all cases given in \cref{tbl:joint_estimation_cases} when estimating $|\symSetA\cup\symSetB\cup\symSetX|$ using the conventional approach and the maximum likelihood method, respectively.}
\label{tbl:joint_estimation_abx}
\csvreader[before reading=\scriptsize, tabular=r|rrr|rrr|r, table head=
 & 
\multicolumn{3}{c|}{conventional approach}
&
\multicolumn{3}{c|}{maximum likelihood method}
&
\multicolumn{1}{c}{improvement}
\\
\# & 
\multicolumn{1}{c}{mean} & 
\multicolumn{1}{c}{stdev} & 
\multicolumn{1}{c|}{rmse} & 
\multicolumn{1}{c}{mean} & 
\multicolumn{1}{c}{stdev} & 
\multicolumn{1}{c|}{rmse} &
\multicolumn{1}{c}{rmse ratio}
\\\hline]
{joint_cardinality_calculation.csv}{
inclExclMeanABX=\inclExclMeanABX,
inclExclStdDevABX=\inclExclStdDevABX,
inclExclRMSEABX=\inclExclRMSEABX,
maxLikeMeanABX=\maxLikeMeanABX,
maxLikeStdDevABX=\maxLikeStdDevABX,
maxLikeRMSEABX=\maxLikeRMSEABX,
improvementRmseABX=\improvementRmseABX
}{
\thecsvrow & 
\numformat{\inclExclMeanABX} & 
\numformat{\inclExclStdDevABX} & 
\numformat{\inclExclRMSEABX} & 
\numformat{\maxLikeMeanABX} & 
\numformat{\maxLikeStdDevABX} & 
\numformat{\maxLikeRMSEABX} &
\numformattwo{\improvementRmseABX}
}
\end{table}

\section{Future work}
\label{sec:future_work}
As described in \cref{sec:depoissonization} an unbiased estimator for the rate in the Poisson model, is also unbiased estimator for the cardinality. We have shown that this approach works well for the improved raw estimator as well as for the maximum likelihood estimator even though both are only approximately unbiased. Therefore, it would be interesting what conditions on an approximately unbiased Poisson rate estimator are sufficient to guarantee approximate unbiasedness, if used as cardinality estimator.

Using the maximum likelihood method we have been able to improve the cardinality estimates for the results of set operations between two HyperLogLog sketches. Unfortunately, joint cardinality estimation is much more expensive than for a single HyperLogLog sketch, because it requires maximization of a multi-dimensional function. Since we have found the improved raw estimator which is almost as precise as the maximum likelihood estimator for the single HyperLogLog case, we could imagine that there also exists a faster algorithm for the two HyperLogLog case. It is expected that such a new algorithm makes use of all the information given by the sufficient statistic \eqref{equ:sufficient_joint_statistic}.

The presented maximum likelihood method could also be used to estimate the cardinality of set operations between more than two HyperLogLog sketches. However, further research is necessary to determine, if this is feasible from a practical point of view. As for the inclusion-exclusion principle, the effort would scale at least exponentially with the number of involved HyperLogLog sketches.

The maximum likelihood method can also be used to estimate distance measures such as the Jaccard distance of two sets that are represented as HyperLogLog sketches. This directly leads to the question whether the HyperLogLog algorithm could be used for locality-sensitive hashing \cite{Leskovec2014, Wang2014}. Various locality-sensitive hashing algorithms have been proposed in the past. Among the most popular ones are the SimHash \cite{Charikar2002} and the minwise hashing \cite{Broder1997} algorithms whose hash collision probabilities are a function of the angular and Jaccard distances, respectively. A generalization of the latter method is $\symNumBitsMinwiseHashing$-bit minwise hashing that improves memory efficiency by only storing the lowest $\symNumBitsMinwiseHashing$ bits of the minimum hash value \cite{Li2011}. The probability that two different sets are mapped to equal hash values $\symRegValVariate_1$ and $\symRegValVariate_2$ is roughly 
\begin{equation}
\label{equ:minwise_hahsing_probability}
\symProbability(\symRegValVariate_1=\symRegValVariate_2)
\approx
1-
(1-\textstyle\frac{1}{2 ^\symNumBitsMinwiseHashing})\symDistanceMeasure.
\end{equation}

\begin{figure}
\centering
\includegraphics[width=0.6\textwidth]{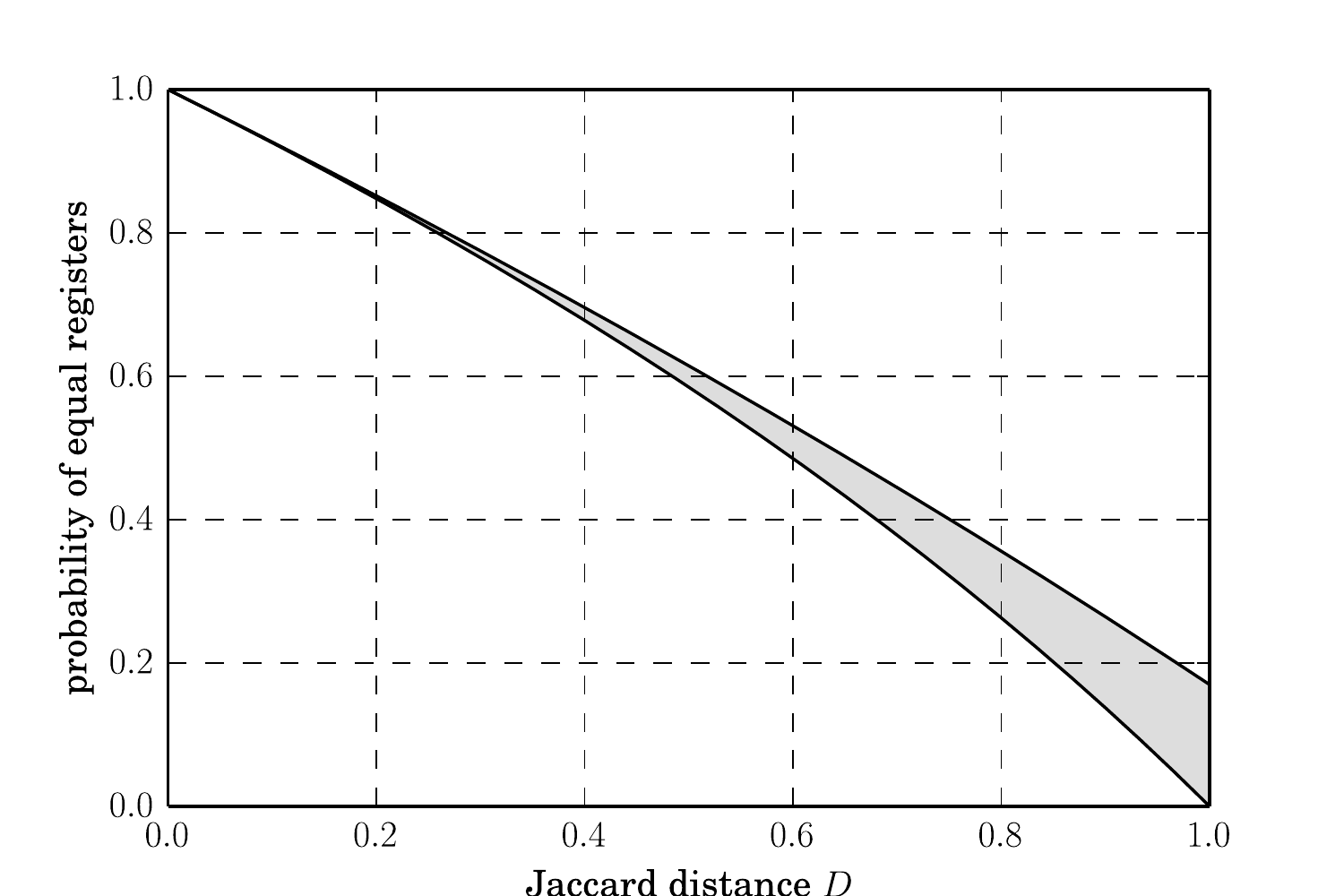}
\caption{The approximate probability range of equal register values as a function of the Jaccard distance.}
\label{fig:equal_register_probability}
\end{figure}

The HyperLogLog algorithm itself can be regarded as hashing algorithm as it maps sets to register values. For sufficiently large cardinalities we can use the Poisson approximation and assume that the number of zero-valued HyperLogLog registers can be ignored. Furthermore, if the HyperLogLog parameter $\symRegRange$ is chosen large enough, the number of saturated registers can be ignored as well. As a consequence, we can assume that the distribution of register values follows \eqref{equ:assumed_register_val_distribution} and the probability that a register has the same value for two different sets is (compare \eqref{equ:register_value_joint_pmf})
\begin{equation}
\symProbability(\symRegValVariate_1=\symRegValVariate_2)
=
\sum_{\symRegVal = -\infty}^{\infty}
e^{-\frac{\symPoissonRate_\symSetASuffix + \symPoissonRate_\symSetBSuffix + \symPoissonRate_\symSetXSuffix}
{\symNumReg 2^\symRegVal}
}
\left(
1
-
e^{-\frac{\symPoissonRate_\symSetASuffix +  \symPoissonRate_\symSetXSuffix}
{\symNumReg 2^\symRegVal}
}
-
e^{-\frac{\symPoissonRate_\symSetBSuffix + \symPoissonRate_\symSetXSuffix}
{\symNumReg 2^\symRegVal}
}
+
e^{-\frac{\symPoissonRate_\symSetASuffix + \symPoissonRate_\symSetBSuffix + \symPoissonRate_\symSetXSuffix}
{\symNumReg 2^\symRegVal}
}
\right).
\end{equation}
Using the approximation 
$\sum_{\symRegVal=-\infty}^\infty
e^{-\frac{\symX}{2^\symRegVal}} - e^{-\frac{\symY}{2^\symRegVal}}
\approx
2\symAlpha_\infty \left(\log(\symY)-\log(\symX)\right)$ (compare \eqref{equ:approx_formula} in \cref{app:analysis_xi}) we get

\begin{equation}
\symProbability(\symRegValVariate_1=\symRegValVariate_2)
\approx
1
+
2\symAlpha_\infty
\log\!\left(
1
-
\textstyle\frac{1}{2}
\symDistanceMeasure
+
\textstyle\frac{1}{4}
\symDistanceMeasure^2
\textstyle\frac{
\symPoissonRate_\symSetASuffix
\symPoissonRate_\symSetBSuffix
}{\left(
\symPoissonRate_\symSetASuffix+
\symPoissonRate_\symSetBSuffix
\right)^2}
\right)
\end{equation}
where $\symDistanceMeasure = \frac{\symPoissonRate_\symSetASuffix
+\symPoissonRate_\symSetBSuffix}
{\symPoissonRate_\symSetASuffix+\symPoissonRate_\symSetBSuffix+
\symPoissonRate_\symSetXSuffix}$
 is the Jaccard distance. Since $
\frac{
\symPoissonRate_\symSetASuffix
\symPoissonRate_\symSetBSuffix
}{\left(
\symPoissonRate_\symSetASuffix+
\symPoissonRate_\symSetBSuffix
\right)^2}
$ is always in the range $[0,\frac{1}{4}]$, the probability for equal register values can be bounded by
\begin{equation}
1
+
2\symAlpha_\infty
\log\!\left(
1
-
\textstyle\frac{1}{2}
\symDistanceMeasure
\right)
\lesssim
\symProbability(\symRegValVariate_1=\symRegValVariate_2)
\lesssim
1
+
2\symAlpha_\infty
\log\!\left(
1
-
\textstyle\frac{1}{2}
\symDistanceMeasure
+
\textstyle\frac{1}{16}
\symDistanceMeasure^2
\right).
\end{equation}
As shown in \cref{fig:equal_register_probability} the bounds are very close, especially for small Jaccard distances, where the probability can be well approximated by
\begin{equation}
\symProbability(\symRegValVariate_1=\symRegValVariate_2)
\approx
1-
\symAlpha_\infty\symDistanceMeasure.
\end{equation}
This dependency on the Jaccard distance is very similar to that of minwise hashing \eqref{equ:minwise_hahsing_probability}, which makes the
HyperLogLog algorithm an interesting candidate for locality-sensitive hashing with respect to the Jaccard similarity. The memory efficiency of different hashing approaches can be measured by a storage factor that is the variance of the distance estimator multiplied by the number of bits used for storing the hash signature \cite{Li2011}. For a hash algorithm that maps two different sets to the same $\symNumBitsMinwiseHashing$-bit hash value with probability $1-\symA\symDistanceMeasure$ with some constant $\symA\in(0,1]$, the storage factor is given by $\symNumBitsMinwiseHashing\symDistanceMeasure\left(\frac{1}{\symA}-\symDistanceMeasure\right)$. 
Particularly, for the case $D=0.3$, this gives \num{6.72} for conventional minwise hashing with 32-bit hash values ($\symNumBitsMinwiseHashing=32$, $\symA\approx 1$), \num{0.51} for 1-bit minwise hashing ($\symNumBitsMinwiseHashing=1$, $\symA=\frac{1}{2}$), and  \num{1.96} for the HyperLogLog algorithm with 6-bit registers ($\symNumBitsMinwiseHashing=6$, $\symA=\symAlpha_\infty$). This means that the HyperLogLog algorithm is not as memory-efficient as 1-bit minwise hashing, but significantly better than 32-bit minwise hashing, when estimating the Jaccard distance by just counting equal registers. However, in contrast to 1-bit minwise hashing the HyperLogLog sketch contains additional information which allows estimating the cardinality or merging two HyperLogLog hash signatures. Furthermore, the method described in \cref{sec:cardinality_estimation_set_intersections} would allow a more accurate estimation of the Jaccard distance by using the estimates for intersection and union sizes. This could be used for additional more precise filtering when searching for similar items.

Since HyperLogLog sketches can be efficiently constructed, because only a single hash function evaluation is needed for each item, the preprocessing step would be very fast. In contrast, preprocessing is very costly for minwise hashing, because of the many required permutations \cite{Li2011}. To overcome this problem, one permutation hashing was proposed \cite{Li2012} which keeps for a predefined number of bins the minimum of all values that are mapped to this bin. Actually, this is very similar to the HyperLogLog algorithm. The bins correspond to HyperLogLog registers which keep the first one-bit position of the minimum value instead of the minimum value itself. This close connection was also pointed out in \cite{Cohen2014}, where both methods have been considered as variants of $k$-partition minwise hashing. The only difference is that the HyperLogLog algorithm uses base-2 ranks while one permutation hashing uses full ranks. 

\section{Conclusion}
\label{sec:conclusion}
We have presented new algorithms for the estimation of cardinalities from HyperLogLog sketches based on the Poisson approximation. For the estimation from a single sketch, we have developed two different algorithms that are inherently unbiased over the full cardinality range without dependence on empirically determined bias correction data. The first uses the original estimator extended by theoretically motivated correction terms. Due to its simplicity we believe that it has the potential to become the standard textbook cardinality estimation algorithm for HyperLogLog sketches. The second is based on the maximum likelihood method and solves the corresponding optimization problem using the secant method. The maximum likelihood method was also successfully applied to the estimation of set operation result sizes where the operands are represented by HyperLogLog sketches. The new approach improves the cardinality estimates of set intersections, relative complements, as well as unions significantly when compared to the conventional approach using the inclusion-exclusion principle.
\appendix

\section{Analysis of $\symPowerSeriesFunc(\symX)$}
\label{app:analysis_xi}
The Fourier series of the periodic function
\begin{equation}
\label{equ:def_xi}
\symPowerSeriesFunc(\symX):= \log(2) \sum_{\symRegVal = -\infty}^\infty
2^{\symRegVal+\symX}
e^{-2^{\symRegVal+\symX}},
\end{equation}
which has a period equal to 1, is
\begin{equation}
\symPowerSeriesFunc(\symX) = \frac{\symA_0}{2} + \Re\!\left(\sum_{\symIndexL=1}^{\infty}
\symA_\symIndexL e^{2\pi\symImaginary\symIndexL\symX}\right)
\end{equation}
with coefficients
\begin{multline}
\symA_\symIndexL = 2\int_0^1 \symPowerSeriesFunc(\symX) e^{-2\pi\symImaginary\symIndexL\symX}d\symX
=
2\log(2)
\int_0^1
\sum_{\symRegVal = -\infty}^\infty
2^{\symRegVal+\symX}
e^{-2^{\symRegVal+\symX}}
e^{-2\pi\symImaginary\symIndexL\symX}
d\symX
=
\\
=
2\log(2)
\sum_{\symRegVal = -\infty}^\infty
\int_{\symRegVal}^{\symRegVal+1}
2^{\symX}
e^{-2^{\symX}}
e^{-2\pi\symImaginary\symIndexL\symX}
d\symX
=
2\log(2) \int_{-\infty}^{\infty} 2^\symX
e^{-2^\symX}
e^{-2\pi\symImaginary\symIndexL\symX}d\symX.
\end{multline}
The variable transformation $\symY = 2^\symX$ yields
\begin{equation}
\symA_\symIndexL
=
2\int_{0}^{\infty} e^{-\symY}
\symY^{-\frac{2\pi\symImaginary\symIndexL}{\log(2)}}d\symY
= 
2\Gamma\!\left(1-\frac{2\pi\symImaginary\symIndexL}{\log(2)}\right)
\end{equation}
where $\Gamma$ denotes the gamma function. Obviously, the constant term in the Fourier series is equal to $1$, because $\symA_0=2$. 

To investigate the maximum deviation of $\symPowerSeriesFunc(\symX)$ from this value we consider all further coefficients $\symA_\symIndexL$ with $\symIndexL\geq 1$. Using the identity $\left|\Gamma(1+\symImaginary\symX)\right| = 
\sqrt{\frac{\pi\symX}{\sinh(\pi\symX)}}
$
we are able to write for the absolute values of the coefficients
\begin{equation}
|\symA_\symIndexL| = 
2\sqrt{\frac{\symB\symIndexL}{\sinh(\symB\symIndexL)}}
\quad
\text{with}
\
\symB:=\frac{2\pi^2}{\log 2}.
\end{equation}
In particular, the amplitude of the first harmonic is 
$|\symA_1|
\approx
\num{9.884e-6}$. Clearly, the maximum deviation of $\symPowerSeriesFunc(\symX)-1$ from the first harmonic must be smaller than $\sum_{\symIndexL=2}^{\infty} 
|\symA_\symIndexL|$. The ratio of subsequent coefficients is given by
\begin{equation}
\frac{|\symA_{\symIndexL+1}|}{|\symA_{\symIndexL}|}
=
\sqrt{\frac{\symIndexL+1}{\symIndexL}}
\sqrt{\frac{
\sinh(\symB \symIndexL)
}{
\sinh(\symB (\symIndexL+1))
}}
=
\sqrt{\frac{\symIndexL+1}{\symIndexL}}
\sqrt{\frac{
1
}{
\cosh(\symB)
+
\frac{
\sinh(\symB)
}
{
\tanh(\symB \symIndexL)
}
}}.
\end{equation}
For $\symIndexL\geq 2$ we have $\sqrt{\frac{\symIndexL+1}{\symIndexL}}\leq \sqrt{\frac{3}{2}}$. Together with $\tanh(x)\leq 1$ we obtain
\begin{equation}
\frac{|\symA_{\symIndexL+1}|}{
|\symA_{\symIndexL}|
}
\leq
\sqrt{\frac{3}{2}}
\sqrt{\frac{
1
}{
\cosh(\symB)
+
\sinh(\symB)
}}
\\
=
\sqrt{\frac{
3
}{
2 e^\symB
}}
\end{equation}
which leads to $|\symA_{\symIndexL}| \leq 
|\symA_{2}|
\left(\sqrt{\frac{
3
}{
2 e^\symB
}}\right)^{\symIndexL-2}$ for $\symIndexL\geq 2$ and further to
\begin{equation}
\sum_{\symIndexL=2}^{\infty} 
|\symA_\symIndexL|
\leq
|\symA_2|
\sum_{\symIndexL=0}^{\infty} 
\left(
\sqrt{\frac{
3
}{
2e^\symB
}}
\right)^\symIndexL
=
2\sqrt{
\frac{2\symB}{\sinh(2\symB)}}
\frac{1}{1-
\sqrt{\frac{
3
}{
2 e^\symB
}}
}
\approx
\num{9.154e-12}.
\end{equation}
As a consequence, the maximum deviation of $\symPowerSeriesFunc(\symX)$ from 1 is bounded by
\begin{equation}
\num{9.884e-6}
\leq
|\symA_1|
-
\sum_{\symIndexL=2}^{\infty} 
|\symA_\symIndexL|
\leq
\max_{\symX}(|\symPowerSeriesFunc(\symX)-1|) 
\leq 
|\symA_1|
+
\sum_{\symIndexL=2}^{\infty} 
|\symA_\symIndexL|
\leq 
\num{9.885e-6}.
\end{equation}

An interesting approximation formula can be derived from $\symPowerSeriesFunc(\symX) \approx 1$ by integrating on both sides: 
\begin{equation}
\label{equ:approx_formula}
\sum_{\symRegVal = -\infty}^\infty
e^{-2^{\symRegVal+\symY}} - e^{-2^{\symRegVal+\symX}} \approx \symX - \symY.
\end{equation}

\section{Numerical stability of recursion formula for $\symHelper(\symX)$}
\label{app:helper_stable}
In order to investigate the error propagation of a single recursion step using \eqref{equ:helper_recursion1} we define $\symHelper_1 := \symHelper(2\symX)$ and $\symHelper_2 := \symHelper(4\symX)$. The recursion formula simplifies to
\begin{equation}
\label{equ:h2}
\symHelper_2 = \frac{\symX+\symHelper_1(1-\symHelper_1)}{\symX+(1-\symHelper_1)}.
\end{equation}
If $\symHelper_1$ is approximated by $\tilde{\symHelper}_1 = \symHelper_1\left(1+\symError_1\right)$ with relative error $\symError_1$, the recursion formula will give an approximation for $\symHelper_2$
\begin{equation}
\label{equ:h2_approx}
\tilde{\symHelper}_2 = 
\frac{\symX+\tilde{\symHelper}_1(1-\tilde{\symHelper}_1)}{\symX+(1-\tilde{\symHelper}_1)}.
\end{equation}
The corresponding relative error $\symError_2$ is given by
\begin{equation}
\label{equ:h2_relative_error}
\symError_2 = \frac{\tilde{\symHelper}_2}{\symHelper_2}-1.
\end{equation}
Combination of \eqref{equ:h2}, \eqref{equ:h2_approx}, and \eqref{equ:h2_relative_error} yields for its absolute value
\begin{equation}
\left|\symError_2\right|
=
\left|\symError_1\right|
\frac{
\left|
\frac{\symHelper_1\left(1-2\symHelper_1\right)}{\symX+\symHelper_1\left(1-\symHelper_1\right)}
+
\frac{\symHelper_1}{\symX+1-\symHelper_1}
-
\symError_1
\frac{\symHelper_1^2}{\symX+\symHelper_1\left(1-\symHelper_1\right)}
\right|
}{
\left|
1-
\symError_1
\frac{\symHelper_1}{\symX+1-\symHelper_1}
\right|
}
\end{equation}
and the triangle inequality leads to
\begin{equation}
\left|\symError_2\right|
\leq
\left|\symError_1\right|
\frac{
\left|
\frac{\symHelper_1\left(1-2\symHelper_1\right)}{\symX+\symHelper_1\left(1-\symHelper_1\right)}
+
\frac{\symHelper_1}{\symX+1-\symHelper_1}
\right|
+
\left|
\symError_1
\right|
\frac{\symHelper_1^2}{\symX+\symHelper_1\left(1-\symHelper_1\right)}
}{
\left|
1-
\symError_1
\frac{\symHelper_1}{\symX+1-\symHelper_1}
\right|
}.
\end{equation}

By numerical means it is easy to show that the inequalities $\left|
\frac{\symHelper_1\left(1-2\symHelper_1\right)}{\symX+\symHelper_1\left(1-\symHelper_1\right)}
+
\frac{\symHelper_1}{\symX+1-\symHelper_1}
\right|\leq 0.517
$,
$\frac{\symHelper_1^2}{\symX+\symHelper_1\left(1-\symHelper_1\right)}
\leq 0.436$,
and
$\frac{\symHelper_1}{\symX+1-\symHelper_1}\leq 0.529$ hold for all $\symX\geq0$. Therefore, if we additionally assume, for example, $\left|\symError_1\right|\leq0.1$, we get
\begin{equation}
\left|\symError_2\right|
\leq
\left|\symError_1\right|
\frac{0.517 + 0.1\cdot 0.436}{1-0.1\cdot 0.529}
\leq
\left|\symError_1\right|
\cdot
0.592,
\end{equation}
which means that the relative error is decreasing in each recursion step and the recursive calculation of $\symHelper$ is numerically stable.

\section{Error caused by approximation of $\symHelper(\symX)$}
\label{app:error_approx}
According to \eqref{equ:func} the exact estimate $\symXEstimate$ fulfills 
\begin{equation}
\label{equ:max_likelihood_2}
\symXEstimate\sum_{\symRegVal=0}^\symRegRange \frac{\symCountVariate_\symRegVal}{2^\symRegVal}+
\sum_{\symRegVal=1}^\symRegRange \symCountVariate_\symRegVal\symHelper\!\left(\frac{\symXEstimate}{2^\symRegVal}\right)
+
\symCountVariate_{\symRegRange+1}\symHelper\!\left(\frac{\symXEstimate}{2^\symRegRange}\right)
=
\symNumReg-\symCountVariate_0.
\end{equation}
If $\symHelper$ is not calculated exactly but approximated by $\symHelperApprox$ with maximum relative error $\symError_\symHelper\ll 1$
\begin{equation}
\label{equ:abs_error_helper}
\left|\symHelperApprox(\symX) - \symHelper(\symX) \right|  \leq \symError_\symHelper\symHelper(\symX)
\end{equation}
the solution of the equation will be off by some relative error $\symError_\symX$:
\begin{equation}
\label{equ:ml_equation}
\symXEstimate\left(1+\symError_\symX\right)\sum_{\symRegVal=0}^\symRegRange \frac{\symCountVariate_\symRegVal}{2^\symRegVal}+
\sum_{\symRegVal=1}^\symRegRange \symCountVariate_\symRegVal\symHelperApprox\!\left(\frac{\symXEstimate\left(1+\symError_\symX\right)}{2^\symRegVal}\right)
+
\symCountVariate_{\symRegRange+1}
\symHelperApprox\!\left(
\frac{\symXEstimate\left(1+\symError_\symX\right)}{2^\symRegRange}\right)
=
\symNumReg-\symCountVariate_0.
\end{equation}
Due to \eqref{equ:abs_error_helper} there exists some $\symAlpha \in [-\symError_\symHelper, \symError_\symHelper]$ for which
\begin{multline}
\symXEstimate\left(1+\symError_\symX\right)\sum_{\symRegVal=0}^\symRegRange \frac{\symCountVariate_\symRegVal}{2^\symRegVal}
+
\left(1+\symAlpha\right)
\sum_{\symRegVal=1}^\symRegRange \symCountVariate_\symRegVal
\symHelper\!\left(\frac{\symXEstimate\left(1+\symError_\symX\right)
}{2^\symRegVal}\right)
+
\\
+
\left(1+\symAlpha\right)
\symCountVariate_{\symRegRange+1}
\symHelper\!\left(
\frac{\symXEstimate\left(1+\symError_\symX\right)}{2^\symRegRange}\right)
=
\symNumReg-\symCountVariate_0.
\end{multline}
Since $\symHelper'(\symX)\in[0,0.5]$ for $\symX \geq 0$, there exists a $\symBeta\in[0,0.5]$ for which
\begin{multline}
\label{equ:appendix3}
\symXEstimate\left(1+\symError_\symX\right)\sum_{\symRegVal=0}^\symRegRange \frac{\symCountVariate_\symRegVal}{2^\symRegVal}+
\left(
1+\symAlpha
\right)
\left(
\sum_{\symRegVal=1}^\symRegRange 
\symCountVariate_\symRegVal
\symHelper\!\left(\frac{\symXEstimate
}{2^\symRegVal}\right)
+
\frac{\symCountVariate_\symRegVal}{2^\symRegVal}
\symXEstimate\symError_\symX
\symBeta
\right)
+
\\
+
\left(
1+\symAlpha
\right)
\left(
\symCountVariate_{\symRegRange+1}
\symHelper\!\left(
\frac{\symXEstimate}{2^\symRegRange}\right)
+
\frac{\symCountVariate_{\symRegRange+1}}{2^\symRegRange}
\symXEstimate\symError_\symX\symBeta
\right)
=
\symNumReg-\symCountVariate_0.
\end{multline}
Subtracting \eqref{equ:max_likelihood_2} multiplied by $\left(1+\symAlpha\right)$ from \eqref{equ:appendix3} and resolving 
$\symError_\symX$ gives
\begin{equation}
\symError_\symX
=
\symAlpha
\frac{
\symXEstimate
\sum_{\symRegVal=0}^\symRegRange \frac{\symCountVariate_\symRegVal}{2^\symRegVal}
-\left(\symNumReg-\symCountVariate_0\right)
}
{
\symXEstimate
\left(
\sum_{\symRegVal=0}^\symRegRange \frac{\symCountVariate_\symRegVal}{2^\symRegVal}
+
\left(1+\symAlpha\right)
\symBeta
\left(
\sum_{\symRegVal=1}^\symRegRange 
\frac{\symCountVariate_\symRegVal
}{2^\symRegVal}
+
\frac{\symCountVariate_{\symRegRange+1}
}{2^\symRegRange}
\right)
\right)
}.
\end{equation}
Using $\left|\symAlpha\right|\leq\symError_\symHelper$, $\symBeta\geq0$, and \eqref{equ:weak_upper_bound} the absolute value of the relative error
can be bounded by
\begin{equation}
\left|\symError_\symX\right| 
\leq
\left|\symError_\symHelper\right| 
\frac{
\left(\symNumReg-\symCountVariate_0\right)-
\symXEstimate
\sum_{\symRegVal=0}^\symRegRange \frac{\symCountVariate_\symRegVal}{2^\symRegVal}
}
{
\symXEstimate
\sum_{\symRegVal=0}^\symRegRange \frac{\symCountVariate_\symRegVal}{2^\symRegVal}
}.
\end{equation}
Furthermore, using \eqref{equ:weak_lower_bound} we finally get
\begin{equation}
\left|\symError_\symX\right| 
\leq
\frac{\left|\symError_\symHelper\right|}{2}
\frac{
\sum_{\symRegVal=1}^\symRegRange \frac{\symCountVariate_\symRegVal}{2^\symRegVal}
+
\frac{\symCountVariate_{\symRegRange+1}
}{2^\symRegRange}
}
{
\sum_{\symRegVal=0}^\symRegRange \frac{\symCountVariate_\symRegVal}{2^\symRegVal}
}
\leq
\frac{\left|\symError_\symHelper\right|}{2}
\left(
1
+
\frac{
\frac{\symCountVariate_{\symRegRange+1}
}{2^\symRegRange}
}
{
\sum_{\symRegVal=0}^\symRegRange \frac{\symCountVariate_\symRegVal}{2^\symRegVal}
}
\right)
\leq
\frac{\left|\symError_\symHelper\right|}{2}
\left(
1
+
\frac{
\symCountVariate_{\symRegRange+1}
}
{
\symNumReg-\symCountVariate_{\symRegRange+1}
}
\right).
\end{equation}
Hence, as long as most registers are not saturated ($\symCountVariate_{\symRegRange+1}\ll\symNumReg$), the relative error $\symError_\symX$ of the calculated estimate using the approximation $\symHelperApprox(\symX)$ for $\symHelper(\symX)$ has the same order of magnitude as $\symError_\symHelper$.

\bibliographystyle{unsrt}
\bibliography{bibliography.bib}

\begin{thebibliography}{10}

\bibitem{Alon1999}
Noga Alon, Yossi Matias, and Mario Szegedy.
\newblock The space complexity of approximating the frequency moments.
\newblock {\em Journal of Computer and System Sciences}, 58(1):137 -- 147,
  1999.

\bibitem{Metwally2008}
Ahmed Metwally, Divyakant Agrawal, and Amr~El Abbadi.
\newblock Why go logarithmic if we can go linear?: Towards effective distinct
  counting of search traffic.
\newblock In {\em Proceedings of the 11th International Conference on Extending
  Database Technology: Advances in Database Technology}, pages 618--629,
  Nantes, France, March 2008.

\bibitem{Ting2014}
Daniel Ting.
\newblock Streamed approximate counting of distinct elements: Beating optimal
  batch methods.
\newblock In {\em Proceedings of the 20th ACM SIGKDD International Conference
  on Knowledge Discovery and Data Mining}, pages 442--451, New York, NY, USA,
  August 2014.

\bibitem{Kane2010}
Daniel~M. Kane, Jelani Nelson, and David~P. Woodruff.
\newblock An optimal algorithm for the distinct elements problem.
\newblock In {\em Proceedings of the 29th ACM SIGMOD-SIGACT-SIGART Symposium on
  Principles of Database Systems}, pages 41--52, Indianapolis, IN, USA, June
  2010.

\bibitem{Indyk2003}
Piotr Indyk and David Woodruff.
\newblock Tight lower bounds for the distinct elements problem.
\newblock In {\em Proceedings of the 44th Annual IEEE Symposium on Foundations
  of Computer Science}, pages 283--288, Cambridge, MA, USA, October 2003.

\bibitem{Flajolet2007}
Philippe Flajolet, Éric Fusy, Olivier Gandouet, and Frédéric Meunier.
\newblock Hyperloglog: The analysis of a near-optimal cardinality estimation
  algorithm.
\newblock In {\em Proceedings of the 13th Conference on Analysis of
  Algorithms}, pages 127--146, Juan des Pins, France, June 2007.

\bibitem{Heule2013}
Stefan Heule, Marc Nunkesser, and Alexander Hall.
\newblock Hyperloglog in practice: Algorithmic engineering of a state of the
  art cardinality estimation algorithm.
\newblock In {\em Proceedings of the 16th International Conference on Extending
  Database Technology}, pages 683--692, Genoa, Italy, March 2013.

\bibitem{Rhodes2015}
Lee Rhodes.
\newblock System and method for enhanced accuracy cardinality estimation,
  September~24 2015.
\newblock US Patent 20,150,269,178.

\bibitem{Sanfilippo2014}
Salvatore Sanfilippo.
\newblock Redis new data structure: The {HyperLogLog}.
\newblock {\url{http://antirez.com/news/75}}, 2014.

\bibitem{Cohen2014}
Edith Cohen.
\newblock All-distances sketches, revisited: {HIP} estimators for massive
  graphs analysis.
\newblock In {\em Proceedings of the 33rd ACM SIGMOD-SIGACT-SIGART Symposium on
  Principles of Database Systems}, pages 88--99, New York, NY, USA, 2014.

\bibitem{Chen2011}
Aiyou Chen, Jin Cao, Larry Shepp, and Tuan Nguyen.
\newblock Distinct counting with a self-learning bitmap.
\newblock {\em Journal of the American Statistical Association},
  106(495):879--890, 2011.

\bibitem{Beyer2007}
Kevin Beyer, Peter~J. Haas, Berthold Reinwald, Yannis Sismanis, and Rainer
  Gemulla.
\newblock On synopses for distinct-value estimation under multiset operations.
\newblock In {\em Proceedings of the 26th ACM SIGMOD International Conference
  on Management of Data}, pages 199--210, Beijing, China, June 2007.

\bibitem{Dasgupta2015}
Anirban Dasgupta, Kevin Lang, Lee Rhodes, and Justin Thaler.
\newblock A framework for estimating stream expression cardinalities.
\newblock {\em arXiv preprint arXiv:1510.01455}, 2015.

\bibitem{Pascoe2013}
Andrew Pascoe.
\newblock Hyperloglog and {M}in{H}ash - {A} union for intersections.
\newblock {\url{http://tech.adroll.com/media/hllminhash.pdf}}, 2013.

\bibitem{Cohen2016}
Reuven Cohen, Liran Katzir, and Aviv Yehezkel.
\newblock A minimal variance estimator for the cardinality of big data set
  intersection.
\newblock {\em arXiv preprint arXiv:1606.00996}, 2016.

\bibitem{Flajolet1985}
Philippe Flajolet and G.~Nigel Martin.
\newblock Probabilistic counting algorithms for data base applications.
\newblock {\em Journal of Computer and System Sciences}, 31(2):182 -- 209,
  1985.

\bibitem{Whang1990}
Kyu-Young Whang, Brad~T. Vander-Zanden, and Howard~M. Taylor.
\newblock A linear-time probabilistic counting algorithm for database
  applications.
\newblock {\em ACM Transactions on Database Systems}, 15(2):208--229, 1990.

\bibitem{Jacquet1998}
Philippe Jacquet and Wojciech Szpankowski.
\newblock Analytical depoissonization and its applications.
\newblock {\em Theoretical Computer Science}, 201(1):1--62, 1998.

\bibitem{Durand2003}
Marianne Durand and Philippe Flajolet.
\newblock Loglog counting of large cardinalities.
\newblock In {\em Proceedings of the 11th Annual European Symposium on
  Algorithms}, pages 605--617, Budapest, Hungary, September 2003.

\bibitem{Casella2002}
George Casella and Roger~L. Berger.
\newblock {\em Statistical inference}.
\newblock Duxbury, Pacific Grove, CA, USA, 2nd edition, 2002.

\bibitem{Clifford2012}
Peter Clifford and Ioana~A. Cosma.
\newblock A statistical analysis of probabilistic counting algorithms.
\newblock {\em Scandinavian Journal of Statistics}, 39(1):1--14, 2012.

\bibitem{Ting2016}
Daniel Ting.
\newblock Towards optimal cardinality estimation of unions and intersections
  with sketches.
\newblock In {\em Proceedings of the 22nd ACM SIGKDD Conference on Knowledge
  and Data Mining}, pages 1195--1204, San Francisco, CA, USA, August 2016.

\bibitem{Press2007}
William~H. Press.
\newblock {\em Numerical recipes 3rd edition: The art of scientific computing}.
\newblock Cambridge University Press, 2007.

\bibitem{King2009}
Davis~E. King.
\newblock Dlib-ml: A machine learning toolkit.
\newblock {\em Journal of Machine Learning Research}, 10:1755--1758, 2009.

\bibitem{Leskovec2014}
Jure Leskovec, Anand Rajaraman, and Jeffrey~David Ullman.
\newblock {\em Mining of massive datasets}.
\newblock Cambridge University Press, 2014.

\bibitem{Wang2014}
Jingdong Wang, Heng~Tao Shen, Jingkuan Song, and Jianqiu Ji.
\newblock Hashing for similarity search: A survey.
\newblock {\em arXiv preprint arXiv:1408.2927}, 2014.

\bibitem{Charikar2002}
Moses~S. Charikar.
\newblock Similarity estimation techniques from rounding algorithms.
\newblock In {\em Proceedings of the 34th Annual ACM Symposium on Theory of
  Computing}, pages 380--388, Montreal, Canada, May 2002.

\bibitem{Broder1997}
Andrei~Z. Broder.
\newblock On the resemblance and containment of documents.
\newblock In {\em Proceedings of the Compression and Complexity of Sequences},
  pages 21--29, Positano, Italy, June 1997.

\bibitem{Li2011}
Ping Li and Arnd~Christian K\"{o}nig.
\newblock Theory and applications of b-bit minwise hashing.
\newblock {\em Communications of the ACM}, 54(8):101--109, 2011.

\bibitem{Li2012}
Ping Li, Art Owen, and Cun hui Zhang.
\newblock One permutation hashing.
\newblock {\em Advances in Neural Information Processing Systems},
  25:3113--3121, 2012.

\end{thebibliography}

\end{document}